\documentclass[prd,preprint,eqsecnum,nofootinbib,amsmath,amssymb,
               tightenlines,dvips,superscriptaddress]{revtex4}
\usepackage{graphicx}
\usepackage{bm}%	bold math
\usepackage{amsfonts}
\usepackage{amssymb}

\usepackage{dsfont}
\def\openone{\mathds{1}}

\def\b{{\bm b}}
\def\x{{\bm x}}
\def\p{{\bm p}}
\def\q{{\bm q}}
\def\k{{\bm k}}
\def\A{{\bm A}}
\def\B{{\bm B}}
\def\C{{\bm C}}
\def\P{{\bm P}}
\def\bkappa{{\bm\kappa}}
\def\angJ{{\bm{\mathcal J}}}
\def\angS{{\bm{\mathcal S}}}
\def\bcalB{{\bm{\mathcal B}}}
\def\bcalP{{\bm{\mathcal P}}}
\def\bcalT{{\bm{\mathcal T}}}
\def\uOmega{\underline{\Omega}}
\def\Ybar{\overline{Y}}

\def\Nc{N_{\rm c}}
\def\CA{C_{\rm A}}
\def\mD{m_{\rm D}}
\def\alphas{\alpha_{\rm s}}
\def\Re{\operatorname{Re}}
\def\Im{\operatorname{Im}}
\def\tr{\operatorname{tr}}
\def\sgn{\operatorname{sgn}}
\def\grad{{\bm\nabla}}

\def\dlangle{\langle\!\langle}
\def\drangle{\rangle\!\rangle}
\def\Bigdlangle{\Big\langle\!\!\Big\langle}
\def\Bigdrangle{\Big\rangle\!\!\Big\rangle}
\def\ssum{{\textstyle\sum}}

\def\ix{{\rm i}}
\def\fx{{\rm f}}
\def\xx{{\rm x}}
\def\xbx{{\bar{\rm x}}}
\def\yx{{\rm y}}
\def\ybx{{\bar{\rm y}}}
\def\zx{{\rm z}}
\def\Bx{\xbx}
\def\Ax{\ybx}
\def\bx{\yx}
\def\ax{\xx}

\def\Hilbert{{\mathbb H}}
\def\mpart{{\cal M}}
\def\tildeV{{\widetilde V}}

\begin {document}

%%%%%%%%%%%%%%%%%%%%%%%%%%%%%%%%%%%%%%%%%%%%%%%%%%%%%%%%%%%%%%%%%%%%%%%%%%%%%%%

%%%%%%%%%%%%%%%%%%%%%%%%%%%%%%%%%%%%%%%%%%%%%%%%%%%%%%%%%%%%%%%%%%%%%%%%%%%%%%%

\title
    {
      The LPM effect in sequential bremsstrahlung%
%\footnote{
%      This preprint incorporates corrections to the corresponding
%      JHEP publication [JHEP 04 (2015) 070],
%      for which a separate erratum will be forthcoming.
%}
    }

\author{Peter Arnold}
\affiliation
    {%
    Department of Physics,
    University of Virginia,
    Charlottesville, Virginia 22904-4714, USA
    \medskip
    }%
\author{Shahin Iqbal}
\affiliation
    {%
    Department of Physics,
    University of Virginia,
    Charlottesville, Virginia 22904-4714, USA
    \medskip
    }%
\affiliation
    {%
    National Centre for Physics, \\
    Quaid-i-Azam University Campus,
    Islamabad, 45320 Pakistan
    \medskip
    }%

\date {\today}

\begin {abstract}%
{%
   The splitting processes of bremsstrahlung and pair production in a medium
   are coherent over large distances in the very high energy limit,
   which leads to a suppression known as the Landau-Pomeranchuk-Migdal
   (LPM) effect.  We analyze the case when the coherence
   lengths of two consecutive splitting processes overlap, which is
   important for understanding corrections to standard treatments
   of the LPM effect in QCD.  Previous authors have analyzed this
   problem in the case of overlapping double bremsstrahlung where at least one
   of the bremsstrahlung gluons is soft.
   Here we show how to generalize to include
   the case where both splittings are hard.  A number of techniques
   must be developed, and so in this paper
   we simplify
   by (i) restricting attention to a subset of the interference
   effects, which we call the ``crossed'' diagrams, and (ii) working in
   the large-$\Nc$ limit.  We first develop some general formulas that
   could in principle be implemented numerically (with substantial difficulty).
   To make more analytic progress, we then focus on the
   case of a thick, homogeneous
   medium and make the multiple scattering
   approximation (also known as the $\hat q$ or harmonic approximation)
   appropriate at high energy.  We show that
   the differential rate $d\Gamma/dx\,dy$
   for overlapping double bremsstrahlung of gluons with momentum
   fractions $x$ and
   $y$ can then be reduced to the calculation of a
   1-dimensional integral, which we perform numerically.
   [Though this paper is unfortunately long, our introduction is
   enough for getting the gist of the method.]
}%
\end {abstract}

\maketitle
\thispagestyle {empty}

{\def\boldmath{}\tableofcontents}
\newpage

%%%%%%%%%%%%%%%%%%%%%%%%%%%%%%%%%%%%%%%%%%%%%%%%%%%%%%%%%%%%%%%%%%%%%%%%%%%%%%%

\section{Introduction}
\label{sec:intro}

\begin {figure}[b]
\begin {center}
  \includegraphics[scale=0.5]{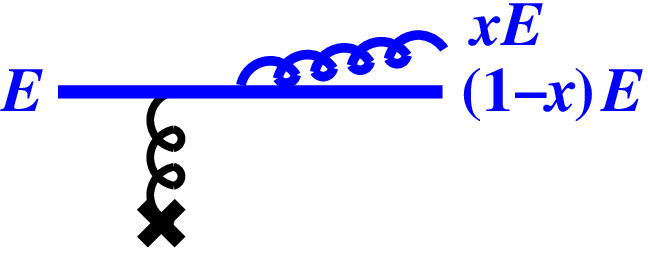}
  \caption{
     \label{fig:brem}
     High-energy bremsstrahlung (blue) during a collision (black)
     with a particle from
     the medium.  The curly line ending in a cross represents the
     electromagnetic or gluonic fields in the medium created by sources,
     such as by a nucleus or a passing thermal parton.
  }
\end {center}

\end {figure}

\begin {figure}[t]
\begin {center}
  \includegraphics[scale=0.35]{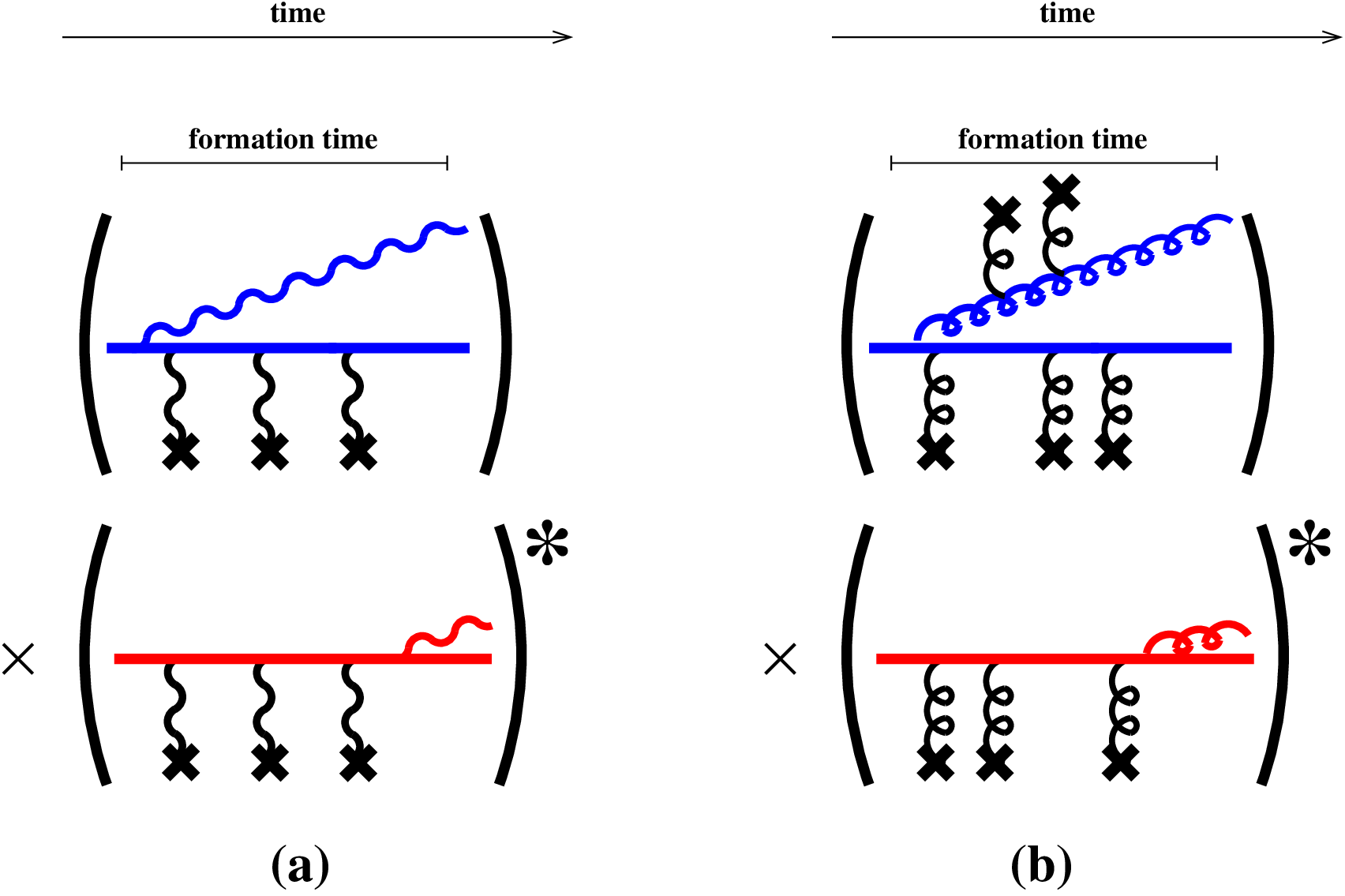}
  \caption{
     \label{fig:lpm1}
     Examples of interference terms contributing to the LPM effect
     in (a) QED and (b) QCD.  For consistency with future figures,
     we have used blue for high-energy particles in the amplitude and
     red for high-energy particles in the conjugate amplitude.
  }
\end {center}
\end {figure}

At high enough energy, particles passing through matter (cosmic
rays through the atmosphere, high-energy partons through a quark-gluon
plasma, electrons through an electromagnetic calorimeter) lose energy
primarily through splitting: hard bremsstrahlung or pair production which,
when repeated, produces a shower of lower energy particles.
Naively, one would calculate the rate $\Gamma$ for each such splitting
by (roughly speaking) computing the cross-section $\sigma$ for
splitting during a collision between the high energy particle and
a particle in the medium, as in the bremsstrahlung process depicted
in fig. \ref{fig:brem}, and then finding $\Gamma \sim n v \sigma$,
where $n$ is the density of things to scatter from and $v$ the
relative velocity.  The flaw in this argument, as known since
the 1950s, is that the duration of this splitting process, called
the formation time, grows with energy.  At high enough energy,
the formation time exceeds the mean free time between collisions
with the medium, and then consecutive scatterings from the medium
may no longer be treated as quantum mechanically independent for
the purpose of computing the splitting rate.  This is known
as the Landau-Pomeranchuk-Migdal (LPM) effect \cite{LP,Migdal},
which dramatically
suppresses the splitting rate at very high energy.  In the language
of Feynman diagrams, the LPM effect represents important interferences between
splitting before and after a sequence of elastic collisions with
the medium, such as shown in fig.\ \ref{fig:lpm1}a for QED and
fig.\ \ref{fig:lpm1}b for QCD.  The analysis of the LPM effect
in QCD was pioneered in the 1990s by Baier et al.\ \cite{BDMPS12,BDMPS3} and
Zakharov \cite{Zakharov}, known collectively as BDMPS-Z.


\begin {figure}[t]
\begin {center}
  \includegraphics[scale=0.5]{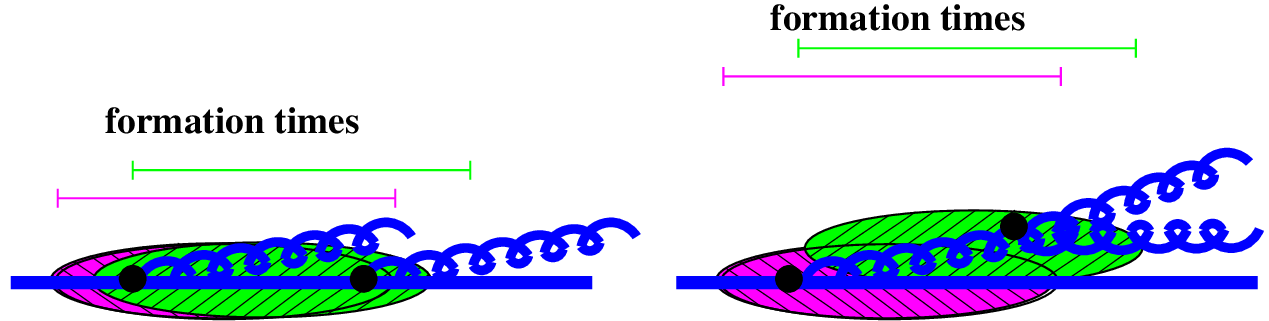}
  \caption{
     \label{fig:overlap}
     Two consecutive splittings that are close enough that their
     formation times overlap.  Each formation time region is
     depicted by a green or blue, hatched oval.
  }
\end {center}
\end {figure}

\begin {figure}[t]
\begin {center}
  \includegraphics[scale=0.5]{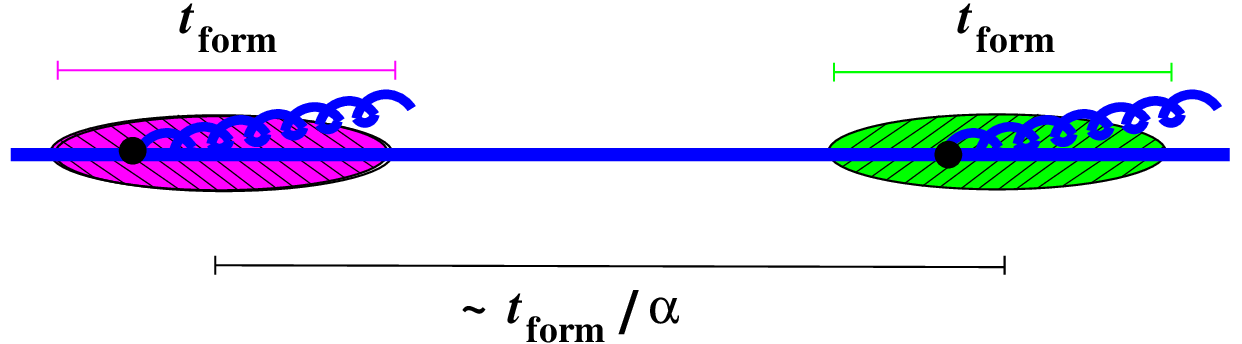}
  \caption{
     \label{fig:typical}
     The hierarchy of scales for typical consecutive splitting, if
     the relevant $\alpha$ is small and if
     one ignores logarithmic enhancements
     in QCD associated with one of the daughters becoming soft.
     (The cartoon in this picture assumes that the momenta of the
     two bremsstrahlung gauge bosons are parametrically similar.
     In QCD, formation times shrink as bremsstrahlung gluons become
     soft.)
  }
\end {center}
\end {figure}

A natural question that arises is whether consecutive splittings of
the high energy particle, and not merely consecutive collisions with
the medium, occur within the formation time.  That is, once we
use the LPM or BDMPS-Z formalism to compute the rate for a single
splitting, can we then treat consecutive splittings as independent and
so simply use the single-splitting rate in a Monte Carlo to compute
the development of the shower (which we need for answering detailed
questions about energy loss)?  Or is there instead a significant
contribution from processes where the formation times associated
with consecutive splittings overlap, as depicted in
fig.\ \ref{fig:overlap}?
Formally, the probability of a splitting is parametrically
of order $\alpha$ ($\alphas$ in QCD) in one formation time,
and so the formation time is one power of $\alpha$ smaller
than the typical time between consecutive splittings, as depicted
in fig.\ \ref{fig:typical}.  This means that the contribution of
overlapping formation times, as in fig.\ \ref{fig:overlap}, is
formally suppressed by one factor of $\alpha$.
However, in QCD applications, (i) this $\alphas$ is at best moderately
small at the scales relevant for the high-energy splitting process
in actual relativistic heavy ion collisions, and (ii) this factor
of $\alphas$ is accompanied by a potentially large double logarithm
\cite{Blaizot,Iancu,Wu}.  It is therefore important to analyze the
effect of overlapping formation times.

Note that in the last figure we
have drawn only the high-energy particles.  This will be our convention
throughout the rest of the paper: the collisions with
the medium are implicit, and the high-energy particle lines should
be understood as decorated with numerous interactions with the medium,
as in fig.\ \ref{fig:lpm1}.

Our goal in this paper is to develop formalism for a full computation of
the LPM effect (to include the treatment of overlapping formation times)
in the case of two consecutive splittings, such as
fig.\ \ref{fig:overlap}.  We will then implement this
formalism to compute results for double-splitting rates in QCD in certain
simplifying limits, to include the multiple-scattering approximation
(also known as the $\hat q$ or harmonic oscillator approximation), which is
appropriate for typical events at high energy when the medium is
much thicker than
the mean free path for collisions with the medium.

Previous authors \cite{Blaizot,Iancu,Wu}
have performed explicit calculations for QCD to leading-log
order in the limit of
$y \ll x \ll 1$, where $x$ and $y$ are the momentum fractions of two
of the final gluons.
[That is, if the initial parton energy is $E$,
the three daughters after the two splittings have energies
$xE$, $yE$, and $(1{-}x{-}y)E$.]  These results have interesting and
important consequences, which we will briefly mention later.
Wu \cite{Wu} has also developed formalism for
studying the somewhat more general case of $x,y \ll 1$ without
assuming $y \ll x$, but so far has only performed explicit calculations
when $y \ll x$.%
\footnote{
  There have also been some attempts to include
  a rough, heuristic treatment of overlapping formation times
  in the Monte Carlo
  generator JEWEL \cite{JEWEL}.
  We have not yet worked out its relationship to the results
  described here or in refs.\ \cite{Blaizot,Iancu,Wu}.
}
In this paper, we study the problem
of general $x$ and $y$, without requiring that either be small,
and we will carry out explicit calculations in that case.
In the context of the multiple-scattering (harmonic oscillator)
approximation, we will be able to go beyond leading logarithms
and develop methods for explicitly
computing the full result for double splitting.

% --------------------------------------------------------------------------

\subsection {What we compute}

In this, our first paper on the case of general $x$ and $y$, we will
simplify the discussion by focusing on the case of the large
$\Nc$ limit of QCD.  This will simplify treatment of the color dynamics
of high-energy partons (a simplification that is
not needed in the $y \ll x \ll 1$ limit considered by
previous authors).%
\footnote{
   Blaizot and Mehtar-Tani \cite{Blaizot} also discuss the simplification
   afforded by the large-$\Nc$ limit for moving the discussion beyond the
   soft-gluon limit.
}
For the sake of a minor, further simplification,
we will also focus on the case where the initial high-energy particle
is a gluon rather than a quark
(and so, in the large-$\Nc$ limit, all the daughter particles
after each splitting are gluons as well).

Though we will develop formal results that
are more general, our explicit calculations in this paper will be
specialized to the case of a thick, static, homogeneous
medium.  ``Thick'' means that the medium
is wide compared to the typical formation length.

\begin {figure}[t]
\begin {center}
  \includegraphics[scale=0.5]{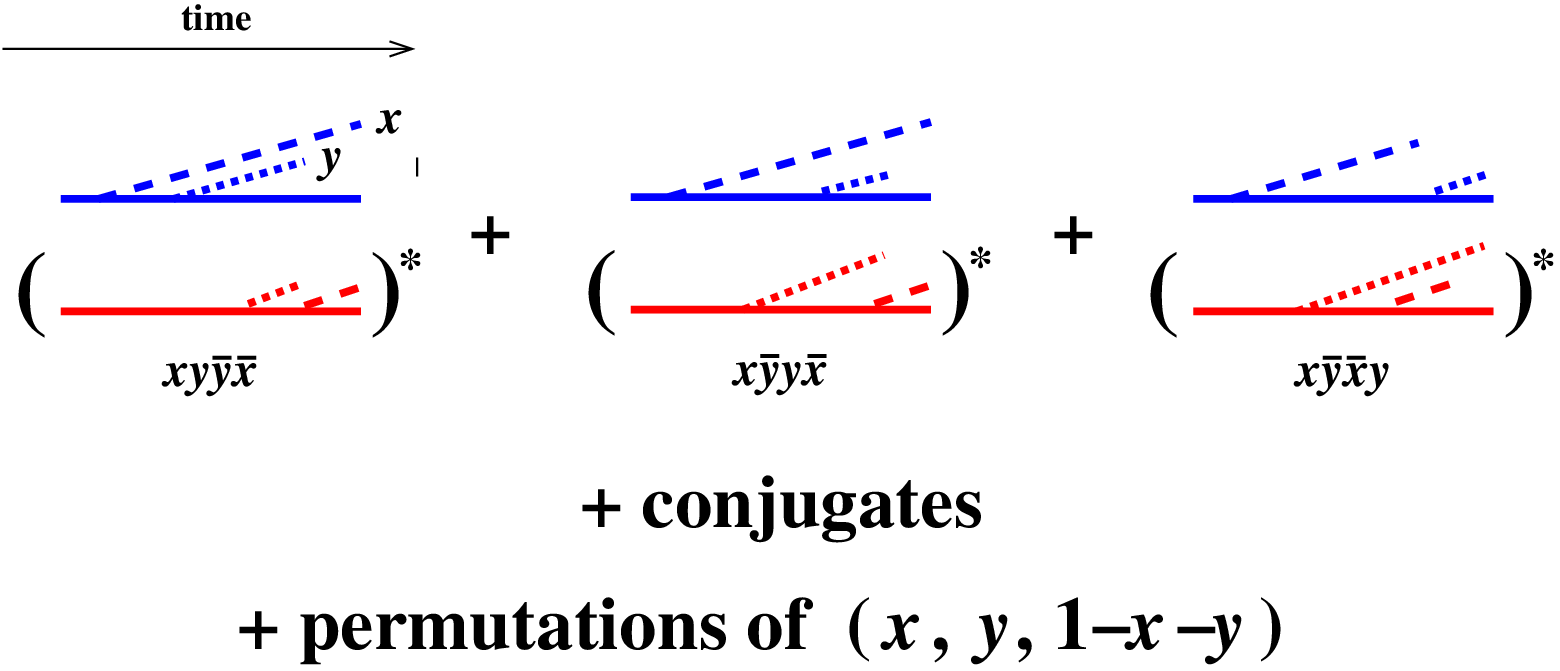}
  \caption{
     \label{fig:subset}
     The subset of interference contributions to double splitting
     that we will evaluate in
     this paper: the ``crossed'' diagrams.
     To simplify the drawing, all particles, including
     bremsstrahlung gluons, are indicated by straight lines.
     The long-dashed and short-dashed lines
     are the daughters with momentum fractions $x$ and $y$
     respectively. 
     The naming of the diagrams indicates the time order
     in which emissions occur in the amplitude and conjugate amplitude.
     For instance, $x\bar y y \bar x$ means first
     (i) $x$ emission in the amplitude, then (ii) $y$ emission in the
     conjugate amplitude, then (iii) $y$ emission in the amplitude,
     and then (iv) $x$ emission in the conjugate amplitude.
  }
\end {center}
\end {figure}

\begin {figure}[t]
\begin {center}
  \includegraphics[scale=0.5]{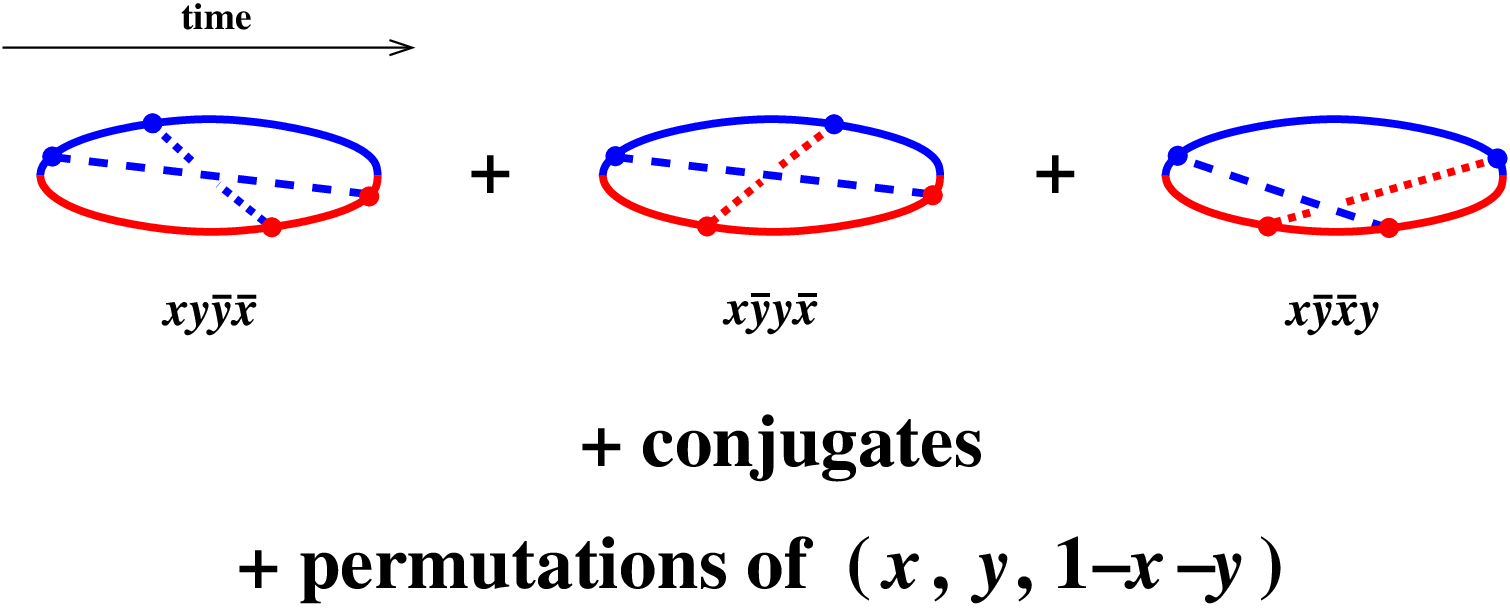}
  \caption{
     \label{fig:subset2}
     An alternative depiction of fig.\ \ref{fig:subset}, with
     amplitudes (blue) and conjugate amplitudes (red) sewn together.
     The dashed lines are colored according to whether they
     were first emitted in the amplitude or conjugate amplitude.
  }
\end {center}
\end {figure}

\begin {figure}[t]
\begin {center}
  \includegraphics[scale=0.5]{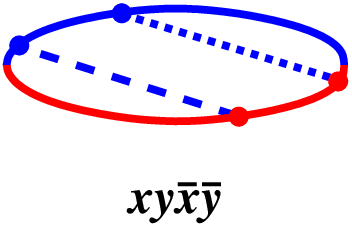}
  \caption{
     \label{fig:uncrossed}
     An example of an interference diagram not evaluated
     in this paper, to be treated in later work.
  }
\end {center}
\end {figure}

\begin {figure}[t]
\begin {center}
  \includegraphics[scale=0.5]{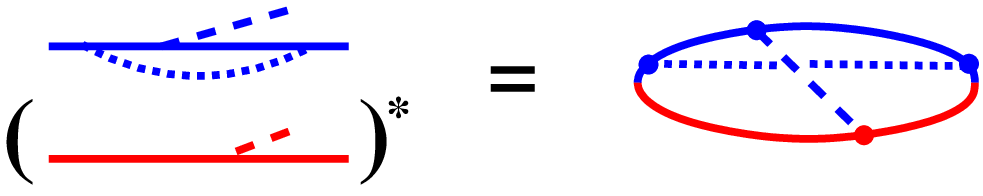}
  \caption{
     \label{fig:virtual}
     An example of a virtual loop correction to single splitting.
  }
\end {center}
\end {figure}

\begin {figure}[t]
\begin {center}
  \includegraphics[scale=0.5]{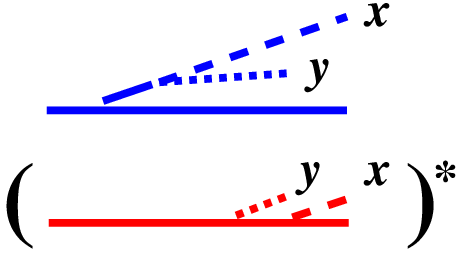}
  \caption{
     \label{fig:zyyz}
     One way to draw the $z y \bar y \bar z$ interference contribution,
     which is the $x \leftrightarrow z \equiv 1{-}x{-}y$ permutation of the
     $x y \bar y \bar x$ interference shown explicitly in
     fig.\ \ref{fig:subset}.
  }
\end {center}
\end {figure}

Finally, in this first paper, we will only compute a
subset of the contributions to the double splitting rate,
depicted in fig.\ \ref{fig:subset}.  It is convenient to also
draw these same interference terms with the amplitudes and conjugate
amplitudes tied together, as in fig.\ \ref{fig:subset2}.  We
will refer to these contributions as the ``crossed'' contributions,
since they involve two crossed lines when drawn as in
fig.\ \ref{fig:subset2}.%
\footnote{
   Although these diagrams are crossed when drawn as time-ordered
   diagrams for the rate, as in fig.\ \ref{fig:subset2}, they are
   not suppressed in the large-$\Nc$ limit.  They are still planar
   diagrams as far as color factors are concerned because one of
   the crossing lines could have instead been connected by drawing
   it routed around the outside of the oval.  (If the solid lines
   in figs.\ \ref{fig:subset} and \ref{fig:subset2} were
   quarks instead of gluons, with $x$ and $y$ being bremsstrahlung gluons,
   then the diagrams explicitly shown would be $1/\Nc$
   suppressed, but some of the ``permutations'' referred to in
   the figure, such as the one
   shown later in fig.\ \ref{fig:zyyz}, would not be.)
}
An example of an interference
contribution that we will not calculate in this paper, but plan
to address in future work, is the un-crossed diagram of
fig.\ \ref{fig:uncrossed}.  (The evaluation of this contribution involves
some additional subtleties beyond what is required for the crossed
contributions, to be briefly mentioned in our conclusion.)
In this paper, we will also, for simplicity, focus
on the rate for a high-energy particle to split into three on-shell
daughters, which all persist after the double-splitting process.
That is, we will not yet consider loop corrections to single
splitting, such as shown in fig.\ \ref{fig:virtual}.
Such loop corrections are necessary for using calculations of
splitting rates to compute energy loss,
as has been analyzed in refs.\ \cite{Blaizot,Iancu,Wu} for the case
$x,y \ll 1$.

The ``$+$ permutations'' line of figs.\ \ref{fig:subset} and
\ref{fig:subset2} contains more types of interference terms than one
might at first realize.  For example, it contains the diagram of
fig.\ \ref{fig:zyyz} corresponding to the interference of (i) separate
emission of $x$ and $y$ with (ii) the emission of an $x{+}y$
gluon that then splits into $x$ and $y$.  This diagram is equivalent
to the $x \leftrightarrow 1{-}x{-}y$
permutation of the $x y \bar y \bar x$ interference shown explicitly
in fig.\ \ref{fig:subset}.  We will correspondingly refer to
fig.\ \ref{fig:zyyz} as the $z y \bar y \bar z$ interference, where
$z \equiv 1{-}x{-}y$.

% -------------------------------------------------------------------------

\subsection {Overview of method and what's new}

Here, in broad outline, are the elements involved in the calculation.
We start by turning the problem into a problem in two-dimensional
non-Hermitian quantum mechanics, slightly generalizing techniques
that have been used by other authors.

% .........................................................................

\subsubsection {3- and 4-particle quantum mechanics}

\begin {figure}[t]
\begin {center}
  \includegraphics[scale=0.5]{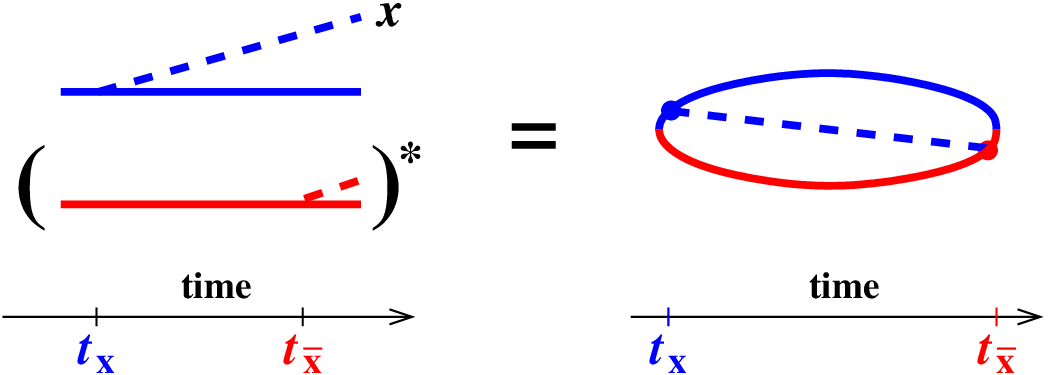}
  \caption{
     \label{fig:xxrate}
     The basic interference contribution for single (independent) splitting,
     together with our labeling convention for the splitting times in
     the amplitude and conjugate amplitude.
  }
\end {center}
\end {figure}

Start by considering an interference contribution to the rate for
single splitting, shown in fig.\ \ref{fig:xxrate}.  In the time interval
$t_\xx < t < t_\xbx$ between splitting in the amplitude and conjugate
amplitude, the number of high-energy particles in the diagram does
not change.  As we shall review later, at high energy
it is possible to reduce the
problem in this time interval to time evolution in
two-dimensional quantum mechanics for three non-relativistic particles,
two representing the two daughter particles in the amplitude in
fig.\ \ref{fig:xxrate} during this time interval
and one representing the parent particle
in the conjugate amplitude in fig.\ \ref{fig:xxrate} during the same
time interval.  The two dimensions of the quantum mechanics problem
are the plane transverse to an axis approximately aligned with
the initial direction of the parent.  The ``non-relativistic'' character
of the transverse dynamics comes from the large-$p_z$
(and small-angle) approximation
to the kinetic energy
$\varepsilon_\p$ of a high-energy particle of mass $\mpart$,
\begin {equation}
   \varepsilon_\p = \sqrt{p_z^2+\p_\perp^2+\mpart^2}
   \simeq p_z + \frac{\p_\perp^2 + \mpart^2}{2 p_z}
   \simeq \frac{\p_\perp^2}{2 p_z}  + {\rm constant} .
\label {eq:varepsilon}
\end {equation}
The right-hand side looks like a two-dimensional non-relativistic kinetic
energy $\p_\perp^2/2m$ with a ``mass'' $m$ with magnitude $|p_z|$, and
(for our purposes) $|p_z|$ may be treated as constant in the above
approximation.%
\footnote{
  The distinction ``for our purposes'' is important.
  In $\epsilon_\p \simeq p_z + \p_\perp^2/2p_z + \cdots$, the variation
  of the first term $p_z$ can be as large as the second term
  $\p_\perp^2/2p_z$ in our context.  However,
  the leading $p_z$ term will cancel in the combinations of
  $\epsilon_\p$'s that will appear in our problem, such as
  (\ref{eq:calHfree}).
}
As we'll discuss, the 3-particle quantum
mechanics problem that describes fig.\ \ref{fig:xxrate} has
the funny feature that the ``mass'' $m$ which describes the particle
in the conjugate amplitude is taken to be negative, which arises from
the sign difference between the evolution $\exp(-i Ht)$ of amplitudes
vs.\ the evolution $\exp(+ i Ht)$ of conjugate amplitudes.  Because
of overall longitudinal momentum conservation in the splitting
process, the two positive
and one negative ``mass'' in this
quantum-mechanics problem satisfy the unusual
property (from the perspective of non-relativistic quantum mechanics)
that
\begin {equation}
   m_1 + m_2 + m_3 = 0 .
\label {eq:m123}
\end {equation}

Another unusual feature of the effective quantum mechanics problem is
that it has a non-Hermitian Hamiltonian.
As the particles propagate, they interact with the background fields in
the plasma.  For a thermal medium (or other random medium), these background
fields are random with well-defined correlations.
When computing a rate in a random background, one should average
the rate over that randomness (e.g.\ average over the thermal
ensemble).  After this average, the interactions of the particles
with the background fields, and their correlations, may be replaced
by a ``potential energy'' term in the quantum mechanics problem,
as we will later review more concretely.%
\footnote{
  \label{foot:scales}
  A potential in non-relativistic quantum mechanics represents an
  instantaneous correlation in our problem.  This is a valid approximation
  in our problem because the time it takes for high-energy
  particles to cross one correlation length of the medium is parametrically
  small compared to the formation time in the high-energy limit.
  The dynamics of the high-energy particles on the time scale of the
  formation time is what is relevant to our calculation.
}
However, unlike a normal quantum mechanics problem, this potential energy
is not real-valued.  The effective quantum mechanics problem that
reproduces
the medium-averaged splitting {\it rate}\/
(rather than the unaveraged splitting {\it amplitude}) has
a non-Hermitian Hamiltonian.
The non-Hermiticity of the Hamiltonian accounts for quantum decoherence
over long times:
it causes the interference contribution
of fig.\ \ref{fig:xxrate} to decay exponentially for time separations
$t_\xbx{-}t_\xx$ large compared to the formation time.

\begin {figure}[t]
\begin {center}
  \includegraphics[scale=0.5]{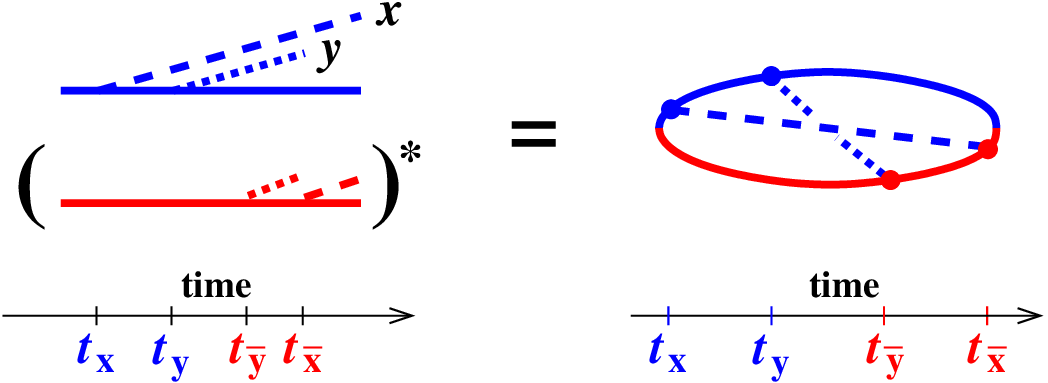}
  \caption{
     \label{fig:xyyxrate}
     The $xy\bar y\bar x$ interference contribution of
     figs. \ref{fig:subset} and \ref{fig:subset2} with
     individual splitting times labeled.
  }
\end {center}
\end {figure}

Now consider a particular interference contribution to the double
splitting rate, such
as the $xy\bar x\bar y$ interference
shown in fig.\ \ref{fig:xyyxrate}.  Within each time interval
between splittings in the figure, such as
(a) $t_\xx < t < t_\yx$, (b) $t_\yx < t < t_\ybx$,
or (c) $t_\ybx < t < t_\xbx$, the number
of high-energy partons does not change.
Using the same methods outlined above, we can reduce the problem of
medium-averaged time evolution during these intervals to
two-dimensional non-Hermitian non-relativistic quantum mechanics
for (a) three, (b) four, and (c) three particles respectively.
For the 4-particle evolution for $t_\yx < t < t_\ybx$, the ``masses''
of the quantum mechanics problem
will satisfy
\begin {equation}
   m_1 + m_2 + m_3 + m_4 = 0 ,
\label {eq:msum}
\end {equation}
in analogy to the 3-particle case (\ref{eq:m123}).

If we can solve the time evolution in the non-Hermitian
quantum mechanics problems,
we can glue the results together using the field-theory matrix elements
for the splittings at $t=t_\xx$, $t_\yx$, $t_\ybx$, and $t_\xbx$ in
fig.\ \ref{fig:xyyxrate} to compute the result that we want for that
interference contribution.  We may similarly compute all the other
interference contributions and add them together.
Fortunately, for the sake of making the calculation practical,
there is a major simplification to the problem, which
we now outline.

% .........................................................................

\subsubsection {Simplification to 1- and 2-particle quantum mechanics}

We will need to solve for two-dimensional
quantum mechanical evolution in problems of the form
\begin {equation}
   {\cal H} =
   \frac{\p_1^2}{2m_1} + \cdots + \frac{\p_N^2}{2m_N}
   + V(\b_1,\cdots\b_N;t) ,
\label {eq:calH}
\end {equation}
where $\b_i$ are the two-dimensional transverse positions of the particles,
$\p_i$ are the corresponding transverse momenta, and $V$ is the
(non-real) potential that incorporates the medium-averaged effects of
scattering from the medium, which may depend on $t \simeq z$
(where $z$ in this context refers to the longitudinal coordinate
$x_3$).
At high energies, the medium will be
(statistically) translation invariant in the transverse plane
over the small transverse distances relevant to the problem, and so
the potential $V$ will be similarly invariant.  For generic masses $m_i$,
that invariance lets us factor
out the analog of the trivial ``center of mass'' motion of the problem
(which turns out not to be relevant)
and so reduce the $N$-particle problem to an effective $N{-}1$ particle
problem.  However, as we shall discuss, something
very special happens when $\sum_i m_i = 0$, which will allow
us to eliminate yet {\it another}\/ two-dimensional degree of freedom
and so reduce the $N$ particle problem to an $N{-}2$ particle problem.
So, 3-particle evolution can be reduced to a simple problem of 1-particle
quantum mechanics, which is equivalent to simplifications originally
used by BDMPS-Z to analyze the problem of the single-splitting rate
via fig.\ \ref{fig:xxrate} in the case of general $x$.
In our case, it will allow us (at the expense of having to develop some
formalism) to reduce the 4-particle two-dimensional
quantum
mechanics problem required for the $t_\yx < t < t_\ybx$ region of
fig.\ \ref{fig:xyyxrate} to a simpler 2-particle two-dimensional quantum
mechanics problem.

% .........................................................................

\subsubsection {Further specialization}

In the multiple scattering (harmonic) approximation,
appropriate at high energy
both for thick media and for ``typical'' events in
relatively thin media,%
\footnote{%
\label{foot:rare}%
Here, thin and thick are defined relative to the typical formation
   length in an infinite medium, which for QCD is
   $\sim \sqrt{\omega/\hat q}$, where $\omega$ is the smallest
   energy of the daughters.  There is often a great deal of confusion
   about the applicability of the harmonic oscillator approximation
   to thin media because interest is often focused on calculations of
   the medium effect
   on the average energy loss $\langle\Delta E\rangle$,
   which in the thin-media case can be dominated
   by {\it rare}\/ events with relatively large $\Delta E$.
   However, the harmonic
   oscillator approximation still describes {\it typical}\/ events in the
   thin-media case,
   provided the thickness of the medium is large compared to the
   mean free path for collisions.  See
   the notes and
   references given for (\ref{eq:qhat}) in appendix \ref{app:details}.
}
the potential $V$ in (\ref{eq:calH}) becomes a
(non-Hermitian) harmonic oscillator potential.
Then 
we will just need to (i) solve for time-evolution of 1-particle and
2-particle harmonic oscillator problems; (ii) glue the results together
with splitting matrix elements;
(iii) integrate over the four
times $t_\xx$, $t_\yx$, $t_\xbx$, and $t_\ybx$ associated with
those splittings in fig.\ \ref{fig:xyyxrate}, for example; and
(iv) integrate over all relevant intermediate states
(i.e.\ transverse positions)
of the
high-energy particles at the intermediate times $t_\yx$ and
$t_\ybx$ in fig.\ \ref{fig:xyyxrate}.
That sounds like a lot of complicated integration.
However, for the case of a thick, homogeneous medium, we will see that it is
possible to do all of these integrations in closed form, except for
a single time integral over the duration $\Delta t$
of the 4-particle evolution
(i.e.\ over $\Delta t \equiv t_\ybx-t_\yx$ for fig.\ \ref{fig:xyyxrate}).
The final $\Delta t$ integral will be performed numerically.

% .........................................................................

\subsubsection {Other aspects worth noting before we begin}

Before we launch into the main body of the paper, we will first
note some other aspects of our problem (double splitting for general
$x$ and $y$) that did not arise in previous work
(specific to small $x$ and $y$).  The first is the necessity for a careful
treatment of the helicity of high-energy particles between splittings.
In the calculation of the single-splitting rate of
fig.\ \ref{fig:xxrate}, it is possible to express the $x$ dependence
of the splitting
matrix elements at $t_\xx$ and $t_\xbx$ in terms of
spin-averaged
Dokshitzer-Gribov-Lipatov-Altarelli-Parisi (DGLAP) splitting functions.
As an example, the differential splitting
rate in the case of a thick, homogeneous
medium is \cite{BDMS}
\begin {subequations}
\label {eq:BDMSggg}
\begin {equation}
   \frac{d\Gamma}{dx}
   = \frac{\alphas}{\pi\sqrt2} \, P(x) \, |\Omega_0| ,
\label {eq:BDMS}
\end {equation}
where $P(x)$ is the relevant spin-averaged DGLAP splitting function
and $\Omega_0$ is the (complex) frequency associated with the
non-Hermitian harmonic oscillator in the multiple scattering
approximation.  For $g \to gg$ splitting for example, $\Omega_0$ turns
out to be
\begin {equation}
   \Omega_0 =
   \sqrt{
     - \frac{i \hat q_{\rm A}}{2 E}
     \left( -1 + \frac{1}{1{-}x} + \frac{1}{x} \right)
   } .
\label {eq:Omega0}
\end {equation}
\end {subequations}
($\hat q_R$, known as the jet-quenching parameter,
is the average squared
transverse momentum per unit length that is transferred
from the medium to a high-energy particle with color
representation $R$.)
When we study double splitting, such as in
figs.\ \ref{fig:subset} and \ref{fig:subset2},
the helicities associated with the various splittings
are tied together in a non-trivial way that cannot be packaged into
spin-averaged DGLAP splitting functions for general values of
$x$ and $y$.  One of our tasks will be to distinguish different
intermediate-state helicity amplitudes in our treatment of the
splitting matrix elements.

The other non-trivial feature
that we want to mention is that
individual interference contributions, such as each of those
in fig.\ \ref{fig:subset}, will turn out to have an unphysical
logarithmic UV divergence arising from coincidence
of three of the four splitting times
in the diagram.%
\footnote{
  This is not the $\int d(\Delta t)/\Delta t$ divergence that is
  part of the double log behavior found for $y \ll x$ in
  refs.\ \cite{Blaizot,Iancu,Wu}.
  For fixed $x$ and $y$ (which is what we consider in this
  paper), that particular logarithm
  is cut off at small
  $\Delta t \sim y \sqrt{E/x \hat q}$,
  as we review in section \ref{sec:logs}.
}
Since the diagrams for the amplitude
are tree level (and similarly for the conjugate amplitude),
there are no loops involved in the amplitude, and so
there should be no actual UV divergences in
the final result for the differential double splitting rate
$d\Gamma/dx\,dy$ once we sum over all contributions.  We will see
that the short-time
UV issue is resolved by being careful about operator ordering
and $i\epsilon$ prescriptions.

% -------------------------------------------------------------------------

\subsection {What we find}

Since we are only computing the ``crossed'' subset of contributions
to the double-splitting rate (fig.\ \ref{fig:subset2}) and not yet all
contributions, the results of this paper lie in the development of
calculational methods rather than a final, total answer for
the differential double-splitting rate
$d\Gamma/dx\,dy$.
That said, it is still interesting to ask if the contribution of
this subset to the result for $d\Gamma/dx\,dy$
has any interesting qualitative features.
First, we will check that for $y \ll x \ll 1$, we find a contribution
with $y$ dependence of the form
\begin {equation}
   \left[ \frac{d\Gamma}{dx\,dy} \right]_{\rm crossed} \propto
   \cdots + 
   \frac{\alphas^2 \ln(y/x)}{x^{3/2} y}
   + \cdots \,.
\label {eq:lny}
\end {equation}
When integrated over $y$, this gives a double log correction
$\propto \alphas^2 \log^2$.
We will
see (with some caveats explained later)
that this result matches
recent $y\ll x$ results
by Blaizot and Mehtar-Tani \cite{Blaizot} and Iancu \cite{Iancu}
and Wu \cite{Wu},
after accounting for the
fact that we have not included virtual corrections like fig.\
\ref{fig:virtual}.
The $x^{-3/2}y^{-1}\ln(y/x)$
contribution to (\ref{eq:lny}) arises from interference
diagrams of the form
$x y \bar y \bar x$, $x \bar y y \bar x$,
$z y \bar y \bar z$ and $z \bar y y \bar z$,
where the $z y \bar y \bar z$ interference was
depicted in fig.\ \ref{fig:zyyz}.
As discussed in refs.\ \cite{Blaizot,Iancu,Wu}, this
correction is due to small-$x$ physics in the collisions of the
high-energy particles with the medium, and it may be absorbed
into a running of the parameter
$\hat q$ that characterizes the strength of interactions of
the high-energy particles with the medium in the multiple-scattering
approximation \cite{Blaizot,Wu0,Iancu2}.

However, in the limit of small $y$ and fixed $x$, we will find an even more
divergent contribution from
crossed diagrams that scales with $y$
as
\begin {equation}
   \left[ \frac{d\Gamma}{dx\,dy} \right]_{\rm crossed} \propto
   \frac{\alphas^2 \ln(y/x)}{x y^{3/2}}
   + \cdots .
\end {equation}
The physical meaning of this soft divergence will be discussed in future
work.

% -------------------------------------------------------------------------

\subsection {Outline}

At the end of this introduction, we will briefly comment on
the relevant scale of the $\alphas$ that controls a perturbative
treatment in the number of overlapping hard splittings.
In section \ref{sec:1gluon}, we review formalism and
results for computing single splitting processes, as in
fig.\ \ref{fig:xxrate}, which will set up some of the language
and formalism we later need for double splitting.
Section \ref{sec:Bs} turns to how
to use symmetry to reduce the problem of $N$ particle
quantum evolution in interference diagrams to an effective
quantum mechanics problem involving only $N{-}2$ particles.
We apply
that technique to the double splitting problem in section
\ref{sec:double}, which also discusses the rules of diagrammatic
perturbation theory in this language, how the large-$\Nc$ limit
simplifies the discussion of color in this paper, and the relationship
of splitting vertex matrix elements to helicity-dependent DGLAP
structure functions.  In that section, we go as far as we can without
either attempting complicated numerics or introducing additional
approximations.  To go further, we turn to the multiple
scattering (harmonic oscillator) and thick-media approximations
in section \ref{sec:xyyx}, where, as a first example,
we reduce the calculation of the
$xy\bar y\bar x$ interference in fig.\ \ref{fig:xyyxrate} to
the form of a single integral over $\Delta t \equiv t_\ybx-t_\yx$.
We also show that this integral is divergent as $\Delta t \to 0$
and isolate the divergence.
In section \ref{sec:other}, we show how all the other
``crossed'' interference diagrams of fig.\ \ref{fig:subset2} can
be related to the $xy\bar y\bar x$ result, and we show that
the small-$\Delta t$ divergences cancel among these diagrams.
We then discuss in section \ref{sec:smalldt} that the canceling
divergences nevertheless leave behind a finite piece associated
with the residue of a pole at $\Delta t=0$.  Getting this right
requires a discussion of the appropriate $i\epsilon$ prescriptions 
associated with $\Delta t \to 0$.
Section \ref{sec:summary} gives a summary of our final results
in terms of a single convergent $\Delta t$ integral which can be
performed numerically.  In section \ref{sec:behavior}, we discuss
our results in the limit where at least one of the radiated
gluons becomes soft and show agreement between the pieces of our
results that match up with pieces of previous authors'
soft-gluon results.  Finally, section \ref{sec:conclusion} offers
our slightly redundant but extremely short conclusion.
A number of details are left for appendices.

% -------------------------------------------------------------------------

\subsection {Note on the scale of \boldmath$\alphas$}

The size of $\alphas$ depends on momentum
scale, and there are two parametrically
different scales relevant to the size of couplings in
splitting processes such as fig.\ \ref{fig:lpm1}.  One is the
strength of the interaction of high-energy particles with the
medium (the couplings to the black, vertical gluon lines
in fig.\ \ref{fig:lpm1}).
The smallest momentum transfers that are important are of order
the inverse Debye screening length, $\mD$, which is a characteristic of the
medium.  In theoretical studies, one may or may not want to
consider $\alphas(\mD)$ as small, but it is certainly not
very small in quark-gluon plasmas created in experiment.
The other important scale is the one associated with the
high-energy splitting
vertex in fig.\ \ref{fig:lpm1}.
The relevant scale there is the typical transverse momentum $Q_\perp$ between
the daughters, which depends on the energy.
(This is the scale of the momenta of each daughter in the
daughters' center-of-momentum
frame.)
In the simple case of hard bremsstrahlung in a thick medium,
the LPM effect causes
$Q_\perp$ to grow slowly with energy as%
\footnote{
   The relative $Q_\perp$ of the transverse momentum between the daughters
   is of order the total momentum transfer from the medium during the
   formation time $t_{\rm form}$, and so $Q_\perp^2 \sim \hat q t_{\rm form}$.
   For hard ($x \sim 1 \sim 1-x$)
   bremsstrahlung in a thick medium, the LPM effect gives
   $t_{\rm form} \sim \sqrt{\omega/\hat q} \sim \sqrt{E/\hat q}$, and so
   $Q_\perp \sim (\hat q E)^{1/4}$.
}
$Q_\perp \sim (\hat q E)^{1/4}$.
If $E$ is large enough, the $\alphas(Q_\perp)$ characterizing the
splitting vertex may therefore be small, regardless of whether
$\alphas(\mD)$ is small.  This approximation---that the amplitude
for each high-energy splitting vertex is small---is one that we will
use implicitly throughout this paper.  The factors of $\alphas$
in this paper will all implicitly be factors of $\alphas(Q_\perp)$
unless explicitly noted otherwise.

% =========================================================================

\section {Review of single splitting}
\label {sec:1gluon}

We will set the stage by summarizing, in our own language,
the starting point for the basic
formalism (used in one form or another since
BDMPS-Z) describing the LPM effect in QCD for
single splitting as in fig.\ \ref{fig:lpm1}.
Everything in this section is equivalent to results already known
in the combined literature on this problem, but it will be useful
for fixing our own conventions and in particular fixing the language
we will use when we move on to discuss double splitting.%
\footnote{
  Our formulation of the problem owes the most to Zakharov \cite{Zakharov},
  together with elements from refs.\ \cite{Migdal, AMY, simple}.
  The notation we use is closest to ref.\ \cite{simple}, whose appendix
  includes a translation table to the notation of several other works.
  Other noteworthy formalisms for studying LPM in QCD include BDMPS
  \cite{BDMPS12,BDMPS3}, but also,
  in various limits or with
  various extensions, GLV \cite{GLV}, ASW \cite{ASW}, and higher
  twist \cite{HT}.
  See ref.\ \cite{brick} for
  comparative discussion and further references.
}

% ------------------------------------------------------------------------

\subsection{Relating bremsstrahlung to 3-particle non-Hermitian
            quantum mechanics}

\begin {figure}[t]
\begin {center}
  \includegraphics[scale=0.5]{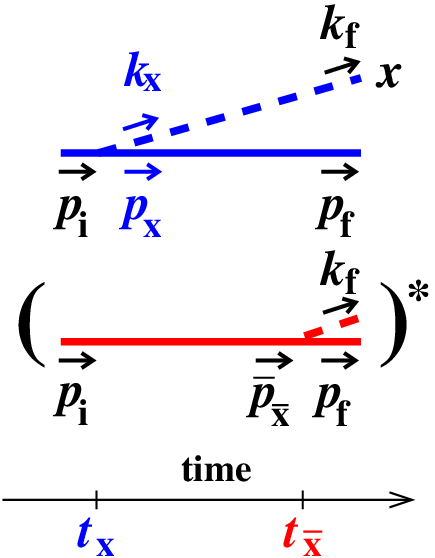}
  \caption{
     \label{fig:xxrate2}
     Our conventions for labeling momenta for the interference
     contribution for single splitting.  In the QED case,
     $\k_\xx = \k_\fx$ since the photon does not directly interact
     with the medium.
  }
\end {center}
\end {figure}

Let's first discuss the case of QED, as in fig.\ \ref{fig:xxrate2}
with the dashed line representing a bremsstrahlung photon.
Let $\delta H$ be the part
of the Hamiltonian that contains the splitting vertices for
the high-energy particles.  Then, working to leading order in
$\delta H$, the differential bremsstrahlung probability $dI/dx$ for
producing a photon with momentum fraction $x$ is%
\footnote{
  In this paper, we do not consider bremsstrahlung associated with
  whatever earlier hard process created our initial high energy parton in
  the medium in the first place.  We consider only late enough times
  ($t \gg 1/\tau_{\rm el}$ where $\tau_{\rm el}$ is the mean free path
  for collisions with the medium) that the high-energy partons are
  already approximately on-shell.  (By the uncertainty principle,
  a particle having collisions roughly every time interval
  $\tau_{\rm el}$ will be off-shell in energy by
  $\Delta E \sim 1/\tau_{\rm el}$, which corresponds to
  $p^\mu p_\mu \sim E \, \Delta E \sim E/\tau_{\rm el}$, which is what
  we mean here by ``approximately on-shell.'')  For a recent interesting
  discussion of the interplay between early vacuum bremsstrahlung
  and late medium-induced bremsstrahlung,
  see Kurkela and Wiedemann \cite{KW}.
}%
\begin {multline}
   \frac{dI}{dx}
   = 2 \Re \Biggl\{
       \frac{E}{2\pi V_\perp}
       \int_{t_\xx < t_\xbx} dt_\xx \> dt_\xbx
       \sum_{\rm pol.}
       \int_{\p_\fx,\k_\fx}
       \Bigdlangle
       \biggl(
         \int_{\p_\xx}
         \langle \p_\fx \k_\fx, t_\xbx | \p_\xx \k_\fx,t_\xx \rangle
         \langle \p_\xx \k_\fx | {-}i \, \delta H | \p_\ix \rangle
       \biggr)
\\
       \times
       \biggl(
         \int_{\bar\p_\xbx}
         \langle \p_\fx \k_\fx| {-}i \, \delta H| \bar\p_\xbx \rangle
         \langle \bar\p_\xbx, t_\xbx| \p_\ix,t_\xx \rangle
       \biggr)^*
       \Bigdrangle
   \Biggr\} ,
\label {eq:1brem1}
\end {multline}
where the $\p$ is the transverse momentum of the charged particle
and $\k$ is the transverse momentum of the bremsstrahlung photon,
labeled as in fig.\ \ref{fig:xxrate2}.  The high-energy particles
propagate through gauge fields sourced by the medium, and the
double angle brackets
denote statistical averaging over those gauge fields.
The factors in the first parenthesis give the amplitude in
fig.\ \ref{fig:xxrate2}, which consists of a splitting matrix element
$\langle \p_\xx \k_\fx | {-}i \, \delta H | \p_\ix \rangle$ at the
first time $t_\xx$, followed by the propagation
$\langle \p_\fx \k_\fx, t_\xbx | \p_\xx \k_\fx,t_\xx \rangle$
in the background field from $t_\xx$ to $t_\xbx$.
In QED, the bremsstrahlung photon ($\k$) is just a spectator
during this propagation.  The factors in the second parenthesis
give the conjugate amplitude in fig.\ \ref{fig:xxrate2}, which has
propagation first, followed by splitting at the later time $t_\xbx$.
The overall $2\Re\{ \cdots \}$ combines (i) the interference term of
fig.\ \ref{fig:xxrate2}, which is the case where emission happens first
in the amplitude, with (ii) its complex conjugate, which is the case
where emission happens first in the conjugate amplitude.  The
sum in (\ref{eq:1brem1}) is over final state polarizations of the daughters.
The symbol $V_\perp$ represents the (infinite)
2-volume (area) of the
transverse plane, upon which final results will not depend.

In this paper, we will delegate additional discussion of some formulas
to appendix \ref{app:details}.
In the case at hand, see appendix \ref{app:details} if wondering why
one need not
include time evolution of the final state $|\p_\fx\k_\fx\rangle$
after $t_\xbx$ in
(\ref{eq:1brem1}),
nor time evolution of the initial state $|\p_\ix\rangle$ before $t_\xx$.
We also comment there on the prefactor $E/2\pi V_\perp$ in
(\ref{eq:1brem1}).

It is useful to rewrite the conjugate amplitude to express
(\ref{eq:1brem1}) as
\begin {multline}
   \frac{dI}{dx}
   = 2 \Re \Biggl\{
       \frac{E}{2\pi V_\perp}
       \int_{t_\xx < t_\xbx} dt_\xx \> dt_\xbx
       \sum_{\rm pol.}
       \int_{\p_\fx,\k_\fx}
       \Bigdlangle
       \biggl(
         \int_{\p_\xx}
         \langle \p_\fx \k_\fx, t_\xbx | \p_\xx \k_\fx,t_\xx \rangle
         \langle \p_\xx \k_\fx | {-}i \, \delta H | \p_\ix \rangle
       \biggr)
\\
       \times
       \biggl(
         \int_{\bar\p_\xbx}
         \langle \p_\ix,t_\xx| \bar\p_\xbx, t_\xbx \rangle
         \langle  \bar\p_\xbx| i \, \delta H |\p_\fx \k_\fx \rangle
       \biggr)
       \Bigdrangle
   \Biggr\} .
\label {eq:1brem2}
\end {multline}
Time-evolution in the amplitude is given by the factor
\begin {equation}
  \langle \p_\fx \k_\fx, t_\xx | \p_\xx \k_\fx,t_\xx \rangle
  =
  \langle \p_\fx \k_\fx| e^{-iH_{(2)} (t_\xbx-t_\xx)} |\p_\xx \k_\fx \rangle ,
\end {equation}
where $H_{(2)}$ is the Hamiltonian for propagating two
high-energy particles through the medium: the charged particle and
the photon.  (Terms $\delta H$
in the full Hamiltonian which change the number of
high-energy particles are not part of
$H_{(2)}$.)
The time evolution in the conjugate amplitude is
analogously in the factor
\begin {equation}
  \langle \p_\ix,t_\xx| \bar\p_\xbx, t_\xbx \rangle
  =
  \langle \p_\ix| e^{+iH_{(1)} (t_\xbx-t_\xx)} |\bar\p_\xbx \rangle ,
\end {equation}
where $H_{(1)}$ is the Hamiltonian for propagating one high-energy
particle through the medium.  Since the interactions with the
medium appear only in $H_{(1)}$ and $H_{(2)}$, the medium-average in
(\ref{eq:1brem2}) can be restricted to the pieces
\begin {equation}
   \Bigdlangle
   e^{-iH_{(2)} (t_\xbx-t_\xx)} |\p_\xx \k_\fx \rangle
   \langle \p_\ix| e^{+iH_{(1)}(t_\xbx-t_\xx)}
   \Bigdrangle .
\label {eq:timeevolve}
\end {equation}
This is just time evolution of an initial
$|\p_\xx \k_\fx \rangle \langle \p_\ix|$ starting from time $t_\xx$.
The object $|\p_\xx \k_\fx \rangle \langle \p_\ix|$ lives in the
space
\begin {equation}
  \bar \Hilbert_{\rm e} \otimes \Hilbert_{{\rm e},\gamma}
  =
  \bar \Hilbert_{\rm e} \otimes \Hilbert_{\rm e} \otimes \Hilbert_\gamma ,
\label {eq:Hilbert}
\end {equation}
where
$\Hilbert_{\rm e}$ is the Hilbert space of states of
a high-energy electron, and
$\Hilbert_{{\rm e},\gamma}$ is the Hilbert space of states with both a high-energy
electron and high-energy photon.  Instead of thinking of
(\ref{eq:Hilbert}) as a product of Hilbert spaces, it is convenient
to formally think of a single Hilbert space of three particles:
one electron, one photon, and one conjugated electron.%
\footnote{
  This picture is at the heart of Zakharov's early work \cite{Zakharov}.
}
Correspondingly, we rewrite (\ref{eq:timeevolve}) in the form
of a 3-particle evolution
\begin {equation}
   \Bigdlangle
   e^{-iH_{(\bar 1+2)} (t_\xbx-t_\xx)} |\p_1,\p_2,\p_3 \rangle
   \Bigdrangle ,
\label {eq:timeevolve2}
\end {equation}
where $H_{(\bar 1+2)} = H_{(2)} - H_{(1)}$.  Here
$H_{(2)}$ operates on the two particles associated with the
amplitude and $H_{(1)}$ on the one associated with the conjugate
amplitude.
It will be convenient to choose the convention that a $\p_i$
above that represents the momentum of
a particle in the {\it conjugate}\/ amplitude
has been negated.  So, for instance,
(\ref{eq:timeevolve}) may be recast in the
form (\ref{eq:timeevolve2}) with the identification
\begin {equation}
   (\p_1,\p_2,\p_3) = (-\p_\ix,\p_\xx,\k)
   \qquad \mbox{at $t=t_\xx$.}
\end {equation}
(Our convention is that the ``first'' of the three particles
is the one from the conjugate amplitude.)
With this sign convention for the $\p_i$, momentum conservation
$\p_\ix = \p_\xx+\k_\fx$ in fig.\ \ref{fig:xxrate2}
means that
\begin {equation}
   \p_1+\p_2+\p_3 = 0 .
\end {equation}

The medium average in (\ref{eq:timeevolve2}) only affects the evolution
operator (which depends on the background fields), and so we may
rewrite (\ref{eq:timeevolve2}) as
\begin {equation}
   e^{-i{\cal H} (t_\xbx-t_\xx)}
   |\p_1,\p_2,\p_3 \rangle
   \equiv
   \Bigdlangle
   e^{-iH_{(\bar 1+2)} (t_\xbx-t_\xx)}
   \Bigdrangle
   |\p_1,\p_2,\p_3 \rangle ,
\end {equation}
where ${\cal H}$ is going to be the Hamiltonian of our effective
3-particle quantum mechanics problem.
Note from this definition that ${\cal H}$ need not be Hermitian, even though
$H_{(\bar 1+2)}$ is.

% ------------------------------------------------------------------------

\subsection{Brief review of the form of \boldmath$\cal H$}

As we'll very briefly review,
the form of ${\cal H}$ turns out to be
(\ref{eq:calH}),
\begin {subequations}
\label {eq:calH2+m}
\begin {equation}
   {\cal H} =
   \frac{\p_1^2}{2m_1} + \frac{\p_2^2}{2m_2} + \frac{\p_3^2}{2m_3}
   + V(\b_1,\b_2,\b_3;t) ,
\label {eq:calH2}
\end {equation}
with non-relativistic ``masses'' given in terms of the initial parton
energy $E$ by
\begin {equation}
   m_i \equiv x_i E .
\end {equation}
\end {subequations}
We have used a similar minus sign convention for defining
longitudinal momentum fractions $x_i$ as we did above for defining
transverse momenta $\p_i$: momentum fractions $x_i$ appearing in the
conjugate amplitude are negated.  In our case here,
\begin {equation}
   (x_1,x_2,x_3) = (-1,1{-}x,x) ,
\label {eq:x123}
\end {equation}
and so $m_1$ is negative while $m_2$ and $m_3$ are positive.  Note
that, with this sign convention,
\begin {equation}
   x_1 + x_2 + x_3 = 0 ,
\end {equation}
which is equivalent to the mass relation (\ref{eq:m123}).

The kinetic terms in (\ref{eq:calH2+m}) can be understood by taking the
limiting case where there is no medium, so that the high-energy particles
are free.  Then
\begin {equation}
   {\cal H} = H_{(2)}-H_{(1)}
    = (\varepsilon_{\p_2}+\varepsilon_{\p_3})-\varepsilon_{\p_1} ,
\label {eq:calHfree}
\end {equation}
where $\varepsilon_\p$ is the energy
of a single free particle.
Using the large-$p_z$ expansion (\ref{eq:varepsilon}) in
(\ref{eq:calHfree}) yields the kinetic term of (\ref{eq:calH2})
plus a term
\begin {equation}
   \frac{{\cal M}_1^2}{2x_1E} + \frac{{\cal M}_2^2}{2x_2E}
   + \frac{{\cal M}_3^2}{2x_3E}
\end {equation}
for the potential.  In this paper, we will assume for simplicity
that the energy is
large enough that effects of parton masses ${\cal M}_i$
(including effective parton masses in the medium) are negligible.%
\footnote{
  For a relativistic plasma and for
  zero-temperature masses small compared to $T$,
  this condition is formally that the energies be $\gg T$.
  But mass effects can be important for thin media in the harmonic
  oscillator approximation
  \cite{ZakharovMass}.
}

The dominant effect of interactions with the medium is to add an imaginary
part $-i \Gamma_{\cal H}$ to ${\cal H}$ related to the
differential rate of scattering from the medium.
For the case of a weakly-coupled, QED plasma, this turns
out to be
\begin {equation}
   V(\b_1,\b_2,\b_3;t) =
      -i \left[
         e^2 \, \bar\Gamma_{\rm el}(0,t)
         - e^2 \, \bar\Gamma_{\rm el}(\b_2{-}\b_1,t)
      \right] ,
\label {eq:Vqed}
\end {equation}
where
$e$ is the charge of the charged high-energy particle, $\b_3$ corresponds
to the photon,  Above, $\Gamma_{\rm el} \equiv e^2 \bar\Gamma_{\rm el}(0)$
is the rate of elastic scattering from the medium, which will vary with
time if the medium properties along the path of the high-energy particles
vary with time.
The generalization
$\Gamma_{\rm el}(\b,t)$ is defined as the Fourier transform of the differential
rate of scattering $d\Gamma_{\rm el}/d^2\q_\perp$ with respect to
the transverse momentum transfer $\q_\perp$:
\begin {equation}
   \bar\Gamma_{\rm el}(\b)
   \equiv \int d^2q_\perp \> \frac{d\bar\Gamma_{\rm el}(t)}{d^2\q_\perp}
          \, e^{i\b\cdot\q_\perp}
   = \frac{1}{\pi} \int d^2 q_\perp \> \frac{d\bar\Gamma_{\rm el}(t)}{d(q_\perp^2)}
          \, e^{i\b\cdot\q_\perp} .
\end {equation}
The second term in (\ref{eq:Vqed}) corresponds to
background field
correlations between amplitude and conjugate amplitude in
fig.\ \ref{fig:lpm1}a.
The first term corresponds to self-energies of the charged particle
lines arising from correlations between the amplitude
and itself, or between the conjugate amplitude and itself.%
\footnote{
  The correlation of two interactions in the amplitude has a real part
  in addition to the imaginary part, but this may be absorbed by replacing
  the mass of the high-energy particle by its effective mass in the medium.
}
The relative sign in (\ref{eq:Vqed}) arises
because the second term corresponds, in the
language of ${\cal H}$, to the interaction of a charged particle
(charge $e$) and a conjugated charged particle (charge $-e$).

To generalize to strongly-coupled plasmas (but weakly coupled
splitting matrix elements $\delta H$ for high-energy particles),
it has been argued that the potential $V$ may instead be related
to the value of real-time Wilson loops
which contain two long, parallel, light-like lines separated
by $\b = \b_2-\b_1$ \cite{WilsonLoops}.
In this paper, we will not commit to weakly or strongly coupled
plasmas.  We will keep $V$ general in the first part of
our analysis, before later specializing to the multiple
scattering approximation.

% ------------------------------------------------------------------------

\subsection{Generalization to QCD}
\label {sec:VQCD}

The case of QCD is similar, but the bremsstrahlung gluon also interacts
with the medium, and so
correlations also arise
between the bremsstrahlung gluon ($\b_3$)
and the other two particles ($\b_1,\b_2$).
For a weakly-coupled plasma, the corresponding potential in
(\ref{eq:calH2}) is then%
\footnote{%
\label{foot:weak}%
The $g^2$ here in section \ref{sec:VQCD} characterizes interactions
  with the medium, and so the relevant scale for this running coupling
  can be as low as the inverse Debye screening length.
  (We promised earlier to alert readers to those few cases where the strong
  coupling in this paper was not evaluated at a scale $Q_\perp$ that
  grows with energy $E$.)
}%
${}^,$%
\footnote{
  Note that $(g T_1,g T_2,g T_3) \to (-e,+e,0)$ turns the QCD result
  (\ref{eq:Vqcd0}) into the QED result (\ref{eq:Vqed}).  Here,
  $gT_3 \to 0$ represents the charge of the photon.
}
%\pagebreak[0]
\begin {multline}
   V(\b_1,\b_2,\b_3;t) =
      -i \biggl[
         \tfrac12 g^2 (T_1^2+T_2^2+T_3^2) \,
             \bar\Gamma_{\rm el}(0,t)
         + g^2 T_2\cdot T_1\,
             \bar\Gamma_{\rm el}(\b_2{-}\b_1,t)
\\
         + g^2 T_3\cdot T_2\,
             \bar\Gamma_{\rm el}(\b_3{-}\b_2,t)
         + g^2 T_1\cdot T_3\,
             \bar\Gamma_{\rm el}(\b_1{-}\b_3,t)
      \biggr] ,
\label {eq:Vqcd0}
\end {multline}
where $T_i^a$ are the color generators that operate on the color space
of particle $i$, and $T_i\cdot T_j \equiv T_i^a T_j^a$.
The differential rate of elastic scattering from the medium
for a particle with color representation
$R$ is normalized here (in weak coupling) as
$\Gamma_{R,\rm el} = g^2 C_R \, \bar\Gamma_{\rm el}$.
The first term in (\ref{eq:Vqcd0}) is the sum of the self-energies of the three
particles.  The $T_i\cdot T_j$ terms involve correlations between
particle $i$ and $j$, with charges $g T_i$ and $g T_j$ respectively.
This means that the $T_2\cdot T_1$ and
$T_3\cdot T_1$ terms correspond to correlations between interactions
in the amplitude and conjugate amplitude, according to our convention
that particle 1 is the particle in the conjugate amplitude.
The $T_2\cdot T_3$ involves correlations between background
interactions of the two daughter
particles in the amplitude.%
\footnote{
  The effects of the imaginary part of this correlation dominate over
  those of the real part.
}
Color conservation implies that
\begin {equation}
   T_1+T_2+T_3 = 0 ,
\label {eq:Tsum}
\end {equation}
similar to $\p_1+\p_2+\p_3=0$ and $x_1+x_2+x_3=0$.
Relation (\ref{eq:Tsum}) can be used%
\footnote{
  For example, (\ref{eq:Tsum}) implies $(T_1+T_2)^2 = T_3^2$,
  which gives $T_1\cdot T_2 = - \tfrac12(T_1^2+T_2^2-T_3^2) =
  - \tfrac12(C_1+C_2-C_3)$.
}
to express $T_i\cdot T_j$ in terms of
fixed Casimirs $C_i \equiv T_i^2$, putting
(\ref{eq:Vqcd0}) into the form
\begin {subequations}
\label{eq:Vqcdwrapper}
\begin {multline}
   V(\b_1,\b_2,\b_3;t) =
      -i g^2 \biggl[
         \tfrac12(C_1{+}C_2{-}C_3) \,
             \Delta\bar\Gamma_{\rm el}(\b_2{-}\b_1,t)
\\
         + \tfrac12(C_2{+}C_3{-}C_1) \,
             \Delta\bar\Gamma_{\rm el}(\b_3{-}\b_2,t)
         + \tfrac12(C_3{+}C_1{-}C_2) \,
             \Delta\bar\Gamma_{\rm el}(\b_1{-}\b_3,t)
      \biggr] ,
\label {eq:Vqcd}
\end {multline}
where
\begin {equation}
   \Delta\bar\Gamma_{\rm el}(\b)
   \equiv
   \bar\Gamma_{\rm el}(0)
   -
   \bar\Gamma_{\rm el}(\b)
   = \int d^2q_\perp \> \frac{d\bar\Gamma_{\rm el}(t)}{d^2\q_\perp}
          \, (1-e^{i\b\cdot\q_\perp})
   .
\label {eq:DGamma2}
\end {equation}
\end {subequations}
Note that the potential $V$ depends only on differences $\b_i{-}\b_j$,
which is a consequence of transverse translation invariance.

It is worth mentioning in passing a simplification in
the case of small $x$: the masses $m_1$ and $m_2$ in (\ref{eq:calH2+m})
then have large magnitudes compared to $m_3$.  As a result, there is little
(transverse)
motion of particles 1 and 2, and so $\b_1$ and $\b_2$ stay close together
compared to $\b_3$.  Then particle 2 and conjugate particle 1
can be thought of as forming a tiny, fixed color dipole which interacts
with particle 3 (e.g.\ a $\bar q q$ or $gg$ dipole interacting with
a softer gluon).
This is one way that the 3-particle problem can be
reduced to a 1-particle problem (of the small-$x$ particle).
However, we will be considering generic $x$ and will not make such
approximations in this work.

% ------------------------------------------------------------------------

\subsection{The (optional) harmonic oscillator approximation}

There is no reason that one cannot work hard to numerically
solve the quantum mechanics problem with the full potential $V$
\cite{ZfullV,SGfullV}.
However, at high energy (for a medium thick compared to
the mean free path for elastic scattering) there is an
oft-used simplification: it is difficult to deflect high-energy
particles, and so (given that the particles start out together),
the values of $\b_i{-}\b_j$ will be small during the formation time.
One might therefore expect that one can replace the potential $V$
above by a quadratic approximation around $\b_i{-}\b_j = 0$.
Formally, the small-$\b$ limit of
(\ref{eq:Vqcdwrapper}) gives
\begin {multline}
   V(\b_1,\b_2,\b_3;t) =
      - \frac{i}8 \Bigl[
         (\hat q_1+\hat q_2-\hat q_3) (\b_2{-}\b_1)^2
\\
         + (\hat q_2+\hat q_3-\hat q_1) \,
             (\b_3{-}\b_2)^2
         + (\hat q_3+\hat q_1-\hat q_2) \,
             (\b_1{-}\b_3)^2
      \Bigr] ,
\label {eq:VqcdHO}
\end {multline}
where $\hat q_R$,
the average squared
transverse momentum transferred from the
medium per unit length, is
\begin {equation}
   \hat q_R(t)
   = \int d^2q_\perp \> \frac{d\Gamma_{R,\rm el}(t)}{d^2\q_\perp}
          \, q_\perp^2 .
\label {eq:qhat}
\end {equation}

In the case of a strongly-interacting plasma, the form of
$V$ may be more complicated than the weakly-interacting
form of (\ref{eq:Vqcdwrapper}), but we give an argument in Appendix\
\ref{app:details} that (\ref{eq:VqcdHO}) holds for any
quadratic approximation to the small $\b_i{-}\b_j$ limit.

See appendix \ref{app:details} for caveats and further discussion
of the definition (\ref{eq:qhat}) of $\hat q_R$.

% ------------------------------------------------------------------------

\subsection{Reduction from 3-particle to 1-particle quantum mechanics}

In our application, $\p_1+\p_2+\p_3 = 0$, and so there are seemingly only
two independent momenta in the Hamiltonian ${\cal H}$
(\ref{eq:calH2}) in the case of interest.
However, we get a further simplification from the 3-dimensional rotation
invariance of our problem.  Specifically, we could choose our
$z$ axis to be in a slightly different direction, while still retaining
the large-$p_z$ approximations that we have made.
In the context of those approximations,
the answer for the splitting rate should accordingly be invariant under
transformations of the form
\begin {equation}
   (\p_i,p_{iz}) \to (\p_i + p_{iz} {\bm\xi},\, p_{iz}) ,
\label {eq:ptransform0}
\end {equation}
where $\bm\xi$ is an arbitrary vector in the transverse plane.
The transformation (\ref{eq:ptransform0}) corresponds to a
3-dimensional rotation by ${\bm\theta} = {\bm e}_z\times{\bm\xi}$
once one linearizes the rotation in $|{\bm\theta}|$.
(Here, ${\bm e}_z$ is the unit vector in the $z$ direction.)
Writing $p_{iz}$ as $m_i \equiv x_i E$, the above transformation is
\begin {equation}
   \p_i \to \p_i + m_i {\bm\xi} = \p_i + x_i E {\bm\xi}.
\label {eq:ptransform}
\end {equation}
(In the language of two-dimensional quantum mechanics, this is
Galilean boost invariance, with $\bm\xi$ playing the role of the
relative velocity of the frames.%
\footnote{
   Three-dimensional rotations mix $(\b,z)$, whereas
   two-dimensional Galilean boost invariance mixes $(\b,t)$.
   The two correspond to each other here because $z \simeq t$ in the
   large-$p_z$ approximation.
}%
)
Because the rate we are computing is independent of
the rotations implemented by (\ref{eq:ptransform}), one
expects that the calculation only depends on invariant
combinations of the $\p_i$, such as $x_i \p_j - x_j \p_i$.
(We will give a more precise argument later.)
Because $\p_1+\p_2+\p_3=0$ and $x_1+x_2+x_3=0$ in our problem,
there is only one independent such combination:
\begin {equation}
   \P \equiv x_2 \p_1 - x_1 \p_2
   = x_3 \p_2 - x_2 \p_3
   = x_1 \p_3 - x_3 \p_1 .
\label {eq:P123}
\end {equation}
And, indeed, the 3-particle kinetic energy term in ${\cal H}$
(\ref{eq:calH2}) can
be rewritten in terms of $\P$ in the form of a 1-particle kinetic energy:
\begin {equation}
   \frac{\p_1^2}{2m_1} + \frac{\p_2^2}{2m_2} + \frac{\p_3^2}{2m_3}
   =
   \frac{\P^2}{2M}
\end {equation}
with
\begin {equation}
   M \equiv - x_1 x_2 x_3 E .
\label {eq:M3}
\end {equation}
From (\ref{eq:x123}) $M = x(1-x)E$, which is positive since
our $x_1$ is negative.

The most natural choice for a variable conjugate to
$x_j \p_i - x_i \p_j$ is $(\b_i-\b_j)/(x_i+x_j)$, noting that
the commutator
\begin {equation}
   \Bigl[ x_j \p_i - x_i \p_j \, , \frac{\b_i-\b_j}{(x_i+x_j)} \Bigr]
   = - i .
\end {equation}
Given the string of equalities in (\ref{eq:P123}), one might
expect that in our problem there is correspondingly a single
independent transverse position variable $\B$ with
\begin {equation}
   \B \equiv \frac{\b_1-\b_2}{(x_1+x_2)}
   = \frac{\b_2-\b_3}{(x_2+x_3)}
   = \frac{\b_3-\b_1}{(x_3+x_1)} \,.
\label {eq:B123}
\end {equation}
This is indeed the case, but we will hold off on giving a more
precise argument until section \ref{sec:Bs}, where we will
simultaneously work out
the generalizations that we need for the double splitting calculation
(where we will additionally
reduce 4-particle quantum mechanics to 2-particle quantum
mechanics).  Note that we could use $x_1+x_2+x_3=0$ to simplify
$x_1+x_2$ to $-x_3$ and so on in (\ref{eq:B123}), but the
combinations in (\ref{eq:B123}) turn out to be the ones that
generalize nicely to the case of more than 3 particles.

Let's now slightly change notation for the translation-invariant
3-particle potential in ${\cal H}$ by writing
\begin {equation}
   V(\b_1,\b_2,\b_3;t) = U(\b_1{-}\b_2,\b_2{-}\b_3,\b_3{-}\b_1;t) .
\end {equation}
Listing three $\b_i{-}\b_j$ combinations on the right-hand side is
redundant, since only two are independent,
but the redundancy makes manifest the permutation symmetries.
Taking (\ref{eq:B123}) at face value for the moment, the
3-particle Hamiltonian (\ref{eq:calH2}) then reduces to
the 1-particle Hamiltonian
\begin {equation}
   {\cal H} =
   \frac{\P^2}{2M}
   + U\bigl( (x_1{+}x_2)\B,(x_2{+}x_3)\B,(x_3{+}x_1)\B;t \bigr) .
\label {eq:calHB}
\end {equation}
As an example,
the potential (\ref{eq:Vqcd}) for weakly-coupled plasmas would
give (now using $x_1+x_2=-x_3$, etc.)%
\footnote{
  Here again, as in footnote \ref{foot:weak}, the $g^2$ characterizes
  interactions with the medium.
}
\begin {multline}
   {\cal H} =
   \frac{\P^2}{2M}
      -i g^2 \biggl[
         \tfrac12(C_1{+}C_2{-}C_3) \,
             \Delta\Gamma_{\rm el}(-x_3\B,t)
\\
         + \tfrac12(C_2{+}C_3{-}C_1) \,
             \Delta\bar\Gamma_{\rm el}(-x_1\B,t)
         + \tfrac12(C_3{+}C_1{-}C_2) \,
             \Delta\bar\Gamma_{\rm el}(-x_2\B,t)
      \biggr] ,
\end {multline}
which is equivalent (with different notation and slight generalization)
to the formalism originally used by BDMPS-Z.
As another example (not restricted to weak coupling),
the general harmonic oscillator approximation of
(\ref{eq:VqcdHO}) gives
\begin {subequations}
\label {eq:calH3}
\begin {equation}
   {\cal H} =
   \frac{\P^2}{2M}
      + \tfrac12 M \, \Omega_0^2(t) \,  \B^2
\end {equation}
with
\begin {equation}
   \Omega_0^2 = 
   -\frac{i}{2E}
    \left(
       \frac{\hat q_1}{x_1} + \frac{\hat q_2}{x_2} + \frac{\hat q_3}{x_3}
    \right) .
\label {eq:Omega0B}
\end {equation}
\end {subequations}

% ------------------------------------------------------------------------

\subsection{Finishing up}

In QCD, bremsstrahlung gluons can interact directly with the medium,
and so
the intermediate and final momenta
$\k_\xx$ and $\k_\fx$ need not be the same in fig.\ \ref{fig:xxrate2}.
The single splitting probability (\ref{eq:1brem2}) for QED
correspondingly generalizes to
\begin {multline}
   \frac{dI}{dx}
   = 2 \Re \Biggl\{
       \frac{E}{2\pi V_\perp}
       \int_{t_\xx < t_\xbx} dt_\xx \> dt_\xbx
       \sum_{\rm pol.}
       \int_{\p_\fx,\k_\fx}
       \Bigdlangle
       \biggl(
         \int_{\p_\xx,\k_\xx}
         \langle \p_\fx \k_\fx, t_\xbx | \p_\xx \k_\xx,t_\xx \rangle
         \langle \p_\xx \k_\xx | {-}i \, \delta H | \p_\ix \rangle
       \biggr)
\\
       \times
       \biggl(
         \int_{\bar\p_\xbx}
         \langle \p_\ix,t_\xx| \bar\p_\xbx, t_\xbx \rangle
         \langle  \bar\p_\xbx| i \, \delta H |\p_\fx \k_\fx \rangle
       \biggr)
       \Bigdrangle
   \Biggr\} .
\label {eq:1brem2qcd}
\end {multline}
(The discussion in this section applies equally well
to $q\to qg$, $g\to gg$, and $g\to q\bar q$,
though in this paper we will eventually focus solely on gluons.)
We will later explicitly discuss the details of reformulating
this formula in terms of $\P$ while
getting all the factors right.  For now, we jump ahead to review the
general expression for the final result, which is
\begin {equation}
   \frac{dI}{dx}
   =
   \frac{\alpha P_{1\to 3}(x)}{[x(1-x)E]^2}
   \Re \int_{t_\xx < t_\xbx} dt_\xbx \> dt_\xx
   \int_{\P_\xbx,\P_\xx}
   \P_\xbx \cdot \P_\xx \,
   \langle \P_\xbx,t_\xbx | \P_\xx,t_\xx \rangle ,
\label {eq:dI123P}
\end {equation}
where $P_{1\to 3}(x)$ is the usual (spin-averaged) DGLAP splitting function.
In this formula, $\langle \P_\xbx,t_\xbx | \P_\xx,t_\xx \rangle$
is the propagator of the 1-particle quantum mechanics problem of
(\ref{eq:calHB}).  The two $\delta H$ matrix elements in
(\ref{eq:1brem2qcd}) are responsible for the two factors of $\P$
in (\ref{eq:dI123P}) and for the factor of $\alpha P_{1\to 3}(x)$.
One may alternatively work in $\B$ space instead of $\P$ space,
in which case the above formula is \cite{Zakharov}
\begin {equation}
   \frac{dI}{dx}
   =
   \frac{\alpha P_{1\to 3}(x)}{[x(1-x)E]^2}
   \Re \int_{t_\xx < t_\xbx} dt_\xbx \> dt_\xx
   \grad_{\B_\xbx} \cdot \grad_{\B_\xx}
   \langle \B_\xbx,t_\xbx | \B_\xx,t_\xx \rangle
   \Bigr|_{\B_\xbx = \B_\xx = 0} .
\label {eq:dI123}
\end {equation}
We will find it convenient to work in $\B$ space when later deriving our
results for double splitting.
In the case of a thick and homogeneous medium, (\ref{eq:dI123}) reduces
to
\begin {equation}
   \frac{d\Gamma}{dx}
   =
   \frac{\alpha P_{1\to 3}(x)}{[x(1-x)E]^2}
   \Re \int_0^\infty d(\Delta t) \>
   \grad_{\B_\xbx} \cdot \grad_{\B_\xx}
   \langle \B_\xbx,\Delta t | \B_\xx,0 \rangle
   \Bigr|_{\B_\xbx = \B_\xx = 0}
\label {eq:drate123}
\end {equation}
for the differential rate of single splitting.

In order to make use of this formula, one could calculate numerical
results for the propagator \cite{ZfullV,SGfullV}.
With the harmonic oscillator
approximation, however, one can be more analytic.  For example,
for a thick, homogeneous medium,
\begin {equation}
   \langle \B,\Delta t | \B',0 \rangle
   =
   \frac{M\Omega_0\csc(\Omega_0 \, \Delta t)}{2\pi i} \,
      \exp\Bigl(
        \tfrac{i}2 M\Omega_0
        \bigl[ (\B^2+\B'^2) \cot(\Omega_0 \Delta t)
              - 2\B\cdot\B' \csc(\Omega_0 \Delta t) \bigr]
      \Bigr) .
\label {eq:1prop}
\end {equation}
Once one takes care of a small technical issue, using this propagator
in (\ref{eq:drate123}) gives the long-known result (\ref{eq:BDMS}).

The technical issue has to do with appropriately handling a
$\Delta t\to 0$ divergence in the integration in (\ref{eq:drate123}),
which we will defer to section \ref{sec:single1eps} where a
thorough discussion will teach us how to handle more substantial
issues in the calculation of double splitting.

% =========================================================================

\section{Reducing \boldmath$N$ particle to
         \boldmath$N{-}2$ particle quantum mechanics}
\label {sec:Bs}

% --------------------------------------------------------------------------

\subsection{Reduction from 2-dimensional point of view}
\label {sec:Bs2}

We now give a general argument about reducing $N$ particle
quantum mechanics to effectively
$N{-}2$ particles in the context of splitting rate calculations.
First, we will identify a special sub-space of the Hilbert space
that the $N$-particle Hamiltonian ${\cal H}$ of (\ref{eq:calH})
leaves invariant, in the special case that
\begin {equation}
   \sum_{i=1}^N m_i = 0 .
\label {eq:mconstraint}
\end {equation}
Then we will show that this is the relevant sub-space for our
problem.

As already noted, transverse translation invariance
means that total
transverse momentum  $\sum_i \p_i$ is
conserved,
\begin {equation}
   \bigl[ {\cal H} \, , \ssum_i \p_i \bigr] = 0 ,
\end {equation}
so that we may focus in particular on the subspace
\begin {equation}
   \ssum_i \p_i = 0
\label {eq:pconstraint}
\end {equation}
relevant to our problem.  We will show that one may simultaneously
impose the constraint that
\begin {equation}
   \ssum_i x_i \b_i = 0 ,
\label {eq:bconstraint}
\end {equation}
and that this also characterizes the relevant sector for our problem.
From the form (\ref{eq:calH}) of ${\cal H}$,
\begin {equation}
   -i \bigl[ {\cal H} \, , \ssum_i m_i \b_i \bigr] = \ssum_i \p_i ,
\label {eq:Hmb}
\end {equation}
which vanishes in the center-of-mass frame (\ref{eq:pconstraint}).
In ordinary quantum mechanics, the relation (\ref{eq:Hmb}) just says
that total mass times the time derivative of center-of-mass
position equals total momentum.  If the total momentum
vanishes, as in (\ref{eq:pconstraint}), then
$[ {\cal H}, \sum_i m_i \b_i ] = 0$.  However, this does not
normally imply that the operator $\sum_i m_i \b_i$ is
constant when in the center-of-mass frame because
\begin {equation}
   -i \bigl[ \ssum_i m_i \b_i \, , \ssum_j \p_j \bigr] = \ssum_i m_i
\label {eq:mbpcomm}
\end {equation}
does not normally vanish, and so we may not simultaneously choose
$\sum_i m_i \b_i$ constant and $\sum_i \p_i = 0$.
In our case (\ref{eq:mconstraint}),
however, (\ref{eq:mbpcomm}) does vanish, and so
both (\ref{eq:pconstraint}) and (\ref{eq:bconstraint}) may be
imposed simultaneously.
Time evolution with ${\cal H}$ will keep us in this sector
if we start there.

Do we start there?  For the sake of concreteness, begin by
considering the case of
single splitting, as in fig.\ \ref{fig:xxrate2} and
eq.\ (\ref{eq:1brem2qcd}).  First consider the state of the
system before any splittings.
We start with $|\p_\ix\rangle\langle\p_\ix|$,
which in our language we would call a ``2-particle'' initial state
\begin {equation}
   |\p_1,\p_2\rangle = |{-}\p_\ix,\p_\ix\rangle
\label {eq:initial1}
\end {equation}
in $\bar\Hilbert_{(1)} \otimes \Hilbert_{(1)}$, with
$(x_{\ix1},x_{\ix2})=(-1,1)$.
Because of the symmetry of the problem, we may formally average over
the symmetry transformations (\ref{eq:ptransform}) without changing
the final answer.
This replaces $|\p_\ix\rangle\langle\p_\ix|$ by something
proportional to
\begin {equation}
   \int_{\p_\ix} |\p_\ix\rangle\langle\p_\ix|
   =
   \int_{\b_\ix} |\b_\ix\rangle\langle\b_\ix| ,
\end {equation}
which is
\begin {equation}
   \int_{\p_\ix} |{-}\p_\ix,\p_\ix\rangle
   =
   \int_{\b_\ix} |\b_\ix,\b_\ix\rangle
\label {eq:initavg}
\end {equation}
in the language of $\bar\Hilbert_{(1)} \otimes \Hilbert_{(1)}$.
From the left-hand side of (\ref{eq:initavg}), note that we start
in the sector $\sum_i\p_i = \p_1+\p_2=0$; from the right-hand side, we
simultaneously start in the sector
$\sum_i x_i\b_i = x_{\ix1} \b_1+x_{\ix2}\b_2 = -\b_1+\b_2=0$.

\begin {figure}[t]
\begin {center}
  \includegraphics[scale=0.5]{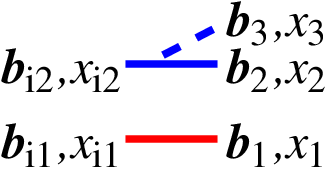}
  \caption{
     \label{fig:xxrateB}
     The notation used in (\ref{eq:preservexb}) for the state of the
     system just before and just after the first splitting in
     fig.\ \ref{fig:xxrate2}.
  }
\end {center}
\end {figure}

Now we need to ask whether having
$x_{\ix1} \b_{\ix1} + x_{\ix2} \b_{\ix 2} = 0$
before the splitting at $t=t_\xx$ in fig.\ \ref{fig:xxrate2}
means that we will have
$x_1 \b_1 + x_2 \b_2 + x_3 \b_3 = 0$ just after the splitting.
At $t=t_\xx$, the conjugate particle
(particle 1) is just a spectator, and so $(\b_1,x_1) = (\b_{\ix 1},x_{\ix 1})$.
(See fig.\ \ref{fig:xxrateB} for reference.)
As to the remaining particles:
the splitting vertex is local, and so $\b_2 = \b_3 = \b_{\ix 2}$.
Conservation of longitudinal momentum gives $x_2+x_3 = x_{\ix 2}$.
Putting this together gives
\begin {equation}
   x_{\ix1} \b_{\ix 1} + x_{\ix2} \b_{\ix 2} =
   x_1 \b_1 + x_2 \b_2 + x_3 \b_3 ,
\label {eq:preservexb}
\end {equation}
and so the condition of $\sum_i x_i \b_i=0$ is preserved by splitting.
This argument works for any splitting with any number of particles,
and so the condition is preserved by all the splittings in
the double-splitting diagrams of figs.\ \ref{fig:subset} and
\ref{fig:subset2} as well.

For the 3-particle case,
the constraint $x_1\b_1 + x_2\b_2 + x_3\b_3 = 0$
(along with $x_1+x_2+x_3=0$) explains the string of equalities
asserted previously in (\ref{eq:B123}),
\begin {equation}
   \frac{\b_1-\b_2}{(x_1+x_2)}
   = \frac{\b_2-\b_3}{(x_2+x_3)}
   = \frac{\b_3-\b_1}{(x_3+x_1)} \,.
\end {equation}
Applying these constraints to the 3-particle Hamiltonian ${\cal H}$
of (\ref{eq:calH2}) allows us to reduce ${\cal H}$ to the effective
1-particle Hamiltonian (\ref{eq:calHB}), as described previously.

% -------------------------------------------------------------------------

\subsection{Reduction from 3-dimensional point of view}

The constraint on $\sum_i x_i \b_i$ may also be understood in 3-dimensional
language as a consequence of invariance under rotations that
change the direction of the $z$ axis by a tiny amount, consistent
with our large-$p_z$ approximations.  The conserved angular momentum
is%
\footnote{
  For our purposes, the medium does not need to be invariant under
  large rotations.  It is enough for it to be invariant over transverse
  distances probed during the formation time.  For example,
  $\Delta b \sim 1/Q_\perp \sim (\hat q \omega)^{-1/4}$ for thick
  media, where $\omega$ is the smallest energy of a daughter.
}
\begin {equation}
   \angJ^{(3)} =
   \Bigl(\sum_i {\bm r}_i^{(3)} \!\times \p_i^{(3)}\Bigr) + \angS^{(3)} ,
\label{eq:frakJ}
\end {equation}
where the superscript ``(3)'' indicates 3-vectors and
$\angS$ is the spin contribution to the angular momentum
from the helicities of the particles.
(In terms of the angular momentum $\bm J$ in the amplitude and $\bar{\bm J}$
in the conjugate amplitude, $\angJ = {\bm J} - \bar{\bm J}$.)
In the high-energy, nearly collinear limit,
(i) the $z$ positions of the particles are
all $\simeq t$, and (ii) $\angS$ will be in the $\pm z$
direction.  Then the transverse piece $\angJ\!_\perp$
of (\ref{eq:frakJ}) is given by
\begin {equation}
   \widetilde \angJ\!_{\perp} \simeq
   \ssum_i \b_i x_i E - t \, \ssum_i \p_i ,
\end {equation}
where $\widetilde\angJ_{\!\perp}^a \equiv \epsilon^{ab}\angJ_{\!\perp}^b$ is
the 2-dimensional dual of $\angJ\!_\perp$.
Since $\ssum_i \p_i =0$, conservation of $\angJ\!_\perp$
then gives conservation of $\ssum_i x_i \b_i$ (in the large-$p_z$
limit).  Since rotational invariance is a property of the
full Hamiltonian (including the $\delta H$ for splitting), this
gives conservation of $\ssum_i x_i \b_i$ over the entire process.

% =========================================================================

\section {Double Splitting using just \boldmath$N{-}2$ ``particles''}
\label{sec:double}

We now turn to writing expressions for contributions to double
splitting, focusing first on the $xy\bar y\bar x$ interference of
fig.\ \ref{fig:xyyxrate}.  Before we start,
we should clarify something about the number of particles involved
at various times in this diagram.

% ------------------------------------------------------------------------

\subsection {\boldmath$N{=}5$
             particle evolution unnecessary}
\label {sec:N5}

\begin {figure}[t]
\begin {center}
  \includegraphics[scale=0.5]{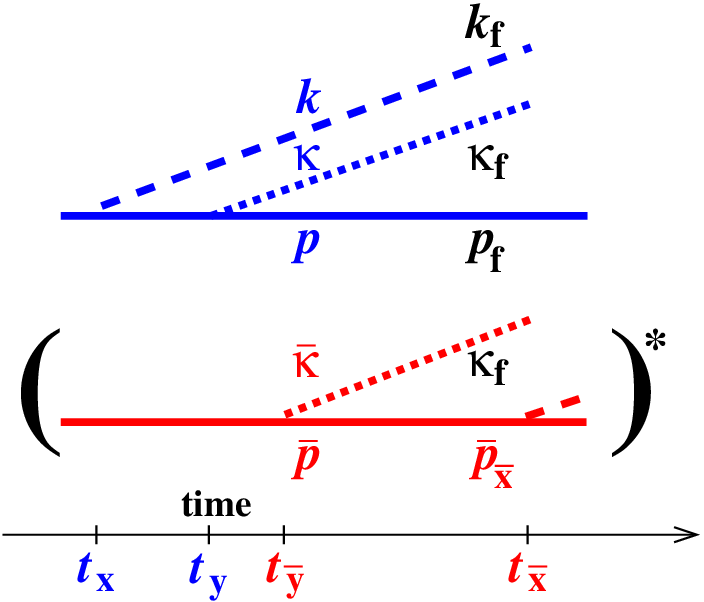}
  \caption{
     \label{fig:5particle}
     Labeling in (\ref{eq:5particle}) for momenta at beginning and end
     of time interval $t_\ybx < t < t_\xbx$.
  }
\end {center}
\end {figure}

The right side of fig.\
\ref{fig:xyyxrate} is drawn in a way that indicates only three
particles (two in the amplitude and one in the conjugate amplitude)
are involved in the final time interval, $t_\ybx < t < t_\xbx$.
However, based on the left-hand side of the figure, one
might think there are five particles in that time region:
all three final daughters ($x$, $y$, and $1{-}x{-}y$) in the amplitude,
and two particles (a final daughter $y$ and an intermediate
daughter/parent $1{-}y$) in the conjugate amplitude.
That is, the terms corresponding to the evolution during
the time period $t_\ybx < t < t_\xbx$ would be of the form
\begin {equation}
   \int_{\bkappa_\fx}
   \Bigdlangle
      \langle \p_\fx \k_\fx \bkappa_\fx, t_\xbx|
              \p \k \bkappa, t_\ybx\rangle \,
      \langle \bar\p_\xbx \bkappa_\fx, t_\xbx|
              \bar\p \bar\bkappa, t_\ybx\rangle^*
   \Bigdrangle ,
\label {eq:5particle}
\end {equation}
with momenta labeled as in fig.\ \ref{fig:5particle} and
where we have also included the integration over the final-state
momentum ${\bm\kappa}_\fx$ of the $y$ daughter.  We can write
this in 5-particle notation as
\begin {equation}
   \int_{\bkappa_\fx}
   \Bigdlangle
      \langle {-}\bar\p_\xbx, {-}\bkappa_\fx,\bkappa_\fx,\p_\fx,\k_\fx, t_\xbx|
              {-}\bar\p, {-}\bar\bkappa,\bkappa,\p,\k, t_\ybx\rangle
   \Bigdrangle .
\end {equation}
The 5-particle problem could be reduced to an effective 3-particle
problem using the constraints of the last section.

Now recall that in the interval between splittings, the high-energy
particles do not directly interact with each other but instead
simply propagate in the background fields of the medium, which
are later averaged with $\dlangle \cdots \drangle$.
So, when computing
$\langle \p_\fx \k_\fx \bkappa_\fx, t_\xbx|
         \p \k \bkappa, t_\ybx\rangle$
in (\ref{eq:5particle})
for a particular background field configuration (before averaging),
the three particles in the amplitude evolve independently with
the 1-particle Hamiltonian $H_{(1)}$ appropriate to that particle.
That is,
\begin {equation}
      \langle \p_\fx \k_\fx \bkappa_\fx, t_\xbx|
              \p \k \bkappa, t_\ybx\rangle
      =
      \langle \p_\fx, t_\xbx|\p, t_\ybx\rangle
      \langle \k_\fx, t_\xbx|\k, t_\ybx\rangle
      \langle \bkappa_\fx, t_\xbx|\bkappa, t_\ybx\rangle ,
\end {equation}
and similarly for the two particles in the conjugate amplitude.
We may therefore rewrite (\ref{eq:5particle}) as
\begin {equation}
   \Bigdlangle
      \langle \p_\fx \k_\fx, t_\xbx|
              \p \k, t_\ybx\rangle \,
      \langle \bar\p_\xbx, t_\xbx|
              \bar\p, t_\ybx\rangle^*
      \int_{\bkappa_\fx}
         \langle \bkappa_\fx, t_\xbx|\bkappa, t_\ybx\rangle \,
         \langle \bkappa_\fx, t_\xbx|\bar\bkappa, t_\ybx\rangle^*
   \Bigdrangle .
\end {equation}
{\it Inside}\/ the double angle brackets
(before averaging), the 1-particle evolution
of the $y$ daughter is unitary, and so
we may use the sum over the final states of the $y$ daughter to
simplify the above expression to
\begin {equation}
   \Bigdlangle
      \langle \p_\fx \k_\fx, t_\xbx|
              \p \k, t_\ybx\rangle \,
      \langle \bar\p_\xbx, t_\xbx|
              \bar\p, t_\ybx\rangle^*
   \Bigdrangle
   \, (2\pi)^2 \delta^{(2)}(\bkappa-\bar\bkappa) .
\end {equation}
What is left is what we would call (medium averaged) 3-particle evolution
(the $x$ and $1{-}x{-}y$ daughters in the amplitude, and the
intermediate $1{-}y$ daughter/parent in the conjugated amplitude),
which can be reduced to effective 1-particle evolution using
the method of the previous section.
The lesson is that the sum over final states of a final-state
daughter may be performed as soon as it has been emitted
from both the amplitude
and conjugate amplitude, and one need not worry about its interactions
after that.%
\footnote{%
\label {foot:scales2}%
There is a hidden assumption in this argument, which is that when we
  take the statistical average $\dlangle\cdots\drangle$ of the background
  fields we may ignore correlations of interactions just before the
  splitting time $t_\ybx$ with interactions just after $t_\ybx$, within
  one correlation length of the plasma.
  This sort of assumption is implicit in many aspects of this
  work and standard treatments of the LPM effect, and its justification
  is that the effects of the background over a single correlation length
  about $t = t_\ybx$ should be parametrically small compared to the
  effects over the entire formation time $t_{\rm form} \propto E^{1/2}$.
  This is similar to the justification of footnote \ref{foot:scales}
  for using an instantaneous potential $V$ in ${\cal H}$, and the
  same sort of argument is implicit for our treatment of all the
  splittings in this paper.
}

% ------------------------------------------------------------------------

\subsection {First expression for \boldmath$xy\bar y\bar x$ interference}
\label{sec:firstxyyx}

As mentioned before, the diagrams in fig.\ \ref{fig:subset2} may
therefore be treated as 3-particle evolution followed by
4-particle evolution followed by 3-particle evolution.
After reducing each $N$ particle problem to an $(N{-}2)$ particle
problem, we may reinterpret each diagram as
effectively 1-particle evolution followed by
2-particle evolution followed by 1-particle evolution.
We will work in $\b$-space with the variables
\begin {equation}
   \B_{ij} \equiv \frac{\b_i-\b_j}{x_i + x_j} ,
\label {eq:Bij}
\end {equation}
of which $N{-}2$ are independent for $N$ particle evolution, once we
impose the constraint $\sum_ix_i\b_i=0$.

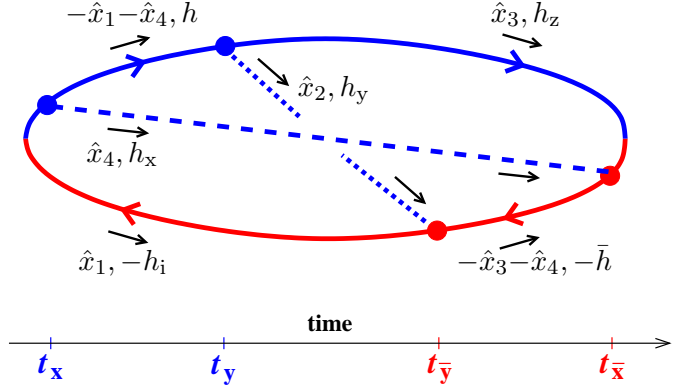
\begin {figure}[t]
\begin {center}
  \begin{picture}(250,150)(0,0)
  \put(0,0){\includegraphics[scale=0.5]{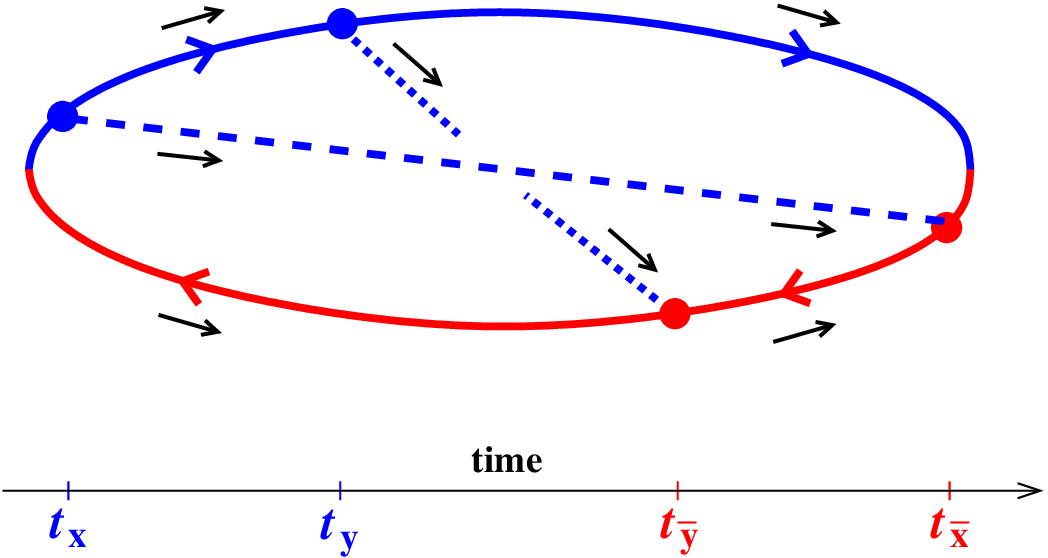}}
  \put(27,44){$\hat x_1,{-}h_{\rm i}$}
  \put(30,85){$\hat x_4,h_{\rm x}$}
  \put(22,138){${-}\hat x_1{-}\hat x_4,h$}
  \put(110,110){$\hat x_2,h_{\rm y}$}
  \put(183,138){$\hat x_3,h_{\rm z}$}
  \put(170,44){${-}\hat x_3{-}\hat x_4,{-}\bar h$}
%  \put(0,0){.}
%  \put(0,150){.}
%  \put(250,0){.}
%  \put(250,150){.}
  \end{picture}
  \caption{
     \label{fig:xyyx}
     Our labeling conventions for longitudinal momenta $x_i$ and
     helicities $h_i$ for the $xy\bar y\bar x$ interference diagram.
  }
\end {center}
\end {figure}

For now, we focus attention on the $xy\bar y\bar x$ interference
depicted by fig.\ \ref{fig:xyyxrate}
and the first diagrams in figs.\ \ref{fig:subset} and
\ref{fig:subset2}.  We will see later that the other interference
diagrams can be related to the $xy\bar y\bar x$ calculation,
and so most of our task will be the evaluation of $xy\bar y\bar x$.
For the $N{=}4$ particle evolution, we label the $x_i$ as
shown in fig.\ \ref{fig:xyyx}, so that
\begin {equation}
   (\hat x_1,\hat x_2,\hat x_3,\hat x_4) = (-1,y,1{-}x{-}y,x) .
\label {eq:xhat}
\end {equation}
(Our numbering convention is not quite arbitrary: avoiding
having $x$ and $y$ cyclically sequential in this list will turn out to
be convenient for discussing the large-$\Nc$ limit.)
For the sake of avoiding confusion, we find it convenient to put
hats atop the $x_i$ when specifically referring to the $x_i$ that
characterize the $N{=}4$ evolution of $xy\bar y\bar x$,
and we will often use these
variables in other time intervals as well.
For example, the three
longitudinal momentum fractions associated with the initial
3-particle evolution for $t_\xx < t < t_\yx$ are
\begin {equation}
   (x_1,x_2,x_3)
   = (\hat x_1,{-}\hat x_1{-}\hat x_4,\hat x_4)
   = (\hat x_1,\hat x_2{+}\hat x_3,\hat x_4) = (-1,1{-}x,x) ,
\end {equation}
and those associated with the final 3-particle evolution for
$t_\ybx < t < t_\xbx$ are
\begin {equation}
   (x_1,x_2,x_3)
   = ({-}\hat x_3{-}\hat x_4,\hat x_3,\hat x_4)
   = (\hat x_1{+}\hat x_2,\hat x_3,\hat x_4)
   = \bigl( {-}(1{-}y), 1{-}x{-}y, x) .
\end {equation}
For the sake of later relating results for $xy\bar y\bar x$ to
other interference contributions, it will be useful to keep
many things in terms of the $\hat x_i$ instead of writing formulas directly
in terms of $x$ and $y$.

In order to help emphasize which variables apply to 4-particle
evolution and which to 3-particle evolution, we will refer
to the $\B_{ij}$ of (\ref{eq:Bij}) by the symbol $\C_{ij}$ in the
4-particle case.
For the splitting matrix elements at the intermediate time
$t_\yx$, it will turn out to be convenient to choose our two
variables describing the 4-particle (effectively 2-particle)
system to be
(i) the
$\C_{ij}$ involving the two spectators to the splitting, and
(ii) the $\C_{ij}$
involving the two particles
involved in the splitting.
The same will be true at the other intermediate time $t_\ybx$.
So, referring to fig.\ \ref{fig:xyyx},
the convenient 4-particle variables will turn out to be
$(\C_{41},\C_{23})$ at $t_\yx$ but $(\C_{34},\C_{12})$ at
$t_\ybx$.  With these conventions, the $xy\bar y\bar x$
interference contribution to double splitting has the form
\begin {align}
   \left[\frac{dI}{dx\,dy}\right]_{xy\bar y\bar x}
   =
   \left( \frac{E}{2\pi} \right)^2
   &
   \int_{t_\xx < t_\yx < t_\ybx < t_\xbx}
   \sum_{\rm pol.}
   \langle|i\,\overline{\delta H}|\B^\Bx\rangle \,
   \langle\B^\Bx,t_\Bx|\B^\Ax,t_\Ax\rangle
\nonumber\\ &\times
   \langle\B^\Ax|i\,\overline{\delta H}|\C_{34}^\Ax,\C_{12}^\Ax\rangle \,
   \langle\C_{34}^\Ax,\C_{12}^\Ax,t_\Ax|\C_{41}^\bx,\C_{23}^\bx,t_\bx\rangle
\nonumber\\ &\times
   \langle\C_{41}^\bx,\C_{23}^\bx|{-}i\,\delta H|\B^\bx\rangle \,
   \langle\B^\bx,t_\bx|\B^\ax,t_\ax\rangle \,
   \langle\B^\ax|{-}i\,\delta H|\rangle ,
\label {eq:xyyx2}
\end {align}
where all intermediate transverse momenta are implicitly integrated
over---i.e.\ there is integration over
$\B^\Bx$, $\B^\Ax$, $\C_{34}^\Ax$, $\C_{12}^\Ax$, $\C_{41}^\bx$, $\C_{23}^\bx$,
$\B^\bx$, and $\B^\ax$ above.
The superscripts are just labels for the different intermediate
states (and are written as superscripts rather than subscripts
for the sake of compactness of names like $\C_{34}^\Ax$).
The initial $N{=}2$ (effectively $N{-}2{=}0$) state is represented by
$|\rangle$ and will be discussed later, along with the final state
$\langle|$.
The specific particles involved
in each $\delta H$ splitting above should be inferred from the
$xy\bar y\bar x$ diagram in fig.\ \ref{fig:xyyx}; we have
written $\overline{\delta H}$ above to indicate splittings
that are in the conjugate amplitude.
For comparison,
the single splitting result, corresponding to
fig.\ \ref{fig:xxrate} and its conjugate,
has the form
\begin {equation}
   \frac{dI}{dx}
   = 2 \Re 
   \left[\frac{dI}{dx}\right]_{x\bar x}
\end {equation}
with
\begin {equation}
   \left[\frac{dI}{dx}\right]_{x\bar x}
   =
   \frac{E}{2\pi}
   \int_{t_\xx < t_\xbx}
   \sum_{\rm pol.}
   \langle|i\,\overline{\delta H}|\B^\Bx\rangle \,
   \langle\B^\Bx,t_\Bx|\B^\ax,t_\ax\rangle \,
   \langle\B^\ax|{-}i\,\delta H|\rangle
\end {equation}
in this notation.

We shall discuss the $\delta H$ matrix elements in sections
\ref{sec:states} and \ref{sec:dH}.
They have the form
\begin {equation}
   \langle \B | \delta H |\rangle
   = -i g \bcalT_{i \to jk} \cdot \grad \delta^{(2)}(\B)
\label {eq:dH32}
\end {equation}
for parent $i$ to split into daughters $j$ and $k$.  Here
the two-dimensional vector ${\bm{\mathcal T}}_{i \to jk}$ has the form
\begin {equation}
  \bcalT_{i \to jk} =
  \frac{ T^{\rm color}_{i \to jk} \bcalP_{i \to jk} }{2 E^{3/2}} \,,
\end {equation}
where $T^{\rm color}$ is the color factor associated with the vertex
[e.g.\ $(T_{\rm A}^a)_{bc} = -i f^{abc}$ for gluon splitting $g \to gg$ with color
indices $a \to bc$].
${\bm{\mathcal P}}$ is related to square roots of
spin-dependent DGLAP splitting functions and depends on the
$x_i$ and the helicities of the parent and daughter particles in
$i \to jk$ (which also determine the helicity of the vector
$\bcalP$).  More details will be given later.

\begin {figure}[t]
\begin {center}
  \begin{picture}(400,110)(0,0)
%  \begin{picture}(140,110)(0,0)
  \put(25,20){\includegraphics[scale=0.7]{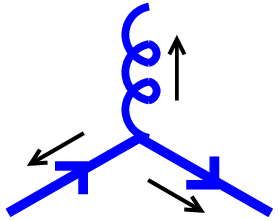}}
  \put(5,10){$\b_i,a_i$}
  \put(115,10){$\b_j,a_j$}
  \put(60,100){$\b_k,a_k$}
  \put(20,48){$x_i,h_i$}
  \put(70,15){$x_j,h_j$}
  \put(88,70){$x_k,h_k$}
  \put(130,55){$\displaystyle{
       = \quad - \frac{g (T_R^{a_k})_{a_j,a_i}}{2 E^{3/2}}
       \> \bcalP_{h_i h_j h_k}\!(x_i,x_j,x_k) \cdot
       \grad \delta^{(2)}({\bcalB}_{ji})
    }$}
%  \put(0,0){.}
%  \put(0,110){.}
%  \put(350,0){.}
%  \put(350,110){.}
  \end{picture}
  \\[10pt]
  \begin{picture}(400,130)(0,0)
%  \begin{picture}(140,110)(0,0)
  \put(25,2){\includegraphics[scale=0.7]{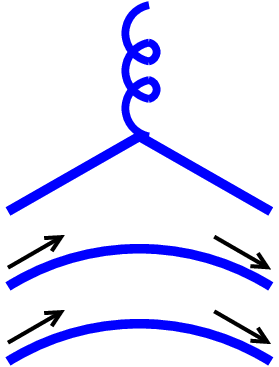}}
  \put(11,0){$\b_n$}
  \put(120,0){$\b_s$}
  \put(11,26){$\b_m$}
  \put(120,26){$\b_r$}
  \put(50,20){$x_n$}
  \put(84,20){$x_s$}
  \put(50,46){$x_m$}
  \put(84,46){$x_r$}
  \put(130,55){$\displaystyle{=}$}
  \put(150,28){\includegraphics[scale=0.7]{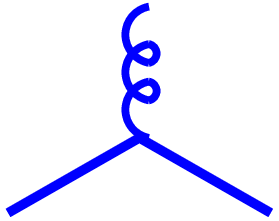}}
  \put(250,55){$\displaystyle{
       \times ~~
       |x_m+x_n|^{-1} \, \delta^{(2)}({\bcalB}_{mn} {-} {\bcalB}_{rs})
    }$}
%  \put(0,0){.}
%  \put(0,130){.}
%  \put(400,0){.}
%  \put(400,130){.}
  \end{picture}
  \caption{
     \label{fig:dH}
     The diagrammatic rules for splittings linking (via
     either $-i\,\delta H$ or $+i\,\overline{\delta H}$)
     the state
     $|\rangle$ to $|\B\rangle$ (top rule) or $|\B\rangle$ to
     $|\C_{34},\C_{12}\rangle$ or permutation thereof (bottom rule). 
     ${\bcalB}_{uv} \equiv (\b_u-\b_v)/(x_u+x_v)$ and may refer, in
     different contexts, to $\pm$ the 3-particle $\B$, or one
     of the 4-particle $\C_{uv}$, or to some mixture.  However,
     note that
     ${\bcalB}_{ji} = {\bcalB}_{kj} = {\bcalB}_{ik}$
     in the top rule, which can be used
     to always write expressions in terms of 3-particle $\B$ and/or
     4-particle $\C_{ij}$'s.
     The blue arrows on the particle line indicate color flow of
     color representation $R$.  (In the case of $R{=}{\rm A}$,
     appropriate to $g \to gg$ splitting, the direction of the color
     flow does not matter.)  $\b_l$, $a_l$, $x_l$, and $h_l$ indicate the
     transverse position, color index, longitudinal momentum, and
     helicity of each particle.  The black arrows give the
     convention for the flow of $x_l$ and $h_l$ in the
     statement of the rule, and these values should be negated if
     they are instead defined by flow in the opposite direction.
     In the bottom rule, color and helicity
     indices and their contractions are not explicitly
     shown for the spectators because they are trivially contracted.
     Conservation of longitudinal momentum means $x_i+x_j+x_k=0$ (top)
     and additionally $x_m=x_r$ and $x_n = x_s$ (bottom).
  }
\end {center}
\end {figure}

The corresponding matrix elements for $\delta H$ linking the
3-particle to 4-particle sector will be reducible to
the previous case (\ref{eq:dH32}).
For example, if the daughters
are particles 2 and 3, we will find that
\begin {align}
   \langle\C_{41},\C_{23}|\delta H|\B\rangle
   &=
     \langle\C_{23}|\delta H|\B\rangle \,
     |\hat x_4+\hat x_1|^{-1} \, \delta^{(2)}(\C_{41}-\B)
\nonumber\\
   &=
     -i g \bcalT_{i \to 23} \cdot \grad \delta^{(2)}(\C_{23}) \,
     |\hat x_4+\hat x_1|^{-1} \, \delta^{(2)}(\C_{41}-\B) .
\label {eq:dH43}
\end {align}
Diagrammatic rules which summarize all the splitting
matrix elements
needed are given in fig.\ \ref{fig:dH}.

Let's now apply these rules to the $xy\bar y\bar x$ interference
depicted in fig.\ \ref{fig:xyyx} for the case where $x$ and $y$
are bremsstrahlung gluons.  In the high-energy limit,
the interactions of high-energy particles
with soft background fields preserve the helicities of
the high-energy particles, and so helicities of individual
high-energy particles are preserved by the propagators in
fig.\ \ref{fig:xyyx}.  The color charges of individual particles
are not preserved, and so propagators like
$\langle\C_{34}^\Ax,\C_{12}^\Ax,t_\Ax|\C_{41}^\bx,\C_{23}^\bx,t_\bx\rangle$
in (\ref{eq:xyyx2}) generically depend on color.
As we'll discuss, the large-$\Nc$ limit that we will take in this
paper simplifies the color dynamics so that we may ignore
this complication.  Applying the rules of fig.\ \ref{fig:dH} to
fig.\ \ref{fig:xyyx},
naively contracting color indices, and using the $\delta$-functions
to perform as many integrals as possible, then yields
(see Appendix\ \ref{app:details} for more detail)
\begin {align}
   \left[\frac{dI}{dx\,dy}\right]_{xy\bar y\bar x}
   &=
   d_R^{-1} \tr(T_R^a T_R^b T_R^a T_R^b)
   \frac{\alphas^2 }{4 E^4}
   |\hat x_1+\hat x_4|^{-1} |\hat x_3+\hat x_4|^{-1}
   \int_{t_\xx < t_\yx < t_\ybx < t_\xbx}
   \sum_{h_\xx,h_\yx,h_\zx,h,\bar h}
   \int_{\B^\Ax,\B^\bx}
\nonumber\\ &\times
   \bcalP_{{-}h_\zx,\bar h,{-}h_\xx}({-}\hat x_3,\hat x_3{+}\hat x_4,{-}\hat x_4)
       \cdot \grad_{\B^\Bx}
   \langle\B^\Bx,t_\Bx|\B^\Ax,t_\Ax\rangle
   \Bigr|_{\B^\Bx=0}
\nonumber\\ &\times
   \bigl( \bcalP_{{-}\bar h,h_\ix,{-}h_\yx}
               (\hat x_1{+}\hat x_2,{-}\hat x_1,{-}\hat x_2)
          \cdot \grad_{\C_{12}^\Ax} \bigr)
   \bigl( \bcalP_{{-}h,h_\zx,h_\yx}({-}\hat x_2{-}\hat x_3,\hat x_3,\hat x_2)
          \cdot \grad_{\C_{23}^\bx} \bigr)
\nonumber\\ &\qquad\qquad
  \langle\C_{34}^\Ax,\C_{12}^\Ax,t_\Ax|\C_{41}^\bx,\C_{23}^\bx,t_\bx\rangle
   \Bigr|_{\C_{12}^\Ax=0=\C_{23}^\bx; ~ \C_{34}^\Ax=\B^\Ax; ~ \C_{41}^\bx=\B^\bx}
\nonumber\\ &\times
   \bcalP_{{-}h_\ix,h,h_\xx}(\hat x_1,{-}\hat x_1{-}\hat x_4,\hat x_4)
        \cdot \grad_{\B^\ax}
   \langle\B^\bx,t_\bx|\B^\ax,t_\ax\rangle
   \Bigr|_{\B^\ax=0} ,
\label {eq:IIxyyx1}
\end {align}
where $d_R$ is the dimension of color representation $R$.
For the case considered in this paper of an initial gluon,
the color factor is
\begin {equation}
   d_A^{-1} \tr(T_{\rm A}^a T_{\rm A}^b T_{\rm A}^a T_{\rm A}^b)
   = \tfrac12 \CA^2 = \tfrac12 \Nc^2 .
\end {equation}

% ------------------------------------------------------------------------

\subsection {Large-\boldmath$\Nc$ limit}

When discussing the single splitting rate in section \ref{sec:1gluon},
we were able to determine the dot products $T_i\cdot T_j$ of color
charges in the effective potential (\ref{eq:Vqcd0}) in
terms of fixed Casimirs $C_i$, and so we avoided having to deal
with color dynamics.  This simplification is no longer possible,
in general, when
analyzing the $N{=}4$ particle evolution that appears in
double-scattering:  $T_1+T_2+T_3+T_4=0$ is not sufficient to
determine all the $T_i\cdot T_j$ for 4 particles.
In this work, we sidestep this complication by focusing
on the large-$\Nc$ limit.

\begin {figure}[t]
\begin {center}
  \includegraphics[scale=0.8]{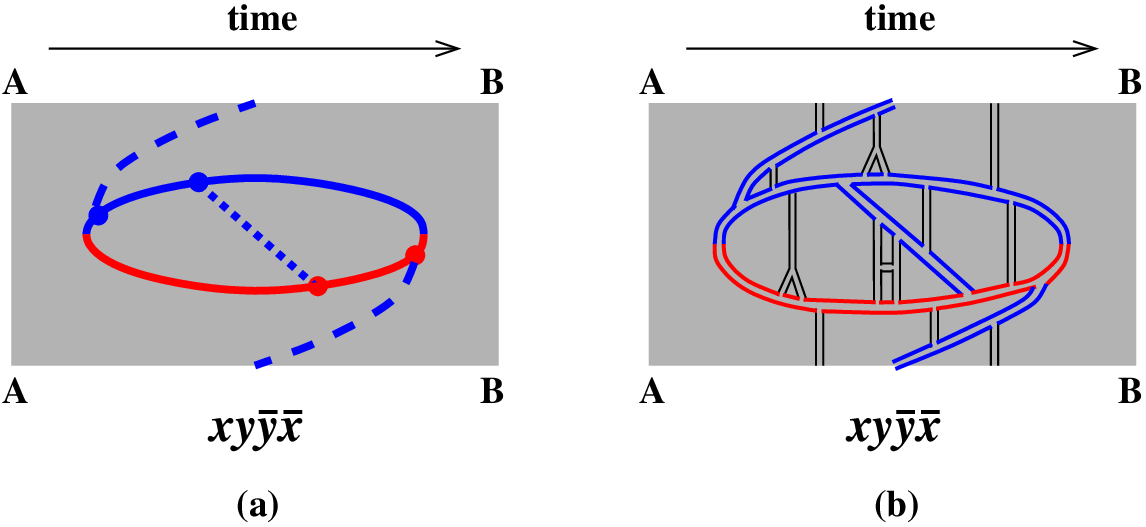}
  \caption{
     \label{fig:cylinder}
     (a) The $xy\bar y\bar x$ interference diagram drawn on a cylinder.
     The top edge AB of the shaded region is to be identified with the
     bottom edge AB.
     (b) The same showing color flow for an explicit example of
     medium background field correlations (black) that gives a planar
     diagram (and so leading-order in $1/\Nc$).
     Bonus craft project: Cut out these figures with children and
     glue into cylinders.  Collect and trade with friends!
  }
\end {center}
\end {figure}

Because of the kinematics of our problem (high energy particles
interacting with the plasma via momentum transfers that are soft
compared to $E$), it has been convenient in our analysis to
time-order our interference diagrams, as in fig.\ \ref{fig:subset2}.
To discuss the large-$\Nc$ limit, however, we would also like to draw
our diagrams as planar diagrams.  In the case that all particles
are gluons, the diagrams of fig.\ \ref{fig:subset2} can be
drawn as time-ordered planar diagrams by drawing them on a
cylinder (since a cylinder can be mapped into a plane),
as shown in fig.\ \ref{fig:cylinder}a for the $xy\bar y\bar x$ interference.

Fig.\ \ref{fig:cylinder}b shows the color flow of
fig.\ \ref{fig:cylinder}a decorated by an explicit example of
background-field interactions and correlations in the large-$\Nc$
limit.  The requirement of planarity means that each
fundamental-charge high-energy color line shown in
fig.\ \ref{fig:cylinder}b only interacts (via medium correlations)
with one other, and those interactions are independent of the
interactions of other color lines.  That is, the general $N$-particle
potential in ${\cal H}$ among our adjoint-charge particles
reduces to
\begin {equation}
  V(\b_1,\cdots\b_N;t) =
  V_{\rm F}(\b_2{-}\b_1;t)
  + V_{\rm F}(\b_3{-}\b_2;t)
  + \cdots
  + V_{\rm F}(\b_N{-}\b_{N-1};t)
  + V_{\rm F}(\b_1{-}\b_{N};t) ,
\end {equation}
where $V_{\rm F}$ is the large-$\Nc$ potential (in the language of
${\cal H}$) between a
fundamental charge and anti-fundamental charge particle,
and $\b_1$, $\b_2$, ... $\b_N$ label the high-energy particles
cyclically as one goes around the cylinder (e.g.\ from bottom to
top in fig.\ \ref{fig:cylinder}a).  In the harmonic oscillator
approximation, this becomes
\begin {equation}
  - \frac{i \hat q_{\rm F}}{4}
  \Bigl[ (\b_2-\b_1)^2 + (\b_3-\b_2)^2 + \cdots + (\b_N-\b_{N-1})^2
         + (\b_1-\b_N)^2 \Bigr] ,
\end {equation}
which (in the large-$\Nc$ limit) is the same as
\begin {equation}
  - \frac{i \hat q_{\rm A}}{8}
  \Bigl[ (\b_2-\b_1)^2 + (\b_3-\b_2)^2 + \cdots + (\b_N-\b_{N-1})^2
         + (\b_1-\b_N)^2 \Bigr] .
\label {eq:VlargeN}
\end {equation}
Note that for $N{=}3$ this agrees with the all-gluon case of
the more general formula
(\ref{eq:VqcdHO}), which did not depend on large $\Nc$.

% ------------------------------------------------------------------------

\subsection {Reduction of states from \boldmath$N$ to
             \boldmath$N{-}2$ particles}
\label{sec:states}

We now turn to an explicit construction of the $N{-}2$ particle
states such as $|\C_{34},\C_{12}\rangle$ appearing earlier.
The utility of an explicit construction is that it will allow us
to find the correct normalization factors associated with $\delta H$
matrix elements,
like the $|x_m+x_n|$ in the
bottom-right of fig.\ \ref{fig:dH}.

For $\sum_i x_i=0$,
we may project any $\b$-space state onto the subspace with
$\sum_i x_i \b_i = 0$ and $\sum_i \p_i =0$ by (i) imposing
$\sum_i x_i \b_i = 0$ in $\b$-space, and (ii) averaging over all
$\b$-space translations in order to project onto $\sum_i \p_i = 0$.
Treating all $N$ particles symmetrically, this procedure defines
\begin {equation}
   |\{\B_{ij}\}\rangle
   = {\cal N} \int_{\Delta\b}
       |\b_1+\Delta\b, \cdots \b_N+\Delta\b\rangle
       \Biggr|_{\sum x_i\b_i=0}  ,
\label {eq:project}
\end {equation}
where
$\{\B_{ij}\}$ represents the (not independent) set of
all of the values of
$\B_{ij} \equiv (\b_i-\b_j)/(x_i+x_j)$.
${\cal N}$ is an $x_i$-independent normalization factor
described in appendix \ref{app:project}.
In the appendix, we show that this particle-symmetric definition
of the projected states normalizes those states as
\begin {subequations}
\begin {align}
  \bigl\langle\{\C_{ij}\}\big|\{\C_{ij}'\}\bigr\rangle
  &= (\hat x_3+\hat x_4)^{-2} \, \delta^{(2)}(\C_{34}{-}\C'_{34}) \,
                     \delta^{(2)}(\C_{12}{-}\C'_{12})
  \qquad (N{=}4),
\label {eq:Csnorm}
\\
  \bigl\langle\{\B_{ij}\}\big|\{\B_{ij}'\}\bigr\rangle
  &= \delta^{(2)}(\B_{12}{-}\B'_{12})
  \qquad (N{=}3),
\label {eq:Bsnorm}
\\
  \langle|\rangle &= x_1^2
  \qquad (N{=}2),
\label {eq:N2norm}
\end {align}
\end {subequations}
where we have used our convention of referring to the 4-particle
variables as $\C_{ij}$ and $\hat x_i$.
One may use any permutation of indices
on the right-hand sides above.
Note that $(\hat x_3+\hat x_4)^2 = (\hat x_1+\hat x_2)^2$ in the
$N{=}4$ case and $x_1^2=x_2^2$ in the $N{=}2$ case.

The notation $|\C_{34},\C_{12}\rangle$ that we used earlier is related
to the above symmetrically-defined states by
\begin {equation}
  |\C_{34},\C_{12}\rangle
  \equiv |\hat x_3+\hat x_4| \, \bigl| \{\C_{ij}\} \bigr\rangle ,
\label{eq:C3412def}
\end {equation}
so that the $N{=}4$ states are more conveniently normalized as
\begin {equation}
  \langle\C_{34},\C_{12}|\C'_{34},\C'_{12}\rangle
  = \delta^{(2)}(\C_{34}{-}\C'_{34}) \,
                     \delta^{(2)}(\C_{12}{-}\C'_{12}) .
\label{eq:C3412norm}
\end {equation}
The $N{=}3$ state $|\B\rangle$ used earlier is simply
\begin {equation}
  |\B\rangle = \bigl| \{\B_{ij}\} \bigr\rangle .
\label {eq:Bdef}
\end {equation}

The normalization (\ref{eq:N2norm}) of $|\rangle$ may seem a bit strange,
but $|\rangle$ is what we want for the final state.  At the end,
after the emission of both the $x$ and $y$ bremsstrahlung gluons,
we want to sum over all possible final states of the emitter,
\begin {equation}
   \int_{\p_\fx} |\p_\fx\rangle \langle \p_\fx |
   =
   \int_{\b_\fx} |\b_\fx\rangle \langle \b_\fx |
\label {eq:end1}
\end {equation}
which in our $N$-particle notation is
\begin {equation}
   |{\rm end}\rangle = \int_{\b_\fx} |\b_\fx,\b_\fx\rangle .
\end {equation}
By a shift of integration variable, this is ${\cal N}^{-1}|\rangle$
with $|\rangle$ defined by (\ref{eq:project}) as
\begin {equation}
   |\rangle \equiv
   {\cal N}
   \int_{\Delta\b} |\b_1{+}\Delta\b,\b_2{+}\Delta\b\rangle
   \Bigr|_{x_1\b_1+x_2\b_2=0} ,
\label {eq:end3}
\end {equation}
noting that $x_1 \b_1 + x_2 \b_2=0$
enforces $\b_1 = \b_2$, given that $x_1+x_2=0$.
A similar argument can be made about the initial state if we first
use the symmetry of the problem to average over small changes in
the directions of the initial momentum $\p_\ix$
as in (\ref{eq:initial1}--\ref{eq:initavg}).
Details of
how the normalization factor ${\cal N}$
cancels in the final result are given in appendix \ref{app:project}.

In the next sub-section, we will discuss in detail the form of the
$\delta H$ matrix elements for individual particles.
For splitting $i \to jk$ in the amplitude, this has the form
\begin {equation}
   \langle \p_j,\p_k | \delta H | \p_i \rangle
   = g \bcalT_{i \to jk} \cdot \P_{jk}
\label {eq:dHamp}
\end {equation}
with $\P_{jk} \equiv x_k\p_j - x_j\p_k$.%
\footnote{
  Note that $\P_{jk} = \P_{ki} = \P_{ij}$ if we were to pick the
  convention that all $x_l$ flow away from the vertex, so
  that $x_i$ is negative, and were to similarly negate $\p_i$.
}
We then need to project this result from $N$ particle language to
$N{-}2$ particle language.
In appendix \ref{app:project}, we
show that (\ref{eq:dHamp}) implies, in our constrained subspace, the
results (\ref{eq:dH32}) and (\ref{eq:dH43}) quoted previously
for the matrix elements $\langle \B | \delta H |\rangle$ and
$\langle\C_{41},\C_{23}|\delta H|\B\rangle$.

% ------------------------------------------------------------------------

\subsection {Splitting matrix elements}
\label {sec:dH}

The matrix elements for nearly-collinear bremsstrahlung (or pair creation)
is textbook material, used, for instance, in discussions of DGLAP
evolution.  In our case, we will need individual helicity-dependent
matrix elements (e.g.\ from ref.\ \cite{Altarelli}) and not just
helicity-averaged matrix elements.  We also need a few conversions
compared to the standard literature.  Our states $|\p\rangle$ have
been normalized with non-relativistic normalization
$\langle \p_i | \p'_i \rangle = (2\pi)^2 \delta^{(2)}(\p_i-\p'_i)$ because
we have cast our problem into the language of non-relativistic
quantum mechanics.  Results in the literature are instead for
states  with relativistic normalization
$\langle \p_i | \p'_i \rangle = 2 E_i (2\pi)^2 \delta^{(2)}(\p_i-\p'_i)$
[times $2\pi\,\delta(p_{iz}-p'_{iz})$],%
\footnote{
   In this paper, we have generally swept the longitudinal momentum
   conservation factors
   $2\pi\,\delta(p_{iz}-p_{iz}')$ under the rug.
   See appendix \ref{app:details}
   on (\ref{eq:dHamp}--\ref{eq:bcalP1}) for a more detailed
   discussion.
}
and so the matrix elements differ by factors of
$\sqrt{2 E_i} = \sqrt{2 |x_i| E}$.
Secondly, results in the literature are typically specialized to
a choice of $z$ axis that is {\it exactly}\/
aligned with the direction of the initial
particle: $\p_i = 0$ for $\p_i \to \p_j\p_k$.  We will want to generalize
to slightly different directions of the $z$ axis and so will
express results in terms of the invariant
$\P_{jk} \equiv x_k \p_j - x_j \p_k$ instead of, say, $\p_k$.
Finally, results in the literature are generally expressed with the
convention that $E$ is the energy of the parent particle.
In our problem, the parent of the second splitting is a daughter
of the first splitting, and so we will write parent
energies more generally as $|x_i| E$ instead of $E$.
Doing this will also make manifest
in our formulas the symmetry between the three particles
(parent and daughters) participating
in a single splitting process.

We are going to just present here the formulas for the matrix
elements.  Readers interested in details about the connection
to standard formulas in the literature should see appendix
\ref{app:dHconnect}.  The matrix elements are given by
(\ref{eq:dHamp}) with%
\footnote{
  See appendix \ref{app:details} for a note on dimensional analysis.
}
\begin {equation}
  \bcalT_{i \to jk} =
  \frac{ T^{\rm color}_{i \to jk} \bcalP_{i \to jk} }{2 E^{3/2}}
\label {eq:bcalT}
\end {equation}
and
\begin {equation}
   \bcalP_{i\to jk} =
   \frac{{\bm e}_{(\pm)}}{ |x_i x_j x_k| }
   \sqrt{\frac{P_{i\to jk}(x_i,x_j,x_k)}{C_R}} \,,
\label {eq:bcalP1}
\end {equation}
where the $P_{i\to jk}$ are helicity-dependent DGLAP splitting
functions and
\begin {equation}
%   {\bm e_{(\pm)}} = \frac{{\bm e}_x \pm i {\bm e}_y}{\sqrt2}
   {\bm e_{(\pm)}} \equiv {\bm e}_x \pm i {\bm e}_y
\end {equation}
are $\pm$ helicity basis vectors in the transverse plane with
sign determined by the conservation of the $z$-component
${\bm J}_z$ of angular momentum in the nearly-collinear limit.
This is easiest to state if we adopt the convention that
each helicity flows away from the vertex.  So, instead of
thinking of the helicities in a splitting as,
for example, ${+} \to {+}{+}$, we
will think of them as $(h_i,h_j,h_k)=({-},{+},{+})$.
With this convention, (\ref{eq:bcalP1}) is, more precisely,
\begin {equation}
   \bcalP_{i\to jk} =
   \frac{{\bm e}_{(h_i+h_j+h_k)}}{ |x_i x_j x_k| }
   \sqrt{\frac{P_{h_i,h_j,h_k}(x_i,x_j,x_k)}{C_R}} \,.
\end {equation}

For $g \to gg$ with $x_i \to x_jx_k$, the DGLAP splitting functions
(for $|x_j|,|x_k| > 0$) are
\begin {align}
   P_{+\to++} = \CA \frac{x_i^4}{|x_i x_j x_k|} \,,
   \qquad
   P_{+\to+-} &= \CA \frac{x_j^4}{|x_i x_j x_k|} \,,
   \qquad
   P_{+\to-+} = \CA \frac{x_k^4}{|x_i x_j x_k|} \,,
\nonumber\\
   P_{+\to--} &= 0 ,
\label{eq:Pexplicit}
\end {align}
corresponding to
\begin {subequations}
\label{eq:calPexplicit}
\begin {align}
   \bcalP_{+\to++} = \bcalP_{-,+,+}
                &= \frac{x_i^2}{|x_i x_j x_k|^{3/2}} \, {\bm e}_{(+)} ,
\\
   \bcalP_{+\to+-} = \bcalP_{-,+,-}
                &= \frac{x_j^2}{|x_i x_j x_k|^{3/2}} \, {\bm e}_{(-)},
\\
   \bcalP_{+\to-+} = \bcalP_{-,-,+}
                &= \frac{x_k^2}{|x_i x_j x_k|^{3/2}} \, {\bm e}_{(-)},
\\[7pt]
   \bcalP_{+\to--} = \bcalP_{-,-,-} &= 0 .
\end {align}
\end {subequations}
The remaining cases are obtained by negating all helicities
and the subscript of ${\bm e}_{(\pm)}$.
Note that the $\bcalP$'s do not depend on the signs of the
$x_l$.
Note also that in the soft bremsstrahlung limit $|x_k| \ll |x_i|$,
helicity is conserved among the harder particles
(i.e.\ $+ \to ++$ and $+ \to +-$ are unsuppressed but
$+ \to -+$ is suppressed) and
$|\bcalP_{+\to++}| \simeq |\bcalP_{+\to+-}|$.
These simplifications account for the fact that a careful
treatment of intermediate helicities was not generally necessary in earlier
work \cite{Blaizot,Iancu,Wu} focused on
the soft bremsstrahlung limit.
(See Appendix \ref{app:Wu} for a caveat.)

% ------------------------------------------------------------------------

\subsection {Helicity sums}
\label {sec:helicity}

It is now time to do the helicity sum in our expression (\ref{eq:IIxyyx1})
for the $x y \bar y \bar x$ interference.  The only things that
depend on helicities are the $\bcalP$ factors.  These factors will
be the same for all of the diagrams in the first row of figs.\
\ref{fig:subset} and \ref{fig:subset2}, and in this case it will
be more convenient to write in terms of $x$ and $y$ explicitly
rather than in terms of the 4-particle $\hat x_i$'s.  Specifically, we need
%\footnote{
%  In (\ref{eq:calPsum}), the ordering of the three particles
%  in each $\bcalP$ comes from applying
%  fig.\ \ref{fig:dH} to fig.\ \ref{fig:xyyx}.
%  When all particles are gluons, however,
%  $\bcalP_{h_i,h_j,h_k}(x_i,x_j,x_K)$ is invariant under permutations
%  of the three particles involved, and this ordering is unimportant.
%}
\begin {multline}
   \sum_{h_\xx,h_\yx,h_\zx,h,\bar h}
   {\cal P}^{\bar n}_{{-}h_\zx,\bar h,{-}h_\xx}\bigl({-}(1{-}x{-}y),1{-}y,{-}x\bigr) \,
   {\cal P}^{\bar m}_{{-}\bar h,h_\ix,{-}h_\yx}\bigl({-}(1{-}y),1,{-}y\bigr)
\\ \times
   {\cal P}^n_{{-}h,h_\zx,h_\yx}\bigl({-}(1{-}x),1{-}x{-}y,y\bigr) \,
   {\cal P}^m_{{-}h_\ix,h,h_\xx}\bigl(-1,1{-}x,x\bigr) ,
\label {eq:calPsum}
\end {multline}
where ${\cal P}^n$ are the Cartesian components of $\bcalP$.
If one prefers, this can be written equivalently as
\begin {multline}
   \sum_{h_\xx,h_\yx,h_\zx}
   \Bigl[
   \sum_{\bar h}
   {\cal P}^{\bar n}_{\bar h \to h_\zx,h_\xx}\bigl(1{-}y \to 1{-}x{-}y,x\bigr) \,
   {\cal P}^{\bar m}_{h_\ix \to \bar h, h_\yx}\bigl(1 \to 1{-}y,y\bigr)
   \Bigr]^*
\\ \times
   \Bigl[
   \sum_h
   {\cal P}^n_{h \to h_\zx,h_\yx}\bigl(1{-}x \to 1{-}x{-}y,y\bigr) \,
   {\cal P}^m_{h_\ix \to h,h_\xx}\bigl(1 \to 1{-}x,x\bigr)
   \Bigr] ,
\label {eq:calPsum2}
\end {multline}
which (reading right to left) shows more clearly how the successive
splittings tie together in both the amplitude and conjugate amplitude.

Because of parity/reflection symmetry in the transverse plane, the rate
(\ref{eq:IIxyyx1}) is
independent of the initial helicity $h_\ix$, and so we may replace
(\ref{eq:calPsum2}) by its helicity average over $h_\ix$.
Then, because of
rotation invariance of the transverse plane, the above helicity sum
(averaged over $h_\ix$) must have the form
\begin {equation}
  \alpha(x,y) \, \delta^{\bar n n} \delta^{\bar m m}
  + \beta(x,y) \, \delta^{\bar n \bar m} \delta^{nm}
  + \gamma(x,y) \, \delta^{\bar n m} \delta^{n \bar m}
\label {eq:abcdef}
\end {equation}
for some functions $\alpha$, $\beta$, and $\gamma$.
Using (\ref{eq:calPexplicit}) in (\ref{eq:calPsum2}), we find
\begin {align}
   \begin{pmatrix} \alpha \\ \beta \\ \gamma \end{pmatrix}
   =
   \phantom{+}
   & \begin{pmatrix} - \\ + \\ + \end{pmatrix} \Biggl[
       \frac{x}{y^3(1{-}x)^3(1{-}y)^3(1{-}x{-}y)}
       + \frac{y}{x^3(1{-}x)^3(1{-}y)^3(1{-}x{-}y)}
\nonumber\displaybreak[3]\\ & \qquad\qquad
       + \frac{1{-}x}{x^3y^3(1{-}y)^3(1{-}x{-}y)}
       + \frac{1{-}y}{x^3y^3(1{-}x)^3(1{-}x{-}y)}
   \Biggr]
\nonumber\\
   + & \begin{pmatrix} + \\ - \\ + \end{pmatrix} \Biggl[
       \frac{x}{y^3(1{-}x)(1{-}y)(1{-}x{-}y)^3}
       + \frac{y}{x^3(1{-}x)(1{-}y)(1{-}x{-}y)^3}
\nonumber\displaybreak[3]\\ & \qquad\qquad
       + \frac{1{-}x{-}y}{x^3y^3(1{-}x)(1{-}y)}
       + \frac{1}{x^3y^3(1{-}x)(1{-}y)(1{-}x{-}y)^3}
   \Biggr]
\nonumber\\
   + & \begin{pmatrix} + \\ + \\ - \end{pmatrix} \Biggl[
        \frac{1{-}x}{xy(1{-}y)^3(1{-}x{-}y)^3}
        + \frac{1{-}y}{xy(1{-}x)^3(1{-}x{-}y)^3}
\nonumber\displaybreak[3]\\ & \qquad\qquad
        + \frac{1{-}x{-}y}{xy(1{-}x)^3(1{-}y)^3}
        + \frac{1}{xy(1{-}x)^3(1{-}y)^3(1{-}x{-}y)^3}
   \Biggr]
\label {eq:abc}
\end {align}
for the case studied in this paper, where all high-energy particles
are gluons.  Substituting (\ref{eq:abcdef}) for (\ref{eq:calPsum})
in the $xy\bar y\bar x$ expression (\ref{eq:IIxyyx1}) gives
\begin {align}
   \left[\frac{dI}{dx\,dy}\right]_{xy\bar y\bar x}
   = &
   \frac{\CA^2 \alphas^2 }{8 E^4} \,
   \frac{
     ( \alpha \delta^{\bar n n} \delta^{\bar m m}
     {+} \beta \delta^{\bar n \bar m} \delta^{nm}
     {+} \gamma \delta^{\bar n m} \delta^{n \bar m} )
   }{
     |\hat x_1 + \hat x_4| |\hat x_3 + \hat x_4|
   }
   \int_{t_\xx < t_\yx < t_\ybx < t_\xbx}
\nonumber\\ &
   \int_{\B^\Ax,\B^\bx}
   \nabla_{\B^\Bx}^{\bar n}
   \langle\B^\Bx,t_\Bx|\B^\Ax,t_\Ax\rangle
   \Bigr|_{\B^\Bx=0}
\nonumber\\ &\qquad\times
   \nabla_{\C_{12}^\Ax}^{\bar m}
   \nabla_{\C_{23}^\bx}^n
  \langle\C_{34}^\Ax,\C_{12}^\Ax,t_\Ax|\C_{41}^\bx,\C_{23}^\bx,t_\bx\rangle
   \Bigr|_{\C_{12}^\Ax=0=\C_{23}^\bx; ~ \C_{34}^\Ax=\B^\Ax; ~ \C_{41}^\bx=\B^\bx}
\nonumber\\ &\qquad\times
   \nabla_{\B^\ax}^m
   \langle\B^\bx,t_\bx|\B^\ax,t_\ax\rangle
   \Bigr|_{\B^\ax=0} .
\label {eq:IIxyyx}
\end {align}
Note that $\alpha$, $\beta$, and $\gamma$ are all symmetric under
$x \leftrightarrow y$.

This is as far as we can go without either doing numerics or making
additional approximations, to which we now turn.

% =========================================================================

\section {\boldmath$xy\bar y\bar x$ in harmonic
          and thick-media approximations}
\label {sec:xyyx}

We can make further analytic progress by (i) restricting attention
to the harmonic approximation, and (ii) assuming that the medium
is sufficiently homogeneous that $\hat q$ does not vary significantly
over distances of order the formation length.
In this section, we will continue to focus for now
on the $xy\bar y\bar x$ interference.
We will tackle the $N{=}3$ particle ends of the evolution
in fig.\ \ref{fig:xyyx} and integrate over the first and last
splitting times, $t_\xx$ and $t_\xbx$.  Then we tackle the
$N{=}4$ particle piece of the evolution by solving the
harmonic oscillator problem for the propagator in the large-$\Nc$
limit,
and then we do the $\B$ integrals remaining in
(\ref{eq:IIxyyx}).  This will leave the result in the form
of a single 1-dimensional integral over
$\Delta t \equiv t_\ybx-t_\yx$.

% ---------------------------------------------------------------------------

\subsection{Integrating over first and last times
            \boldmath$t_\xx$ and \boldmath$t_\xbx$}
\label{sec:firstlast}

During the 3-particle phases of the evolution of the
$xy\bar y\bar x$ interference of fig.\ \ref{fig:xyyx},
our ${\cal H}$ is given by (\ref{eq:calH3}) and (\ref{eq:M3}).
For the initial 3-particle evolution corresponding to
$t_\xx < t < t_\yx$, we'll write this as
\begin {equation}
   {\cal H} =
   \frac{\P^2}{2M_\ix}
      + \tfrac12 M_\ix \, \Omega_\ix^2 \B^2
\end {equation}
with
\begin {subequations}
\label {eq:MOmi}
\begin {equation}
   M_\ix = \hat x_1 \hat x_4 (\hat x_1{+}\hat x_4) E
   = x(1-x) E
\label {eq:Mi}
\end {equation}
and (when all particles are gluons)
\begin {equation}
   \Omega_\ix
   = \sqrt{ 
     -\frac{i \hat q_{\rm A}}{2E}
     \left( \frac{1}{\hat x_1} 
            + \frac{1}{\hat x_4} - \frac{1}{\hat x_1{+}\hat x_4} \right)
   }
   = \sqrt{ 
     -\frac{i \hat q_{\rm A}}{2E}
     \left( -1 
            + \frac{1}{x} + \frac{1}{1{-}x} \right)
   } .
\end {equation}
\end {subequations}
For the final 3-particle evolution corresponding to
$t_\ybx < t < t_\xbx$,
\begin {equation}
   {\cal H} =
   \frac{\P^2}{2M_\fx}
      + \tfrac12 M_\fx \, \Omega_\fx^2 \B^2
\end {equation}
with
\begin {subequations}
\label {eq:MOmf}
\begin {equation}
   M_\fx = (\hat x_3{+}\hat x_4) \hat x_4 \hat x_3 E
   = (1-y)x(1-x-y) E
\label {eq:Mf}
\end {equation}
and
\begin {equation}
   \Omega_\fx
   = \sqrt{ 
     -\frac{i \hat q_{\rm A}}{2E}
     \left( - \frac{1}{\hat x_3 + \hat x_4}
            + \frac{1}{\hat x_4} + \frac{1}{\hat x_3}
            \right)
   }
   = \sqrt{ 
     -\frac{i \hat q_{\rm A}}{2E}
     \left( - \frac{1}{1{-}y} + \frac{1}{x} + \frac{1}{1{-}x{-}y}
            \right)
   } .
\end {equation}
\end {subequations}
(Throughout this paper, one should understand $\sqrt{\pm i}$ to
refer to the root $e^{\pm i\pi/4}$.)

The propagator (\ref{eq:1prop}) for a harmonic oscillator is
\begin {equation}
   \langle\B,t|\B',0\rangle
   =
      \frac{M\Omega\csc(\Omega t)}{2\pi i} \,
      \exp\Bigl(
        \tfrac{i}{2} \, M \Omega \bigl[ (\B^2+\B'^2) \cot(\Omega t)
        - 2 \B\cdot\B' \csc(\Omega t) \bigr]
      \Bigr) .
\end {equation}
The factor corresponding to the initial 3-particle evolution in
(\ref{eq:IIxyyx}) is of the form
\begin {equation}
   \grad_{\B'} \langle\B,t|\B',t'\rangle \Bigr|_{\B'=0}
   =
      -
      \frac{M^2\Omega^2\csc^2\bigl(\Omega(t-t')\bigr)}{2\pi} \, \B
      \exp\Bigl(
        \tfrac{i}{2} \, M \Omega \cot\bigl(\Omega(t-t')\bigr) \B^2
      \Bigr) .
\end {equation}
Integrating over the initial time $t' < t$ above gives
\begin {equation}
   \int_{-\infty}^{t} dt' \>
   \grad_{\B'} \langle\B,t|\B',t'\rangle
   \biggr|_{\B'=0}
   =
   \frac{i M \B}{\pi B^2} \,
   \exp\Bigl[
      \tfrac{i}{2} M \Omega B^2 \cot\bigl(\Omega(t-t')\bigr)
   \Bigr]
   \biggr|_{t'=-\infty}^{t'=t} .
\label{eq:t1int2a}
\end {equation}
The term with $t'=t$ has a divergent exponent, corresponding to infinitely
oscillatory behavior in $\B$.  Naively, one might expect to be able to
drop it, since an infinitely oscillatory function vanishes if later
integrated against a smooth function.  We will drop it here but
will later have to return and be more careful about what happens when
times become coincident.  [In particular, if a third time in the double
splitting problem becomes coincident with $t$ and $t'$ above, then
(\ref{eq:t1int2a}) is not integrated against a smooth function
of $\B$.]

In our problem, $\Omega$ has a negative imaginary part, so that
$\cot(\Omega\infty) = i$, and so the $t'=-\infty$ term in
(\ref{eq:t1int2a}) leaves us with
\begin {equation}
   \int_{-\infty}^{t} dt' \>
   \grad_{\B'} \langle\B,t|\B',t'\rangle
   \biggr|_{\B'=0}
   =
   - \frac{i M \B}{\pi B^2} \,
   \exp\bigl(
      - \tfrac12 M \Omega B^2
   \bigr) .
\label {eq:3int0}
\end {equation}
For the sake of evaluating other interference diagrams in the future, it
will be useful to generalize to include cases where $\Omega$ may
have a positive imaginary part.  See appendix \ref{app:details} for
an argument that the more general result may be cast in the convenient form
\begin {subequations}
\label {eq:3ints}
\begin {equation}
   \int_{-\infty}^{t} dt' \>
   \grad_{\B'} \langle\B,t|\B',t'\rangle
   \biggr|_{\B'=0}
   =
   - \frac{i M \B}{\pi B^2} \,
   \exp\bigl(
      - \tfrac12 |M| \Omega B^2
   \bigr) .
   \label{eq:3intsA}
\end {equation}
A similar analysis, relevant to the final 3-particle evolution,
gives (with the same caveats)
\begin {equation}
   \int_{t'}^{+\infty} dt \>
   \grad_{\B} \langle\B,t|\B',t'\rangle
   \biggr|_{\B=0}
   =
   - \frac{i M \B'}{\pi B'^2} \,
   \exp\bigl(
      - \tfrac12 |M| \Omega B'^2
   \bigr) .
\end {equation}
\end {subequations}

Using (\ref{eq:3ints}) on the expression (\ref{eq:IIxyyx}) for the rate
gives
\begin {align}
   \left[\frac{d\Gamma}{dx\,dy}\right]_{xy\bar y\bar x}
   = &
   - \frac{\CA^2 \alphas^2 M_\ix M_\fx}{8 \pi^2 E^4}
   \frac{
     ( \alpha \delta^{\bar n n} \delta^{\bar m m}
     {+} \beta \delta^{\bar n \bar m} \delta^{nm}
     {+} \gamma \delta^{\bar n m} \delta^{n \bar m} )
   }{
     |\hat x_1 + \hat x_4| |\hat x_3 + \hat x_4|
   }
\nonumber\\ & \times
   \int_0^{\infty} d(\Delta t)
   \int_{\B^\Ax,\B^\bx}
   \frac{B^{\ybx}_{\bar n}}{(B^\ybx)^2} \,
   \frac{B^{\yx}_m}{(B^\yx)^2} \,
   \exp\bigl(
      - \tfrac12 |M_\fx| \Omega_\fx (B^\ybx)^2
      - \tfrac12 |M_\ix| \Omega_\ix (B^\yx)^2
   \bigr)
\nonumber\\ &\qquad\times
   \nabla_{\C_{12}^\Ax}^{\bar m}
   \nabla_{\C_{23}^\bx}^n
   \langle\C_{34}^\Ax,\C_{12}^\Ax,\Delta t|\C_{41}^\bx,\C_{23}^\bx,0\rangle
   \Bigr|_{\C_{12}^\Ax=0=\C_{23}^\bx; ~ \C_{34}^\Ax=\B^\Ax; ~ \C_{41}^\bx=\B^\bx}
   ,
\label {eq:IIxyyx2}
\end {align}
where $\Delta t \equiv t_\ybx - t_\yx$, and we have used time translation
invariance (in our thick-media approximation) to convert the
expression for the probability $I$ into an expression for the rate
$\Gamma$.

% -----------------------------------------------------------------------

\subsection{4-particle normal modes and frequencies in large \boldmath$\Nc$}

We now turn to the $N{=}4$ particle propagator
$\langle\C_{34}^\Ax,\C_{12}^\Ax,\Delta t|\C_{41}^\bx,\C_{23}^\bx,0\rangle$.

% .......................................................................

\subsubsection {Reduced Lagrangian}

Consider the 4-particle Lagrangian
\begin {equation}
   L = \tfrac12 m_1 \dot\b_1^2
     + \tfrac12 m_2 \dot\b_2^2
     + \tfrac12 m_3 \dot\b_3^2
     + \tfrac12 m_4 \dot\b_4^2
     - V
\label {eq:L4}
\end {equation}
in the translation invariant case that $V$ depends only on
differences $\b_{ij} \equiv \b_i-\b_j$ of the coordinates.
In the particular case where $\sum_i m_i = 0$, we now want
to impose the constraint $\sum_i m_i\b_i = 0$ on $L$ above.
The result can be written in the form
\begin {equation}
   L =
     \left[
         \tfrac12 m_1 \dot\b_{14}^2
       + \tfrac12 m_2 \dot\b_{24}^2
       + \tfrac12 m_3 \dot\b_{34}^2
       - V(\b_{14},\b_{24},\b_{34})
     \right]_{\sum_i m_i \b_i = 0} ,
\end {equation}
as can be seen by expanding the factors of $(\dot\b_i - \dot\b_4)^2$
and using the constraint and $\sum_i m_i = 0$ to recover (\ref{eq:L4}).
The constraint may be rewritten in the form
$m_1\b_{14}{+}m_2\b_{24}{+}m_3\b_{34}=0$, and so we may rewrite $L$ in terms of
just two variables as, for example,%
\footnote{
  For readers wondering how we could get from the 4-particle description
  to the 2-particle description without ever explicitly
  imposing the condition $\p_1{+}\p_2{+}\p_3{+}\p_4=0$ for our subspace:
  Whenever we can consistently project to $\sum_i m_i\b_i =0$
  for all times (as demonstrated by
  the earlier arguments of section \ref{sec:Bs2}),
  then $\sum_i\p_i=0$ follows automatically from (\ref{eq:Hmb}).
}
\begin {subequations}
\label {eq:L2a}
\begin {align}
   L = &
   \left[
   \tfrac12 \, m_1 \dot\b_{14}^2
   + \tfrac12 \, m_2 \dot\b_{24}^2
   + \tfrac12 \, m_3 \dot\b_{34}^2
   - V(\b_{14},\b_{24},\b_{34})
   \right]_{\b_{24} = - (m_1\b_{14} + m_3\b_{34})/m_2}
\nonumber\\
   = &
   \tfrac12
   \begin{pmatrix} \dot\b_{14} \\ \dot\b_{34} \end{pmatrix}^{\!\top}
   \!M_{(2)}
   \begin{pmatrix} \dot\b_{14} \\ \dot\b_{34} \end{pmatrix}
   - V(\b_{14},\b_{24},\b_{34})\Bigr|_{\b_{24} = - (m_1\b_{14} + m_3\b_{34})/m_2}
\end {align}
with
\begin {equation}
   M_{(2)}
   \equiv
   \begin{pmatrix}
      m_1 & \\ & m_3
   \end {pmatrix}
   + \frac1{m_2}
   \begin{pmatrix} m_1^2 & m_1 m_3 \\ m_3 m_1 & m_3^2 \end{pmatrix}
   =
   \begin{pmatrix}
      \frac1{m_4}+\frac1{m_1} & \frac1{m_4} \\[2pt]
      \frac1{m_4} & \frac1{m_4}+\frac1{m_3}
   \end {pmatrix}^{-1} .
\end {equation}
\end {subequations}

We could now solve this 2-particle problem.  However, it will be
convenient to first change to one of the sets of variables that
we actually use in our rate expression (\ref{eq:IIxyyx2}).
We will trade the $\b_{14}$ and $\b_{34}$ used above for
$\C_{12} \equiv \b_{12}/(x_1+x_2)$ and
$\C_{34} \equiv \b_{34}/(x_3+x_4)$.
Using $\sum_i x_i = 0$ and $\sum_i x_i \b_i = 0$, one can show that
\begin {subequations}
\label {eq:brelations}
\begin {align}
   \b_{12} &= (x_1+x_2) \C_{12} ,
\\
   \b_{13} &= x_2 \C_{12} - x_4 \C_{34} ,
\\
   \b_{14} &= x_2 \C_{12} + x_3 \C_{34} ,
\\
   \b_{23} &= - x_1 \C_{12} - x_4 \C_{34} ,
\\
   \b_{24} &= - x_1 \C_{12} + x_3 \C_{34} , 
\\
   \b_{34} &= (x_3+x_4) \C_{34} .
\end {align}
\end {subequations}
Then (\ref{eq:L2a}) becomes
\begin {equation}
   L
   =
   \tfrac12
   \begin{pmatrix} \dot\C_{34} \\ \dot\C_{12} \end{pmatrix}^{\!\top}
   \!{\mathfrak M}
   \begin{pmatrix} \dot\C_{34} \\ \dot\C_{12} \end{pmatrix}
   - V\bigl( \b_{ij} \to \mbox{\small eqs.\ (\ref{eq:brelations})} \bigr)
\label {eq:L3}
\end {equation}
with
\begin {equation}
   {\mathfrak M}
   =
   \begin{pmatrix}
      x_3 x_4 (x_3+x_4) & \\ & x_1 x_2 (x_1+x_2)
   \end {pmatrix} E
   =
   \begin{pmatrix}
      x_3 x_4 & \\ & -x_1 x_2
   \end {pmatrix} (x_3+x_4) E .
\label {eq:frakM}
\end {equation}
A nice feature of these variables is that the kinetic term is
diagonal, unlike in (\ref{eq:L2a}).
Note that, for the $x_i$ of (\ref{eq:xhat}) relevant to the
$xy\bar y\bar x$ interference, ${\mathfrak M}$
above is positive definite.

In harmonic approximation, (\ref{eq:L3}) will reduce to the form
\begin {equation}
   L
   =
   \tfrac12
   \begin{pmatrix} \dot\C_{34} \\ \dot\C_{12} \end{pmatrix}^{\!\top}
   \!{\mathfrak M}
   \begin{pmatrix} \dot\C_{34} \\ \dot\C_{12} \end{pmatrix}
   - \tfrac12 
   \begin{pmatrix} \C_{34} \\ \C_{12} \end{pmatrix}^{\!\top}
   \!{\mathfrak K}
   \begin{pmatrix} \C_{34} \\ \C_{12} \end{pmatrix} ,
\label {eq:L2}
\end {equation}
with ${\mathfrak K}$ depending on specific details that will
be given in a moment.
The squares $\Omega_j^2$ of the normal mode frequencies will be given
by the eigenvalues of
${\mathfrak M}^{-1/2} {\mathfrak K} \, {\mathfrak M}^{-1/2}$.
The corresponding normal modes $(\C^j_{34},\C^j_{12})$ will be
orthogonal with respect to ${\mathfrak M}$, and we will
normalize them as
\begin {equation}
   \begin{pmatrix} \C^j_{34} \\ \C^j_{12} \end{pmatrix}^{\!\top}
   \!{\mathfrak M}
   \begin{pmatrix} \C^{j'}_{34} \\ \C^{j'}_{12} \end{pmatrix}
   = \delta^{jj'} .
\label {eq:NMnorm}
\end {equation}

We will now turn to specifics by deriving normal mode results
in the large-$\Nc$ approximation when all particles are gluons.
But we will then package our results so that we can continue to
simplify our rate expression (\ref{eq:IIxyyx2}) in a way that is
independent of those details, so that our results can easily be
adapted at a future date to other situations.

% .........................................................................

\subsubsection {Specifics for large-$\Nc$ gluons}
\label {sec:normal}

Combining the large-$\Nc$ harmonic potential
\begin {equation}
  - \frac{i \hat q_{\rm A}}{8}
  ( \b_{12}^2 + \b_{23}^2 + \b_{34}^2 + \b_{41}^2 )
\label {eq:Vlarge4}
\end {equation}
of (\ref{eq:VlargeN}) for gluons with (\ref{eq:brelations}) gives
\begin {equation}
   {\mathfrak K}
   = -\frac{i\hat q_{\rm A}}{4}
   \begin{pmatrix}
      2(x_3^2+x_3 x_4+x_4^2) & x_2 x_3+x_1 x_4 \\
      x_2 x_3+x_1 x_4 & 2(x_1^2+x_1 x_2+x_2^2)
   \end {pmatrix} .
\end {equation}
in (\ref{eq:L2}).
Note that, since this is a harmonic oscillator problem,
the two transverse dimensions decouple, and so this may be treated
as a problem in one transverse dimension rather than two.
The result for the normal mode frequencies $\Omega_\pm$ should have the same
symmetry with respect to exchanging particle labels as
(\ref{eq:Vlarge4}) has, and (mostly for aesthetic reasons)
we use $\sum_i x_i=0$ to algebraically
manipulate the result into a form where this is manifest.
We find
\begin {equation}
  \Omega_\pm =
  \left[ - \frac{i\hat q_{\rm A}}{4E} \left(
    \frac1{x_1} + \frac1{x_2} + \frac1{x_3} + \frac1{x_4} \pm \sqrt\Delta
  \right) \right]^{1/2} ,
\label {eq:Omegapm}
\end {equation}
where
\begin {equation}
  \Delta \equiv
   \frac1{x_1^2} + \frac1{x_2^2} + \frac1{x_3^2} + \frac1{x_4^2}
   - \frac1{x_1 x_2} - \frac1{x_2 x_3} - \frac1{x_3 x_4} - \frac1{x_4 x_1}
   - \frac2{x_1 x_3} - \frac2{x_2 x_4} > 0 \,.
\end {equation}
A more compact (but slightly less manifestly symmetric)
expression is
\begin {equation}
  \Delta =
   \frac1{x_1^2} + \frac1{x_2^2} + \frac1{x_3^2} + \frac1{x_4^2}
   + \frac{(x_3{+}x_4)^2+(x_1{+}x_4)^2}{x_1 x_2 x_3 x_4} \,.
\end {equation}
With normalization (\ref{eq:NMnorm}),
the normal modes $(C^+_{34},C^+_{12})$ and
$(C^-_{34},C^-_{12})$ are%
\footnote{
  The right-hand sides of (\ref{eq:Cmodes}) are in fact symmetric under
  $12,34 \to 21,43$ (which takes $N_\pm \to -N_\mp$),
  but we have not found an appealing way
  (i.e.\ other than brute force symmetrization)
  to write them that makes this symmetry manifest.
  [Also, one might think that $12,34 \to 21,43$ should take
  $(C^\pm_{34},C^\pm_{12}) \to (C^\pm_{43},C^\pm_{21})=(-C^\pm_{34},-C^\pm_{12})$
  instead of to $(+C^\pm_{34},+C^\pm_{12})$, but the
  overall sign of a normal mode is an arbitrary normalization choice,
  reflected by the ambiguity of the square root in (\ref{eq:Cmodes}).]
}
\begin {subequations}
\label {eq:Cmodes}
\begin {align}
   C^\pm_{34}
   &=
   \frac{x_2}{x_3+x_4} \sqrt{\frac{x_1 x_3}{2 N_\pm E}}
   \left[
      \frac1{x_3}-\frac1{x_1}+\frac1{x_4}+\frac{x_1}{x_3 x_2} \pm \sqrt\Delta
   \right] ,
\\
   C^\pm_{12}
   &=
   - \frac{x_4}{x_1+x_2} \sqrt{\frac{x_1 x_3}{2 N_\pm E}}
   \left[
      \frac1{x_1}-\frac1{x_3}+\frac1{x_2}+\frac{x_3}{x_1 x_4} \pm \sqrt\Delta
   \right] ,
\end {align}
\end {subequations}
with
%%\footnote{
%%   We note that $N_\pm$ is invariant under $2 \leftrightarrow 4$ but
%%   $N_\pm \to -N_\mp$ under the cyclic permutation
%%   $1234 \to 2341$.
%%}
\begin {equation}
   N_\pm \equiv
   - x_1 x_2 x_3 x_4 (x_1+x_3)\Delta
   \pm (x_1 x_4 + x_2 x_3)(x_1 x_2 + x_3 x_4) \sqrt\Delta .
\label {eq:Npm}
\end {equation}
For the $x_i$ of (\ref{eq:xhat}) relevant to the
$xy\bar y\bar x$ interference, the $x_1 x_3/2N_\pm E$ under
the square root in (\ref{eq:Cmodes}) is positive.

% .........................................................................

\subsubsection {General form}
\label {sec:GeneralForm}

Whatever the specific formulas for the normal modes are in a given
problem of interest, we can write any $(\C_{34},\C_{12})$ as
a superposition
\begin {equation}
   \begin{pmatrix} \C_{34} \\ \C_{12} \end{pmatrix}
   =
   \A_+ \begin{pmatrix} C^+_{34} \\ C^+_{12} \end{pmatrix}
   +
   \A_- \begin{pmatrix} C^-_{34} \\ C^-_{12} \end{pmatrix}
\label {eq:Adef}
\end {equation}
of normal modes with superposition coefficients
$\A_+$ and $\A_-$.  We will write this in matrix form as
\begin {equation}
   \begin{pmatrix} \C_{34} \\ \C_{12} \end{pmatrix}
   = a_\ybx \begin{pmatrix} \A_+ \\ \A_- \end{pmatrix}
\label {eq:changef}
\end {equation}
with
\begin {equation}
   a_\ybx \equiv
   \begin{pmatrix} C^+_{34} & C^-_{34} \\ C^+_{12} & C^-_{12} \end{pmatrix} .
\label {eq:af}
\end {equation}
We use the subscript $\ybx$ because $(\C_{34},\C_{12})$ are the
variables we want to use at $t_\ybx$ in (\ref{eq:IIxyyx2}).

Because of the normalization (\ref{eq:NMnorm})
of our modes, the Lagrangian (\ref{eq:L2}) in terms
of $\A_\pm$ is simply
\begin {equation}
   L =
   \sum_\pm \left[
      \tfrac12 \, \dot\A_\pm^2
      - \tfrac12 \, \Omega_\pm^2 \A_\pm^2
   \right] .
\label {eq:AL}
\end {equation}
The corresponding propagator is
\begin {multline}
   \langle \A_+,\A_-,t | \A_+',\A_-',0 \rangle
   =
\\
   \prod_\pm \left[
      \frac{\Omega_\pm\csc(\Omega_\pm t)}{2\pi i} \,
      \exp\Bigl(
        i \bigl[ \tfrac12 (\A_\pm^2+\A_\pm'^2) \Omega_\pm\cot(\Omega_\pm t)
        - \A_\pm\cdot\A_\pm' \Omega_\pm \csc(\Omega_\pm t) \bigr]
      \Bigr)
   \right].
\label {eq:Aprop}
\end {multline}

At $t=t_\yx$ in (\ref{eq:IIxyyx2}), the variables we want to use
are $(\C_{41},\C_{23})$ instead of $(\C_{34},\C_{12})$.  However, it
is easy to convert: (\ref{eq:brelations}) and
$\C_{ij} \equiv \b_{ij}/(x_i{+}x_j)$ give
\begin {equation}
   \begin{pmatrix} \C_{41} \\ \C_{23} \end{pmatrix}
   =
   \frac{1}{(x_1+x_4)}
   \begin{pmatrix}
       -x_3 & -x_2 \\
        \phantom{-}x_4 &  \phantom{-}x_1
   \end {pmatrix}
   \begin{pmatrix} \C_{34} \\ \C_{12} \end{pmatrix} ,
\end {equation}
and so
\begin {equation}
   \begin{pmatrix} \C_{41} \\ \C_{23} \end{pmatrix}
   = a_\yx \begin{pmatrix} \A_+ \\ \A_- \end{pmatrix} ,
\label {eq:changei}
\end {equation}
with
\begin {equation}
   a_\yx \equiv
   \frac{1}{(x_1+x_4)}
   \begin{pmatrix}
       -x_3 & -x_2 \\
       \phantom{-}x_4 &  \phantom{-}x_1
   \end {pmatrix}
   a_\ybx .
\label {eq:ai}
\end {equation}
The changes of variables (\ref{eq:changef}) and (\ref{eq:changei})
give%
\footnote{
  See Appendix \ref{app:details} for comments.
}
\begin {equation}
   \langle\C_{34}^\Ax,\C_{12}^\Ax,\Delta t|\C_{41}^\bx,\C_{23}^\bx,0\rangle
   =
   \frac{
     \langle \A_+^\Ax,\A_-^\Ax,\Delta t | \A_+^\bx,\A_-^\bx,0 \rangle
   }{
     | \det a_\Ax | | \det a_\bx |
   } \,.
\label {eq:CvsAprop}
\end {equation}

From (\ref{eq:af}) and the normalization (\ref{eq:NMnorm}),
$a_\ybx^\top {\mathfrak M} a_\ybx = \openone$, and so
\begin {equation}
   {a_\ybx^{-1}}^\top a_\ybx^{-1} = {\mathfrak M}
\label{eq:aaybar}
\end {equation}
(which we will find useful later)
and
\begin {equation}
   |\det a_\ybx|^{-1} = |\det{\mathfrak M}|^{1/2}
   = |x_1 x_2 x_3 x_4|^{1/2} |x_3{+}x_4| E .
\end {equation}
By symmetry, or from (\ref{eq:ai}), similarly
\begin {equation}
   |\det a_\yx|^{-1}
   = |x_1 x_2 x_3 x_4|^{1/2} |x_1{+}x_4| E ,
\end {equation}
so that
\begin {equation}
   \langle\C_{34}^\Ax,\C_{12}^\Ax,\Delta t|\C_{41}^\bx,\C_{23}^\bx,0\rangle
   = 
     |x_1 x_2 x_3 x_4| |x_1{+}x_4| |x_3{+}x_4| E^2
     \langle \A_+^\Ax,\A_-^\Ax,\Delta t | \A_+^\bx,\A_-^\bx,0 \rangle .
\label {eq:Cprop1}
\end {equation}
Combining with (\ref{eq:changef}), (\ref{eq:Aprop}), and (\ref{eq:changei}),
\begin {align}
   \langle\C_{34}^\Ax,\C_{12}^\Ax,\Delta t|&\C_{41}^\bx,\C_{23}^\bx,0\rangle
   =
\nonumber\\ &
   (2\pi i)^{-2}
     ({-}x_1 x_2 x_3 x_4)
     |x_1{+}x_4| |x_3{+}x_4| E^2
   \Omega_+\Omega_- \csc(\Omega_+\Delta t) \csc(\Omega_-\Delta t)
\nonumber\\ &\times
   \exp\Biggl[
     \frac{i}2
     \begin{pmatrix} \C^\bx_{41} \\ \C^\bx_{23} \end{pmatrix}^\top
       a_\bx^{-1\top} \uOmega \cot(\uOmega\,\Delta t) \, a_\bx^{-1}
       \begin{pmatrix} \C^\bx_{41} \\ \C^\bx_{23} \end{pmatrix}
\nonumber\\ &\qquad\quad
     +
     \frac{i}2
     \begin{pmatrix} \C^\Ax_{34} \\ \C^\Ax_{12} \end{pmatrix}^\top
       a_\Ax^{-1\top} \uOmega \cot(\uOmega\,\Delta t) \, a_\Ax^{-1}
       \begin{pmatrix} \C^\Ax_{34} \\ \C^\Ax_{12} \end{pmatrix}
\nonumber\\ &\qquad\quad
     - i
     \begin{pmatrix} \C^\bx_{41} \\ \C^\bx_{23} \end{pmatrix}^\top
       a_\bx^{-1\top} \uOmega \csc(\uOmega\,\Delta t) \, a_\Ax^{-1}
       \begin{pmatrix} \C^\Ax_{34} \\ \C^\Ax_{12} \end{pmatrix}
   \Biggr] ,
\label {eq:Cprop}
\end {align}
where
\begin {equation}
   \uOmega \equiv \begin{pmatrix} \Omega_+ & \\ & \Omega_- \end{pmatrix} .
\label {eq:uOmega}
\end {equation}
We have used (\ref{eq:xhat}) to rewrite $|x_1 x_2 x_3 x_4|$ above
as $-x_1 x_2 x_3 x_4$
for reasons that are described
in appendix \ref{app:sign}, having to do with relating
the result for the $xy\bar y\bar x$ interference to other interference
contributions.

% --------------------------------------------------------------------------

\subsection{Doing the \boldmath$B$ integrals}

We will now assemble our results and then
carry out all remaining integrals except for the
$\Delta t$ integral.

We want to use the propagator (\ref{eq:Cprop}) in the expression
(\ref{eq:IIxyyx2}) for the rate.  It is convenient to first combine
all of the exponential factors (before taking derivatives and
setting various $\C_{ij}$ to zero) and rewrite them as
\begin {multline}
   \exp\Biggl[
     - \frac12
     \begin{pmatrix} \B^\bx \\ \C^\bx_{23} \end{pmatrix}^\top \!
       \begin{pmatrix} X_\bx & Y_\bx \\ Y_\bx & Z_\bx \end{pmatrix}
       \begin{pmatrix} \B^\bx \\ \C^\bx_{23} \end{pmatrix}
     -
     \frac12
     \begin{pmatrix} \B^\Ax \\ \C^\Ax_{12} \end{pmatrix}^\top \!
       \begin{pmatrix} X_\Ax & Y_\Ax \\ Y_\Ax & Z_\Ax \end{pmatrix}
       \begin{pmatrix} \B^\Ax \\ \C^\Ax_{12} \end{pmatrix}
\\
     +
     \begin{pmatrix} \B^\bx \\ \C^\bx_{23} \end{pmatrix}^\top \!
       \begin{pmatrix} X_{\bx\Ax} & Y_{\bx\Ax} \\
                       \Ybar_{\bx\Ax} & Z_{\bx\Ax} \end{pmatrix}
       \begin{pmatrix} \B^\Ax \\ \C^\Ax_{12} \end{pmatrix}
   \Biggr],
\label {eq:rewrite}
\end {multline}
where%
\footnote{
  Using (\ref{eq:aaybar}), one may rewrite the factors of
  $a_\ybx^{-1}$ in (\ref{eq:XYZdef}) as
  $a_\ybx^{-1} = a_\ybx^\top \mathfrak{M}$, if desired.
  A similar relation between $a_\yx^{-1}$ and $a_\yx$ holds
  if one cyclically permutes
  $(x_1,x_2,x_3,x_4) \to (x_2,x_3,x_4,x_1)$
  in the expression (\ref{eq:frakM})
  for $\mathfrak{M}$.
}
\begin {subequations}
\label {eq:XYZdef}
\begin {align}
   \begin{pmatrix} X_\bx & Y_\bx \\ Y_\bx & Z_\bx \end{pmatrix}
   &\equiv
   \begin{pmatrix} |M_\ix|\Omega_\ix & 0 \\ 0 & 0 \end{pmatrix}
     - i a_\bx^{-1\top} \uOmega \cot(\uOmega\,\Delta t)\, a_\bx^{-1} ,
\\
   \begin{pmatrix} X_\Ax & Y_\Ax \\ Y_\Ax & Z_\Ax \end{pmatrix}
   &\equiv
   \begin{pmatrix} |M_\fx|\Omega_\fx & 0 \\ 0 & 0 \end{pmatrix}
     - i a_\Ax^{-1\top} \uOmega \cot(\uOmega\,\Delta t)\, a_\Ax^{-1} ,
\\
   \begin{pmatrix} X_{\bx\Ax} & Y_{\bx\Ax} \\ \Ybar_{\bx\Ax} & Z_{\bx\Ax} \end{pmatrix}
   &\equiv
   - i a_\bx^{-1\top} \uOmega \csc(\uOmega\,\Delta t) \, a_\Ax^{-1} .
\end {align}
\end {subequations}
With this notation, the rate (\ref{eq:IIxyyx2}) becomes
\begin {align}
   \left[\frac{d\Gamma}{dx\,dy}\right]_{xy\bar y\bar x}
   = &
   \frac{\CA^2 \alphas^2 M_\ix M_\fx}{32\pi^4 E^2} \,
     ({-}\hat x_1 \hat x_2 \hat x_3 \hat x_4)
     ( \alpha \delta^{\bar n n} \delta^{\bar m m}
     {+} \beta \delta^{\bar n \bar m} \delta^{nm}
     {+} \gamma \delta^{\bar n m} \delta^{n \bar m} )
\nonumber\\ & \times
   \int_0^{\infty} d(\Delta t) \>
   \Omega_+\Omega_- \csc(\Omega_+\Delta t) \csc(\Omega_-\Delta t)
\nonumber\\ & \times
   \int_{\B^\Ax,\B^\bx}
   \frac{B^{\ybx}_{\bar n}}{(B^\ybx)^2} \,
   \frac{B^{\yx}_m}{(B^\yx)^2} \,
   \bigr[
     (Y_\bx \B^\bx - \Ybar_{\bx\Ax} \B^\Ax)_n
     (Y_\Ax \B^\Ax - Y_{\bx\Ax} \B^\bx)_{\bar m}
     + Z_{\bx\Ax} \delta_{n\bar m}
   \bigl]
\nonumber\\ & \times
   \exp\Bigl[
     - \tfrac12
     X_\bx (B^\bx)^2
     -
     \tfrac12
     X_\Ax (B^\Ax)^2
     +
     X_{\bx\Ax} \B^\bx \cdot \B^\Ax
   \Bigr] .
\end {align}
The $\B$ integrals may be performed using
\begin {subequations}
\label {eq:I}
\begin {align}
   I_0 &\equiv
   \int_{\B^\bx,\B^\Ax}
   \exp\Bigl[
     - \tfrac12 X_\bx (B^\bx)^2
     - \tfrac12 X_\Ax (B^\Ax)^2
     + X_{\bx\Ax} \B^\bx \cdot \B^\Ax
   \Bigr]
\nonumber\\ &\qquad
   =
   \frac{4\pi^2}{(X_\bx X_\Ax - X_{\bx\Ax}^2)} \,,
\displaybreak[0]\\
   I_1 &\equiv
   \int_{\B^\bx,\B^\Ax}
   \frac{\B^\bx\cdot\B^\Ax}{(B^\bx)^2(B^\Ax)^2}
   \exp\Bigl[
     - \tfrac12 X_\bx (B^\bx)^2
     - \tfrac12 X_\Ax (B^\Ax)^2
     + X_{\bx\Ax} \B^\bx \cdot \B^\Ax
   \Bigr]
\nonumber\\ &\qquad
   =
   - \frac{2\pi^2}{X_{\bx\Ax}}
   \ln\left( 1 - \frac{X_{\bx\Ax}^2}{X_\bx X_\Ax} \right) ,
\displaybreak[0]\\
   I_2 &\equiv
   \int_{\B^\bx,\B^\Ax}
   \frac{(\B^\bx\cdot\B^\Ax)^2}{(B^\bx)^2(B^\Ax)^2}
   \exp\Bigl[
     - \tfrac12 X_\bx (B^\bx)^2
     - \tfrac12 X_\Ax (B^\Ax)^2
     + X_{\bx\Ax} \B^\bx \cdot \B^\Ax
   \Bigr]
\nonumber\\ &\qquad
   =
   \frac{2\pi^2}{X_{\bx\Ax}^2}
     \ln\left( 1 - \frac{X_{\bx\Ax}^2}{X_\bx X_\Ax} \right)
   + \frac{4\pi^2}{(X_\bx X_\Ax - X_{\bx\Ax}^2)} \,,
\displaybreak[0]\\
   I_3 &\equiv
   \int_{\B^\bx,\B^\Ax}
   \frac{\B^\bx\cdot\B^\Ax}{(B^\bx)^2}
   \exp\Bigl[
     - \tfrac12 X_\bx (B^\bx)^2
     - \tfrac12 X_\Ax (B^\Ax)^2
     + X_{\bx\Ax} \B^\bx \cdot \B^\Ax
   \Bigr]
\nonumber\\ &\qquad
   =
   \frac{4\pi^2 X_{\bx\Ax}}{X_\Ax(X_\bx X_\Ax - X_{\bx\Ax}^2)} \,,
\displaybreak[0]\\
   I_4 &\equiv
   \int_{\B^\bx,\B^\Ax}
   \frac{\B^\bx\cdot\B^\Ax}{(B^\Ax)^2}
   \exp\Bigl[
     - \tfrac12 X_\bx (B^\bx)^2
     - \tfrac12 X_\Ax (B^\Ax)^2
     + X_{\bx\Ax} \B^\bx \cdot \B^\Ax
   \Bigr]
\nonumber\\ &\qquad
   =
   \frac{4\pi^2 X_{\bx\Ax}}{X_\bx(X_\bx X_\Ax - X_{\bx\Ax}^2)} \,,
\end {align}
\end {subequations}
giving
\begin {align}
   \left[\frac{d\Gamma}{dx\,dy}\right]_{xy\bar y\bar x} = &
   \frac{\CA^2 \alphas^2 M_\ix M_\fx}{32\pi^4 E^2} \, 
   ({-}\hat x_1 \hat x_2 \hat x_3 \hat x_4)
   \int_0^{\infty} d(\Delta t) \>
   \Omega_+\Omega_- \csc(\Omega_+\Delta t) \csc(\Omega_-\Delta t)
\nonumber\\ &\times
   \Bigl\{
     (\beta Y_\bx Y_\Ax + \alpha \Ybar_{\bx\Ax} Y_{\bx\Ax}) I_0
     + (\alpha+\beta+2\gamma) Z_{\bx\Ax} I_1
\nonumber\\ &\quad
     + \bigl[
         (\alpha+\gamma) Y_\bx Y_\Ax
         + (\beta+\gamma) \Ybar_{\bx\Ax} Y_{\bx\Ax}
        \bigr] I_2
     - (\alpha+\beta+\gamma)
       (\Ybar_{\bx\Ax} Y_\Ax I_3 + Y_\bx Y_{\bx\Ax} I_4)
   \Bigl\} .
\label {eq:IGamma}
\end {align}

% -------------------------------------------------------------------------

\subsection{Small-time divergence}
\label {sec:smalldt1}

The $\Delta t$ integration in the $xy\bar y\bar x$ result
(\ref{eq:IGamma}) has both a
linear and a log UV divergence associated with $\Delta t \to 0$.
Specifically, if we expand the integrand in powers of $\Delta t$
(see appendix \ref{app:dtxyyx} for a little more detail),
we find something of the form
\begin {multline}
   \left[\frac{d\Gamma}{dx\,dy}\right]_{xy\bar y\bar x} =
   \int_0^\infty d(\Delta t)
   \Biggr\{
     \frac{\rm stuff}{(\Delta t)^2}
   + \frac{i \CA^2 \alphas^2}{16\pi^2 \, \Delta t} \, 
       \bigl( \Omega_\ix \sgn M_\ix + \Omega_\fx \sgn M_\fx \bigr)
\\ \times
       \hat x_1^2 \hat x_2 \hat x_3^2 \hat x_4
       (\hat x_1+\hat x_4)^2(\hat x_3+\hat x_4)^2
       \left[
         (\alpha + \beta)
         - \frac{(\alpha + \gamma) \hat x_2 \hat x_4}
                {(\hat x_1 {+} \hat x_4)(\hat x_3 {+} \hat x_4)}
       \right]
%\\
     + O\bigl( (\Delta t)^0 \bigr)
  \Biggl\} .
\label {eq:div1}
\end {multline}
We have not bothered to explicitly show the coefficient ``stuff''
of $1/(\Delta t)^2$ because that divergence is easy to dispense with.
Here is one argument that is easy to make now.
The coefficient ``stuff'' turns out not to depend on any of the
frequencies $\Omega$ and so does not depend on $\hat q$.  It is
therefore the contribution to (\ref{eq:div1}) that comes from
radiation in vacuum; it gives the result we would have gotten
if we had set $\hat q$ to zero.  However, the total rate for
bremsstrahlung in vacuum must be zero by energy and momentum conservation,
and so%
\footnote{
  The vacuum rate is zero because we are analyzing the thick medium
  limit, which can be thought of as an infinite medium.  If we
  had instead restricted splitting time integrals in (\ref{eq:1brem1})
  and (\ref{eq:xyyx2}) to start at some
  initial time $0$, then there would be vacuum radiation associated
  with the initial appearance of the particle, related to the vacuum
  radiation associated with the hard process that created our
  initial particle (except lacking an ultraviolet cut-off).
  The usual analysis technique in this case would
  be to subtract off the vacuum result in order to isolate the part
  of $dI/dx$ (or in our case $dI/dx\,dy$) due to medium effects.
}
\begin {equation}
   \left[\frac{d\Gamma}{dx\,dy}\right]_{\rm total} =
   \left[\frac{d\Gamma}{dx\,dy}\right]_{\rm total} -
   \left[\frac{d\Gamma}{dx\,dy}\right]^{\rm vacuum}_{\rm total} .
\label {eq:vac1}
\end {equation}
Thus no harm is done if we subtract the vacuum piece for each
individual contribution that we compute and so compute
\begin {equation}
   \left[\frac{d\Gamma}{dx\,dy}\right]_{xy\bar y\bar x} -
   \left[\frac{d\Gamma}{dx\,dy}\right]^{\hat q\to 0}_{xy\bar y\bar x}
\label {eq:vac2}
\end {equation}
instead of $[d\Gamma/dx\,dy]_{xy\bar y\bar x}$.
This simply subtracts out the $1/(\Delta t)^2$ piece of (\ref{eq:div1}).
(We will give an alternative argument later that the $1/(\Delta t)^2$
divergences cancel without
relying on {\it a priori}\/ knowledge that the total vacuum contribution must
vanish.)

\begin {figure}[t]
\begin {center}
  \includegraphics[scale=0.8]{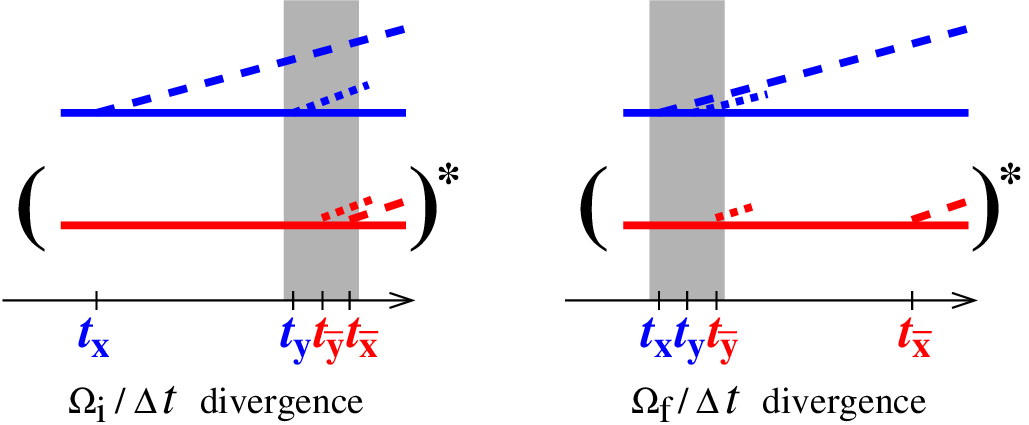}
  \caption{
     \label{fig:xyyxsing}
     A depiction of the (i)
     $\Omega_\ix/\Delta t$ and (ii) $\Omega_\fx/\Delta t$ divergence of
     the $xy\bar y\bar x$ interference as arising from
     the simultaneous approach of (i) $t_\yx$, $t_\ybx$, $t_\xbx$ versus
     (ii) $t_\xx$, $t_\yx$, $t_\ybx$.
     These divergences correspond to the short-time approximation of vacuum
     evolution in the shaded regions.
  }
\end {center}
\end {figure}

In contrast,
the $1/\Delta t$ divergences in (\ref{eq:div1}) do depend on the
medium, but in a very specific
way: they are the sum of terms which depend on
the medium (i.e.\ depend on $\hat q$) either (i) only via $\Omega_\ix$
or (ii) only via $\Omega_\fx$.  In the first case, regarding
the $\Omega_\ix/\Delta t$ terms, this means that
only the contribution from the evolution before $t_\yx$
in fig.\ \ref{fig:xyyx} depends
on the medium, and so (regarding the same terms) the 3- and 4-particle
evolution after $t_\yx$ is equivalent to vacuum evolution.
One expects the medium to be irrelevant over very short times, and indeed we
show explicitly in appendix \ref{app:tripletime} that the
$\Omega_\ix/\Delta t$ divergence arises specifically from the
limit where $t_\yx$, $t_\ybx$, and $t_\xbx$ approach each other
simultaneously in fig.\ \ref{fig:xyyx}.  Similarly, the
$\Omega_\fx/\Delta t$ divergence arises when $t_\xx$, $t_\yx$, and
$t_\ybx$ approach each other simultaneously.
These two situations are depicted in fig.\ \ref{fig:xyyxsing} for
future reference.

We will later see that the $1/\Delta t$ divergences cancel among
different interference contributions, and so we will be able to carry out
the $\Delta t$ integral numerically if we first combine results for
different interference contributions before integrating.  However,
as will be discussed later,
we will nonetheless have to be careful to account for an additional
contribution associated with the pole at $\Delta t=0$.

% =========================================================================

\section {Other crossed interference contributions}
\label {sec:other}

We now turn to relating the other crossed interference diagrams in
fig.\ \ref{fig:subset2} to the result just derived for
the $xy\bar y\bar x$ interference.  Details are given in
appendix \ref{app:relate}, and here we will instead loosely motivate
and then describe the results.

% -------------------------------------------------------------------------

\subsection{\boldmath$x \bar y y \bar x$ interference}
\label {sec:xyyx2}

\begin {figure}[t]
\begin {center}
  \includegraphics[scale=0.5]{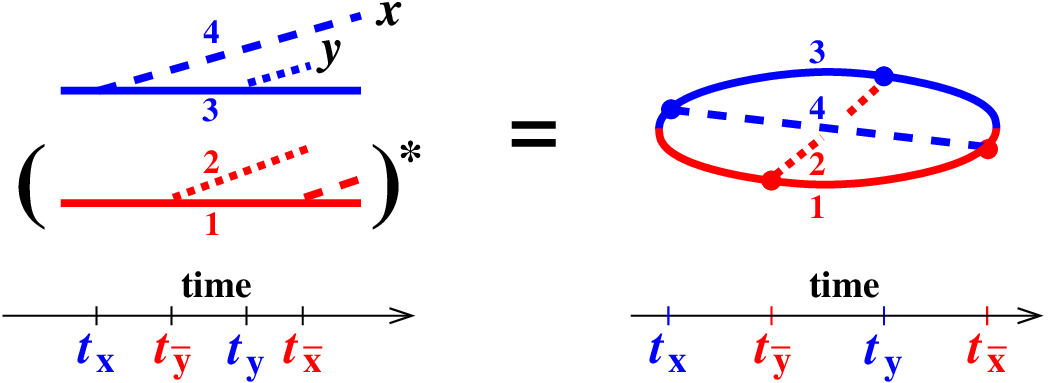}
  \caption{
     \label{fig:xyyx2rate}
     The $x\bar y y\bar x$ interference contribution of
     figs.\ \ref{fig:subset} and \ref{fig:subset2} with
     individual splitting times labeled.  The numbers
     depict our labeling of the 4-particle $x_i$ in (\ref{eq:xprime}).
  }
\end {center}
\end {figure}

The second diagram of figs.\ \ref{fig:subset} and \ref{fig:subset2}
describes what we call the
$x \bar y y \bar x$ interference contribution, which we depict with
time labels in fig.\ \ref{fig:xyyx2rate}.
One difference with the $x y \bar y \bar x$ case analyzed previously
is that the $x_i$ during the period of 4-particle evolution are
different.  In contrast to (\ref{eq:xhat}), we have
\begin {equation}
   (x'_1,x'_2,x'_3,x'_4) = \bigl(-(1{-}y),{-}y,1{-}x,x\bigr) ,
\label {eq:xprime}
\end {equation}
numbered (for the sake of large-$\Nc$) in the
cyclic order one would get if drawing the diagram on a cylinder,
analogous to the $xy\bar y\bar x$ case of fig.\ \ref{fig:cylinder}.
Note that there are now two negative $x_i$ because two of the
high-energy particles are in the conjugate amplitude during the
4-particle evolution ($t_\ybx < t < t_\yx$).
We have chosen to indicate the 4-particle $x_i$ for $x\bar y y \bar x$
using primes, as in (\ref{eq:xprime}), and we will reserve the
hats for the $xy\bar y \bar x$ case, as in (\ref{eq:xhat}).
The two are related by
\begin {equation}
   (x'_1,x'_2,x'_3,x'_4) =
   \bigl(
     -(\hat x_3{+}\hat x_4),{-}\hat x_2,{-}(\hat x_1{+}\hat x_4),\hat x_4
   \bigr) .
\label {eq:xprimeconvert}
\end {equation}
In this notation, one of the changes in going from $xy\bar y\bar x$ to
$x\bar y y\bar x$ can then be summarized as
$\hat x_i \to x'_i$.

Another difference between $x\bar y y\bar x$ and $x y\bar y\bar x$ can be
seen by considering the splittings that occur at the start
($t_\ybx$) and end ($t_\yx$) of
the 4-particle evolution.
At $t{=}t_\ybx$ in fig.\ \ref{fig:xyyx2rate}, particles 3 and
4 are spectators to the splitting, and so the natural variables for
describing the system (following our discussion in section
\ref{sec:firstxyyx})
are $(\C_{34}^\ybx,\C_{12}^\ybx)$.  Similarly, the natural variables at
the end of the 4-particle evolution, $t{=}t_\yx$, are
$(\C_{41}^\yx,\C_{23}^\yx)$.  Looking only at the variable names, this
seems like the same choices as for the $xy\bar y\bar x$ analysis in
(\ref{eq:xyyx2}).  The difference is that $(\C_{34}^\ybx,\C_{12}^\ybx)$
are the variables at the {\it start}\/ of the 4-particle evolution of
$x\bar y y\bar x$ but the {\it end}\/ of the 4-particle evolution of
$xy\bar y\bar x$.  As a result, $(\C_{34}^\ybx,\C_{12}^\ybx)$ will be
tied by the splitting matrix element to the initial 3-particle
evolution ($t_\xx < t < t_\ybx$) for $x\bar y y\bar x$ but was tied
to the final 3-particle evolution ($t_\ybx < t < t_\xbx$) for
$xy\bar y\bar x$.  The treatment of the initial and final 3-particle
evolution was fairly symmetric in section \ref{sec:xyyx}, and the
effect of the 3-particle evolutions in the final result
(\ref{eq:IGamma}) appeared only through the parameters
$(M_\ix,\Omega_\ix)$ and $(M_\fx,\Omega_\fx)$.  So one might expect
that one can account for this difference between $xy\bar y\bar x$
and $x\bar y y\bar x$ simply by interchanging those parameters:
$(M_\ix,\Omega_\ix) \leftrightarrow (M_\fx,\Omega_\fx)$.
In fact, this change is automatic under $\hat x_i \to x'_i$
if one chooses to write
3-particle $M$ and $\Omega$ in terms of the 4-particle $x_i$,
as in the middle expressions in (\ref{eq:MOmi}) and (\ref{eq:MOmf}):
\begin {equation}
   M_\ix = \hat x_1 \hat x_4 (\hat x_1{+}\hat x_4) E
   \underset{\hat x_i {\to} x'_i}\longrightarrow
   x_1' x_4' (x'_1{+}x'_4) E =
   (\hat x_3{+}\hat x_4) \hat x_4 \hat x_3 E = M_\fx ,
\end {equation}
and similarly $M_\fx \to M_\ix$ and
$\Omega_\ix \leftrightarrow \Omega_\fx$.

\begin {figure}[t]
\begin {center}
  \includegraphics[scale=0.5]{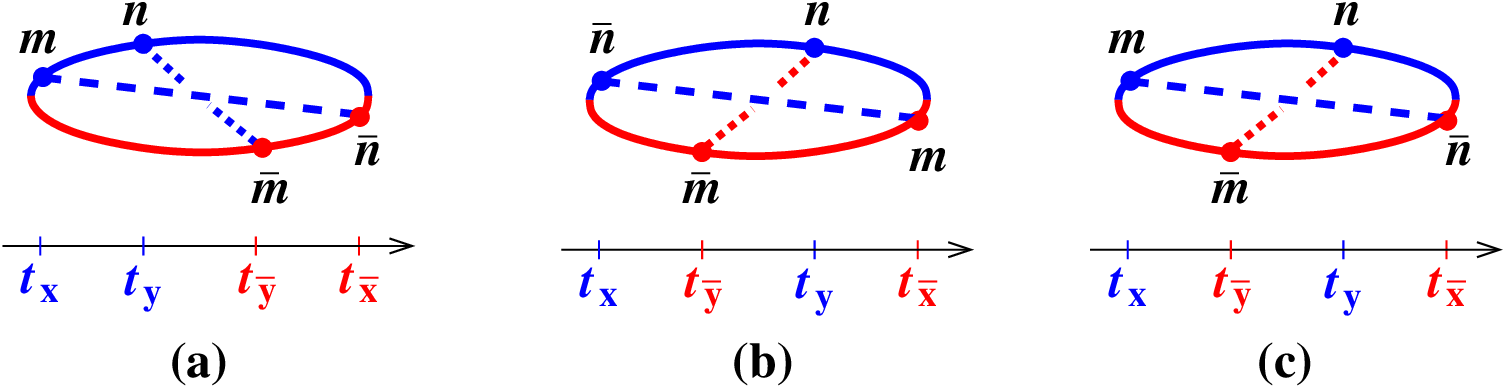}
  \caption{
     \label{fig:relate}
     Relation between labeling of splitting vertices and
     (\ref{eq:abcdef2}) for (a) our previous $xy\bar y\bar x$ calculation,
     (b) what we would get attempting to relate that calculation to
     $x\bar y y \bar x$ by
     only taking $\hat x_i \to x'_i$
     [and so also $(M,\Omega)_\ix \leftrightarrow (M,\Omega)_\fx$],
     and (c) how the labeling should actually be for $x\bar y y\bar x$.
  }
\end {center}
\end {figure}

However, there is
one additional element to the relation between $xy\bar y\bar x$ and
$x\bar y y\bar x$, which are the contractions of the helicity
amplitudes associated with each splitting matrix to make the
combination (\ref{eq:abcdef}),
\begin {equation}
  \alpha(x,y) \, \delta^{\bar n n} \delta^{\bar m m}
  + \beta(x,y) \, \delta^{\bar n \bar m} \delta^{nm}
  + \gamma(x,y) \, \delta^{\bar n m} \delta^{n \bar m} ,
\label {eq:abcdef2}
\end {equation}
discussed for $xy \bar y\bar x$ back in section \ref{sec:helicity},
where the indices $(m,n,\bar m,\bar n)$ were associated with the
splitting vertices as in fig.\ \ref{fig:relate}a.
Holding $x$ and $y$ fixed in $\alpha$, $\beta$, and $\gamma$, the
rules that we have discussed so far for relating $xy \bar y\bar x$
to $x\bar y y\bar x$, which swap the initial and final 3-particle
evolutions, would correspond to (\ref{eq:abcdef2}) contracted
with vertices for $x\bar y y\bar x$ as in fig.\ \ref{fig:relate}b.
However, as discussed earlier and exemplified by (\ref{eq:calPsum2}),
the helicity sums should be the same for all of the diagrams in
the first line of figs.\ \ref{fig:subset} and \ref{fig:subset2}.
Reading from (\ref{eq:calPsum2}), the vertices should be labeled as
in fig.\ \ref{fig:relate}c instead of fig.\ \ref{fig:relate}b.
That is, we need to swap $m \leftrightarrow \bar n$.
From (\ref{eq:abcdef2}), this is equivalent to instead swapping
$\alpha \leftrightarrow \beta$.

Readers who find the above arguments a little too impressionistic should
refer to appendix \ref{app:relate}.  In particular, the discussion above
completely sweeps under the rug a subtlety concerning the overall sign
of the $x \bar y y \bar x$ contribution, which is related to why
we presciently replaced $|x_1 x_2 x_3 x_4|$ by $-x_1 x_2 x_3 x_4$ back
in (\ref{eq:Cprop}).  (Though these are the same for $xy\bar y\bar x$,
they differ by a sign for $x\bar y y \bar x$.)

In summary, the steps necessary to change the result (\ref{eq:IGamma})
for $xy\bar y\bar x$ to a result for $x\bar y y\bar x$ are
\begin {itemize}
\item
   $\hat x_i \to x'_i$ (to include the formulas for $\Omega_\pm$,
   $a_\yx$, and $a_\ybx$);
\item
   $(M_\ix,\Omega_\ix) \leftrightarrow (M_\fx,\Omega_\fx)$ if this
   change was overlooked when applying the $\hat x_i \to x'_i$ rule;
\item
   $\alpha(x,y) \leftrightarrow \beta(x,y)$ .
\end {itemize}
The interpretation of the 4-particle evolution time
$\Delta t$ in the resulting formula is now
$\Delta t = t_\yx - t_\ybx$ rather than $\Delta t = t_\ybx-t_\yx$,
so that $\Delta t$ is still positive.

Applying these rules on the $1/\Delta t$ divergence of
(\ref{eq:div1}), the corresponding divergence for
$x\bar y y\bar x$ is the $\Delta t$ integral of
\begin {equation}
   \frac{i \CA^2 \alphas^2}{16\pi^2 \, \Delta t} \, 
       \bigl( \Omega_\fx \sgn M_\fx + \Omega_\ix \sgn M_\ix \bigr)
       {x'_1}^{\!2} x'_2 {x'_3}^{\!2} x'_4
       (x'_1+x'_4)^2(x'_3+x'_4)^2
       \left[
         (\beta + \alpha)
         - \frac{(\beta + \gamma) x'_2 x'_4}
                {(x'_1 {+} x'_4)(x'_3 {+} x'_4)}
       \right]
    .
\label {eq:div2a}
\end {equation}
Using (\ref{eq:xprimeconvert}), this can be rewritten as
\begin {equation}
   \frac{i \CA^2 \alphas^2}{16\pi^2 \, \Delta t} \, 
       \bigl( \Omega_\ix \sgn M_\ix + \Omega_\fx \sgn M_\fx \bigr)
       \hat x_1^2 \hat x_2 \hat x_3^2 \hat x_4
       (\hat x_1+\hat x_4)^2(\hat x_3+\hat x_4)^2
       \left[
         - (\alpha + \beta)
         - \frac{(\beta + \gamma) \hat x_2 \hat x_4}
                {\hat x_1 \hat x_3}
       \right]
    .
\label {eq:div2b}
\end {equation}
Note that the $(\alpha{+}\beta)$ term will cancel when (\ref{eq:div1})
and (\ref{eq:div2b}) are added together.  To see the complete cancellation
of all $1/\Delta t$ divergences, we must go on to analyze other
interference contributions.

% -------------------------------------------------------------------------

\subsection{\boldmath$x \bar y \bar x y$ interference}
\label {sec:xyxy}

\begin {figure}[t]
\begin {center}
  \includegraphics[scale=0.5]{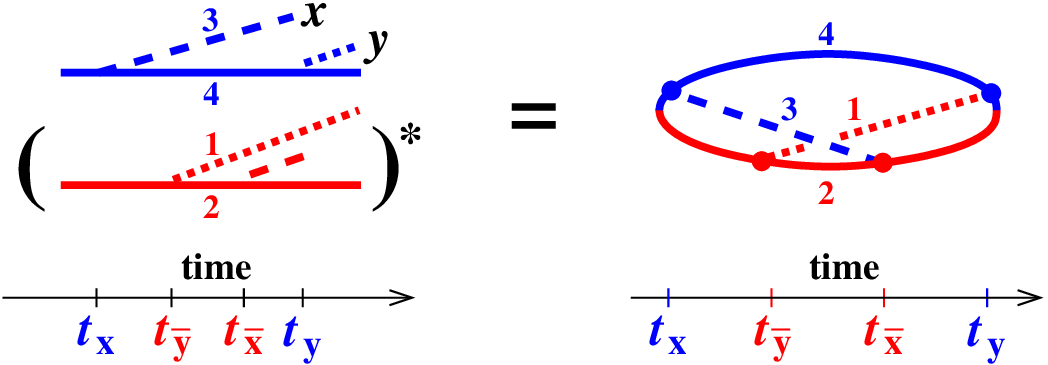}
  \caption{
     \label{fig:xyxyrate}
     The $x\bar y \bar x y$ interference contribution of
     figs.\ \ref{fig:subset} and \ref{fig:subset2} with
     individual splitting times labeled.  The numbers
     depict our labeling of the 4-particle $x_i$ as
     in (\ref{eq:xtilde}).
  }
\end {center}
\end {figure}

The $x\bar y \bar x y$ interference is our name for the third diagram
in figs.\ \ref{fig:subset} and \ref{fig:subset2}.  We will label the
particles during the 4-particle evolution as in fig.\ \ref{fig:xyxyrate}
and designate the 4-particle $x_i$ with tildes:
\begin {equation}
   (\tilde x_1,\tilde x_2,\tilde x_3,\tilde x_4) =
   \bigl({-}y,-(1{-}y),x,1{-}x\bigr) .
\label {eq:xtilde}
\end {equation}
This is just a permutation of
the $x'_i$ in (\ref{eq:xprime}),
\begin {equation}
   (\tilde x_1,\tilde x_2,\tilde x_3,\tilde x_4) =
   (x'_2,x'_1,x'_4,x'_3) .
\label{eq:xtilde1}
\end {equation}
It is a convenient permutation of the labels because, as seen from
comparing figs.\ \ref{fig:xyyx2rate} and \ref{fig:xyxyrate}, it
preserves the labels of which particles are involved in the
splitting at the beginning and end of the 4-particle evolution:
particles 3 and 4 are the spectators at the beginning ($t{=}t_\ybx$), and
1 and 3 are the spectators at the end ($t{=}t_\yx$ for $x\bar y y\bar x$
and $t{=}t_\xbx$ for $x\bar y \bar x y$).  So the structure of the
two diagrams is the same, and the first rule of converting the
results for $x\bar y y\bar x$ is simply $x'_i \to \tilde x_i$, or
equivalently $1\leftrightarrow 2$ and $3\leftrightarrow 4$.
Note that if
we also apply this rule to the formulas for $M_\ix$ and $M_\fx$ in terms
of $x'_i$, it gives the appropriate initial and final 3-particle masses
$M$ associated with fig.\ \ref{fig:xyxyrate}:
$M_\ix = (x'_3{+}x'_4) x'_4 x'_3 E$ does not change under
$13 \leftrightarrow 24$, and
\begin {equation}
   M_\fx = x'_1 x'_4 (x'_1{+}x'_4) E
   \underset{x'_i {\to} \tilde x_i}\longrightarrow
   \tilde M_\fx \equiv \tilde x_1 \tilde x_4 (\tilde x_1{+}\tilde x_4) E =
   x'_2 x'_3 (x'_2{+}x'_3) E .
\end {equation}
Here,
\begin {equation}
  \tilde M_\fx = - y(1{-}x)(1{-}x{-}y) E
\end {equation}
is different from our previous $M$'s because the
particles in the final 3-particle evolution of $x\bar y \bar x y$
are different than those for $xy\bar y\bar x$ and $x\bar y y\bar x$.
Unlike $M_\ix$ and $M_\fx$,
this $\tilde M_\fx$ is negative.  Similarly, the transformation
gives us the appropriate
\begin {equation}
   \tilde\Omega_\fx
   = \sqrt{ 
     -\frac{i \hat q_{\rm A}}{2E}
     \left( \frac{1}{\tilde x_1}
            + \frac{1}{\tilde x_4} - \frac{1}{\tilde x_1 {+} \tilde x_4}
            \right)
   }
   = \sqrt{ 
     -\frac{i \hat q_{\rm A}}{2E}
     \left( -\frac{1}{y} + \frac{1}{1{-}x} - \frac{1}{1{-}x{-}y}
            \right)
   } .
\end {equation}
Unlike the $\Omega_\ix$ and $\Omega_\fx$ discussed previously,
$\tilde\Omega_\fx$ is proportional to $\sqrt{+i}$ rather than
$\sqrt{-i}$ and so has a positive rather than negative imaginary
part.  This is the reason that we earlier made the (at that time
unnecessary) generalization
from (\ref{eq:3int0}) to (\ref{eq:3ints}) in our discussion of
integrating out the first and last splitting times.

We can now express the transformation directly from $xy\bar y\bar x$
to $x\bar y\bar x y$ in one step by rewriting (\ref{eq:xtilde1})
in terms of the $\hat x_i$:
\begin {equation}
   (\tilde x_1,\tilde x_2,\tilde x_3,\tilde x_4) =
   \bigl(
     {-}\hat x_2,-(\hat x_3{+}\hat x_4),\hat x_4,{-}(\hat x_1{+}\hat x_4)
   \bigr) .
\label {eq:xtilde2}
\end {equation}
Now remember that, by itself, the transformation from $\hat x_i$ to
$x'_i$ gave us fig.\ \ref{fig:relate}b instead of what we needed
for $x\bar y y\bar x$,
fig.\ \ref{fig:relate}c.  If we now make the further transformation
$12 \leftrightarrow 34$ that takes us from $x'_i$ to $\tilde x_i$,
fig.\ \ref{fig:relate}b will become fig.\ \ref{fig:relate2}b
for $x\bar y\bar x y$.%
\footnote{
  Again, see Appendix \ref{app:relate} is this be less than clear.
}
But what we need, as read from
(\ref{eq:calPsum2}), is fig.\ \ref{fig:relate2}c.  This requires
$mn\bar n \to n\bar nm$, which, from (\ref{eq:abcdef2}), is
equivalent to $\alpha\beta\gamma \to \gamma\alpha\beta$.

\begin {figure}[t]
\begin {center}
  \includegraphics[scale=0.5]{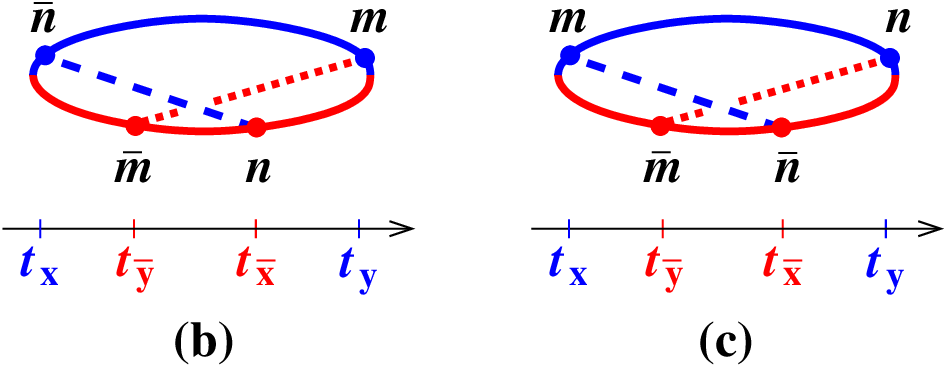}
  \caption{
     \label{fig:relate2}
     As figs.\ \ref{fig:relate}(b,c) but here for $x\bar y\bar x y$.
  }
\end {center}
\end {figure}

In summary, the steps necessary to change the result (\ref{eq:IGamma})
for $xy\bar y\bar x$ to a result for $x\bar y \bar x y$ are
\begin {itemize}
\item
   $\hat x_i \to \tilde x_i$ (to include the formulas for $\Omega_\pm$,
   $a_\yx$, and $a_\ybx$);
\item
   $(M_\ix,\Omega_\ix;M_\fx,\Omega_\fx)
    \to (\tilde M_\fx,\tilde\Omega_\fx;M_\ix,\Omega_\ix)$
    if this
    change was overlooked when applying the $\hat x_i \to \tilde x_i$ rule;
\item
   $\bigl(\alpha(x,y),\beta(x,y),\gamma(x,y)\bigr) \to
    \bigl(\gamma(x,y),\alpha(x,y),\beta(x,y)\bigr)$ .
\end {itemize}
The interpretation of the 4-particle evolution time
$\Delta t$ in the resulting formula is now
$\Delta t = t_\xbx - t_\ybx > 0$.

Applying these rules on the $1/\Delta t$ divergence of
(\ref{eq:div1}), the corresponding divergence for
$x\bar y y\bar x$ is the $\Delta t$ integral of
\begin {equation}
   \frac{i \CA^2 \alphas^2}{16\pi^2 \, \Delta t} \, 
       \bigl( \tilde\Omega_\fx \sgn \tilde M_\fx + \Omega_\ix \sgn M_\ix \bigr)
       \tilde x_1^2 \tilde x_2 \tilde x_3^2 \tilde x_4
       (\tilde x_1+ \tilde x_4)^2(\tilde x_3+\tilde x_4)^2
       \left[
         (\gamma + \alpha)
         - \frac{(\gamma + \beta) \tilde x_2 \tilde x_4}
                {(\tilde x_1 {+} \tilde x_4)(\tilde x_3 {+} \tilde x_4)}
       \right]
    .
\label {eq:div3a}
\end {equation}
Using (\ref{eq:xtilde2}), this can be rewritten as
\begin {multline}
   \frac{i \CA^2 \alphas^2}{16\pi^2 \, \Delta t} \, 
       \bigl( \Omega_\ix \sgn M_\ix + \tilde \Omega_\fx \sgn \tilde M_\fx \bigr)
       \hat x_1^2 \hat x_2 \hat x_3^2 \hat x_4
       (\hat x_1+\hat x_4)^2(\hat x_3+\hat x_4)^2
\\ \times
       \left[
         \frac{(\alpha + \gamma) \hat x_2 \hat x_4}
                {(\hat x_1 {+} \hat x_4)(\hat x_3 {+} \hat x_4)}
         + \frac{(\beta + \gamma) \hat x_2 \hat x_4}
                {\hat x_1 \hat x_3}
       \right]
    .
\label {eq:div3b}
\end {multline}
Note that all of the $\Omega_\ix$ terms now cancel between
(\ref{eq:div1}), (\ref{eq:div2b}), and (\ref{eq:div3b})---that is,
between the three diagrams shown in the first line of
figs.\ \ref{fig:subset} and \ref{fig:subset2}.

% =========================================================================

\section {\boldmath$1/\Delta t$ divergences and
          \boldmath$i\epsilon$ prescriptions}
\label {sec:smalldt}

\begin {figure}[t]
\begin {center}
  \includegraphics[scale=0.5]{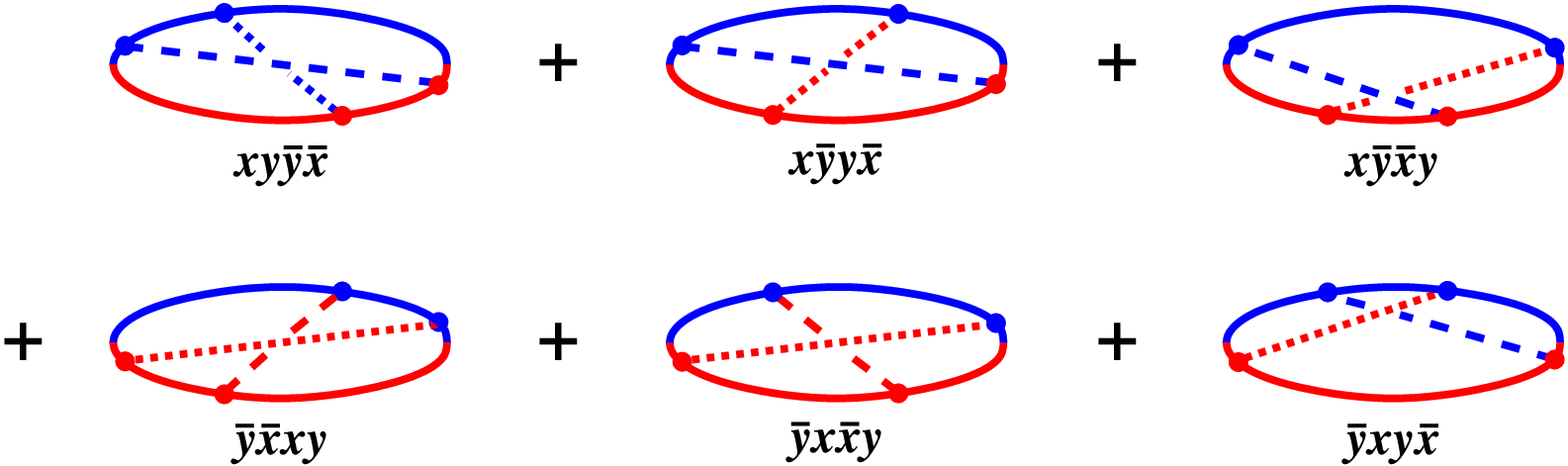}
  \caption{
     \label{fig:cancel}
     A subset of crossed diagrams for which all $1/\Delta t$ divergences
     cancel.
  }
\end {center}
\end {figure}

We have now seen a subset of $1/\Delta t$ divergences cancel in the
first line of fig.\ \ref{fig:subset2}.  To get the rest of the
cancellations just requires a slightly bigger subset
of diagrams: All of the $1/\Delta t$ divergences cancel
between the six diagrams shown in fig.\ \ref{fig:cancel}.
The second line of fig.\ \ref{fig:cancel} is just the conjugate
of the first line permuted by $x \leftrightarrow y$.
One may check that adding together the $1/\Delta t$ terms of
(\ref{eq:div1}), (\ref{eq:div2b}),
(\ref{eq:div3b}) and their conjugates with $x\leftrightarrow y$
gives zero.

The complete set of crossed diagrams depicted by fig.\ \ref{fig:subset2}
can be written as the sum of fig.\ \ref{fig:cancel} and its conjugate,
plus the remaining possible permutations of the three final daughters
$(x,y,1{-}x{-}y)$.
So $1/\Delta t$ divergences will cancel for the complete set of
crossed diagrams as well.
That means that, if we combine results for the crossed
diagrams before doing the $\Delta t$ integral, and we subtract
the corresponding vacuum results as in (\ref{eq:vac1}) and (\ref{eq:vac2}),
the $\Delta t$ integral will be finite and can be performed numerically.
Before we summarize formulas for the final result, however, we need
to take care of a subtlety with the contribution from
$\Delta t{=}0$ to which we have
previously alluded.
And before we do that, it will be enlightening to first review
some technical details of the calculation of single (as opposed to double)
splitting.

\bigskip
{\it Note added}: The analysis of this section misses some terms
in the contribution from $\Delta t=0$.  A corrected analysis,
which includes the missing terms, may be found in ref.\ \cite{dimreg}.

% --------------------------------------------------------------------------

\subsection {\boldmath$i\epsilon$ prescription in single splitting}
\label {sec:single1eps}

Return to the formula (\ref{eq:drate123}) for the single splitting
rate.  Using the propagator (\ref{eq:1prop}), this
formula becomes
\begin {equation}
   \frac{d\Gamma}{dx}
   =
   - \frac{\alpha P_{1\to 3}(x)}{\pi} \,
   \Re \int_0^\infty d(\Delta t) \>
   \Omega_0^2 \csc^2(\Omega_0 \, \Delta t) ,
\label {eq:1bremB}
\end {equation}
where $\Delta t \equiv t_\xbx-t_\xx$.
The integrand blows up like $1/(\Delta t)^2$ as $\Delta t \to 0$.
A simple way to avoid this problem altogether is to subtract out
the vanishing rate of splitting in vacuum, analogous to
(\ref{eq:vac1}):
\begin {multline}
   \frac{d\Gamma}{dx}
   =
   \frac{d\Gamma}{dx}
   - \left[ \frac{d\Gamma}{dx}\right]^{\rm vacuum}
   =
   \frac{d\Gamma}{dx}
   - \lim_{\Omega_0\to0} \frac{d\Gamma}{dx}
\\
   =
   - \frac{\alpha P_{1\to 3}(x)}{\pi} \,
   \Re \int_0^\infty d(\Delta t) \>
   \left[ \Omega_0^2 \csc^2(\Omega_0 \, \Delta t) - \frac1{(\Delta t)^2} \right].
\label {eq:1bremcnvg}
\end {multline}
This integral is convergent and gives the result
\begin {equation}
   \frac{d\Gamma}{dx}
   =
   \frac{\alpha P_{1\to 3}(x)}{\pi} \,
   \Re(i \Omega_0) ,
\end {equation}
which in turn gives (\ref{eq:BDMS}).

However, it will be useful to see how to get the result
without making the vacuum subtraction.
We'll see how to deal with the
$1/(\Delta t)^2$ singularity in (\ref{eq:1bremB}) directly,
but the argument will perhaps look more familiar if we
first transform the $\Delta t$ integral from 0 to $\infty$ to
an integral from $-\infty$ to $+\infty$.

% .......................................................................

\subsubsection {An integral from $-\infty$ to $+\infty$}

\begin {figure}[t]
\begin {center}
  \includegraphics[scale=0.5]{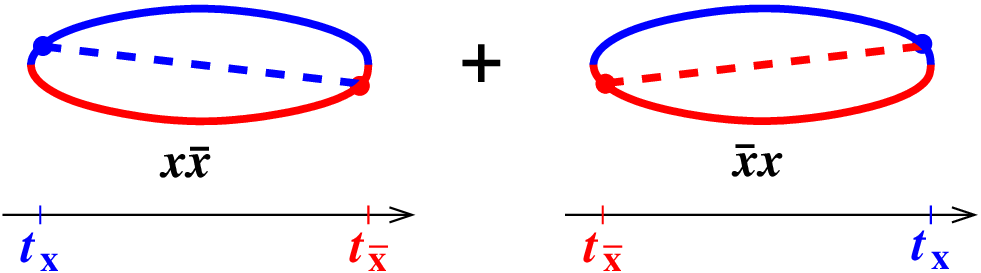}
  \caption{
     \label{fig:xxrate3}
     The interference diagrams contributing to single splitting.
  }
\end {center}
\end {figure}

For the sake of interpretation of $\Delta t$, it will be useful
to consider how we would have arrived at (\ref{eq:1bremB}) if we
had computed the $x\bar x$
interference diagram of fig.\ \ref{fig:xxrate} and
its conjugate $\bar x x$ separately, as depicted in fig.\ \ref{fig:xxrate3}.
The 3-particle $\Omega$ given by (\ref{eq:Omega0B}),
\begin {equation}
   \Omega = 
   \sqrt{
   -\frac{i}{2E}
    \left(
       \frac{\hat q_1}{x_1} + \frac{\hat q_2}{x_2} + \frac{\hat q_3}{x_3}
    \right)
   } ,
\end {equation}
gives (\ref{eq:Omega0}),
\begin {equation}
   \Omega_0 =
   \sqrt{
     - \frac{i \hat q_{\rm A}}{2 E}
     \left( -1 + \frac{1}{1{-}x} + \frac{1}{x} \right)
   } ,
\end {equation}
for the $x\bar x$ interference diagram
[for which $(x_1,x_2,x_3)=(-1,1{-}x,x)$].
But it gives $\Omega_0^*$ for the $\bar x x$ diagram
[for which $(x_1,x_2,x_3)=\bigl(1,-(1{-}x),-x\bigr)$].
The results for these two diagrams, after following the same
steps as in section \ref{sec:1gluon}, are
\begin {subequations}
\label {eq:intpole0}
\begin {equation}
   \left[ \frac{d\Gamma}{dx} \right]_{x\bar x}
   =
   - \frac{\alpha P_{1\to 3}(x)}{2\pi} \,
   \int_0^\infty d(\Delta t) \>
   \Omega_0^2 \csc^2(\Omega_0 \, \Delta t)
\label {eq:dGxxb}
\end {equation}
with $\Delta t \equiv t_\xbx - t_\xx$ and
\begin {equation}
   \left[ \frac{d\Gamma}{dx} \right]_{\bar x x}
   =
   - \frac{\alpha P_{1\to 3}(x)}{2\pi} \,
   \int_0^\infty d(\Delta t) \>
   {\Omega^*}_0^2 \csc^2(\Omega_0^* \, \Delta t)
\label {eq:dGxbx}
\end {equation}
\end {subequations}
with $\Delta t \equiv t_\xx - t_\xbx$.
Adding them together reproduces (\ref{eq:1bremB}), as it should.
Now let's take $\Delta t \to -\Delta t$ in (\ref{eq:dGxbx}) to write
the sum of (\ref{eq:intpole0}) as
\begin {equation}
   \frac{d\Gamma}{dx}
   =
   - \frac{\alpha P_{1\to 3}(x)}{2\pi} \,
   \int_{-\infty}^{+\infty} d(\Delta t) \>
   \begin {cases}
      {\Omega^*_0}^2 \csc^2(\Omega_0^* \, \Delta t), & \Delta t < 0; \\
      \Omega_0^2 \csc^2(\Omega_0 \, \Delta t), & \Delta t > 0 ,
   \end {cases}
\label {eq:intpole}
\end {equation}
where now $\Delta t \equiv t_\xbx - t_\xx$ in both cases.

If the integrand in (\ref{eq:intpole}) were an analytic function of
$\Delta t$, we could now say that the problem of the singularity at
$\Delta t=0$ is just a problem of understanding how one should
integrate around the pole.  The integrand is not analytic because
of the various high-energy approximations that have been made, but
the divergent piece at $\Delta t=0$ is analytic.  We can rewrite
(\ref{eq:intpole}) as the sum of a convergent integral
\begin {equation}
   \frac{d\Gamma}{dx}
   =
   \frac{\alpha P_{1\to 3}(x)}{2\pi} \,
   \int_{-\infty}^{+\infty} d(\Delta t) \>
   \begin {cases}
      {\Omega^*_0}^2 \csc^2(\Omega_0^* \, \Delta t) - (\Delta t)^{-2}, &
          \Delta t < 0; \\
      \Omega_0^{\,\,2\kern2pt} \csc^2(\Omega_0 \, \Delta t) - (\Delta t)^{-2}, &
          \Delta t > 0
   \end {cases}
\end {equation}
[which is equivalent to (\ref{eq:1bremcnvg})] and a potentially divergent
integral
\begin {equation}
   \left[ \frac{d\Gamma}{dx} \right]_{\rm div}
   \propto
   \int_{-\infty}^{+\infty} \frac{d(\Delta t)}{(\Delta t)^2} \,.
\label {eq:intdiv}
\end {equation}
It doesn't really matter how we integrate around the double pole in
(\ref{eq:intdiv}) because we will get zero however we do it.
However, getting straight the prescription will help
deal with the $1/\Delta t$ divergences that arise
in the double splitting case.
So we now discuss the $i\epsilon$ prescription that should be used
for $\Delta t$.

% .......................................................................

\subsubsection {The $i\epsilon$ prescription}
\label {sec:Wightman}

Consider the $\delta H$ matrix elements that appeared
in the formula (\ref{eq:1brem2}) for single splitting,
\begin {equation}
       \int_{\p_\fx,\k_\fx}
       \biggl(
         \int_{\p_\xx}
         \langle \p_\fx \k_\fx, t_\xbx | \p_\xx \k_\fx,t_\xx \rangle
         \langle \p_\xx \k_\fx | {-}i \, \delta H | \p_\ix \rangle
       \biggr)
       \biggl(
         \int_{\bar\p_\xbx}
         \langle \p_\ix,t_\xx| \bar\p_\xbx, t_\xbx \rangle
         \langle  \bar\p_\xbx| i \, \delta H |\p_\fx \k_\fx \rangle
       \biggr) .
\label {eq:dHdH0}
\end {equation}
Using completeness relations, (\ref{eq:dHdH0}) can be rewritten as
\begin {equation}
    \langle \p_\ix | \delta H(t_\xbx) \, \delta H(t_\xx) | \p_\ix \rangle .
\end {equation}
This is a Wightman correlator: the $\delta H$ for the
splitting at $t_\xbx$ is always
ordered to the left of the $\delta H$ for the splitting at $t_\xx$,
regardless of the time order of $t_\xbx$ and $t_\xx$.
The $i\epsilon$
prescription for the small-$\Delta t$ divergence of a
Wightman correlator is
\begin {equation}
   \mbox{$t{-}t' \to t{-}t'{-}i\epsilon$ for
         $\langle A(t) \, B(t') \rangle$.}
\label {eq:WightmanEps}
\end {equation}
See, for example, the argument in Wilson \cite{Wilson} with regard
to the singularity behavior of $A(x) \, B(y)$ in the operator product
expansion.
(We provide our own summary of the argument in appendix \ref{app:details}.)
In our context, we can summarize this rule as follows:
\begin {quote}
  Regard times in the conjugate amplitude as displaced by
  $-i\epsilon$ with respect to times in the amplitude.
\end {quote}

% -------------------------------------------------------------------------

\subsubsection {Other ways to use the prescription}

In our analysis of double splitting, we have always defined our
$\Delta t$ to be positive.  In single splitting, the analog
is to use the $0 < \Delta t < \infty$ integrals of
(\ref{eq:1bremB}) or (\ref{eq:intpole0})
instead of the $-\infty < \Delta t < \infty$
integral of (\ref{eq:intpole}).  It is useful to notice that
the $i\epsilon$ prescription works just as well in this form.
In the case of (\ref{eq:1bremB}), $\Delta t \equiv t_\xbx - t_\xx$.
Since $t_\xx$ is in the amplitude and $t_\xbx$ in the conjugate amplitude,
(\ref{eq:1bremB}) becomes
\begin {equation}
   \frac{d\Gamma}{dx}
   =
   - \frac{\alpha P_{1\to 3}(x)}{\pi} \,
   \Re \int_0^\infty d(\Delta t) \>
   \Omega_0^2 \csc^2\bigl(\Omega_0 \, (\Delta t-i\epsilon)\bigr) .
\end {equation}
The divergent piece is proportional to
\begin {equation}
   \Re \int_0^\infty \frac{d(\Delta t)}{(\Delta t-i\epsilon)^2}
   = \Re \frac1{-i\epsilon}
   = 0 .
\end {equation}
Equivalently (and most like what we will do for double
splitting) we can get the same result from
(\ref{eq:intpole0})
by realizing that $\Delta t \equiv t_\xbx - t_\xx \to \Delta t - i\epsilon$
in the $x\bar x$ piece (\ref{eq:dGxxb}) but
$\Delta t \equiv t_\xx - t_\xbx \to \Delta t + i\epsilon$ in the
$\bar x x$ piece (\ref{eq:dGxbx}), so that the divergent piece of
the sum is proportional to
\begin {equation}
   \int_0^\infty \frac{d(\Delta t)}{(\Delta t-i\epsilon)^2}
   +
   \int_0^\infty \frac{d(\Delta t)}{(\Delta t+i\epsilon)^2}
   = 0.
\label {eq:sumdiv}
\end {equation}

% --------------------------------------------------------------------------

\subsection {Consequences for double splitting}

Just like in the single splitting case, the vacuum $1/(\Delta t)^2$ divergences
in double splitting
will be canceled by adding diagrams to their conjugates, or can be
discarded by simply
subtracting the (vanishing) vacuum result.
However, the sub-leading $1/\Delta t$ divergences give something
non-trivial.

% ........................................................................

\subsubsection {A warm-up example}

As a simple, warm-up example from calculus, imagine the integral
\begin {equation}
  I \equiv \int_{-R}^{+R} d(\Delta t) \> \frac{\Omega}{\Delta t - i\epsilon}
  = i \pi \Omega ,
\label {eq:Iwarmup}
\end {equation}
for some complex frequency $\Omega$ and some bound $R$.
We may replace $\Delta t \to -\Delta t$ for $\Delta t < 0$ to rewrite the
integral as
\begin {equation}
  I = \int_0^R d(\Delta t) \> \frac{\Omega}{\Delta t - i\epsilon}
    + \int_0^R d(\Delta t) \> \frac{-\Omega}{\Delta t + i\epsilon} \,,
\label {eq:sumdiv2}
\end {equation}
which is a $1/\Delta t$ analogy to the $1/(\Delta t)^2$ integrals we had
in (\ref{eq:sumdiv}).  Now combine the two integrands in (\ref{eq:sumdiv2}).
If we ignored the important $i\epsilon$
prescriptions in (\ref{eq:sumdiv2}), we would naively conclude that
the $1/\Delta t$ terms cancel each other, leading to
the incorrect answer $I = 0$.

\begin {figure}[t]
\begin {center}
  \includegraphics[scale=0.5]{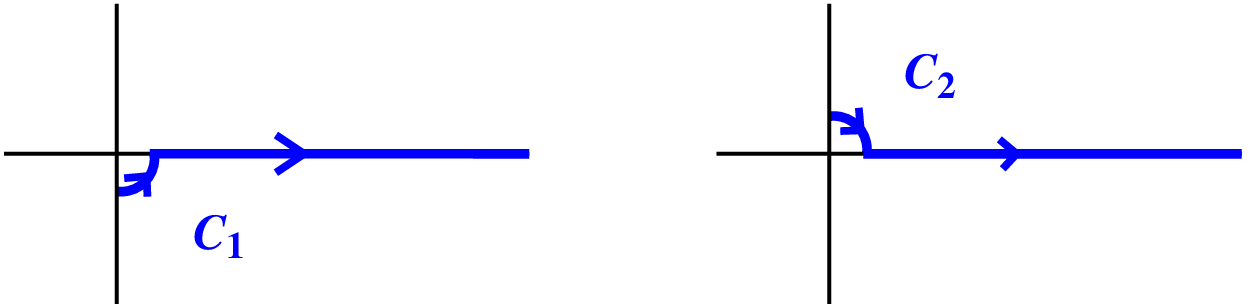}
  \caption{
     \label{fig:contours}
     The contours $C_1$ and $C_2$ of (\ref{eq:contours}) in the
     complex $\Delta t$ plane.
  }
\end {center}
\end {figure}

One can equivalently write
(\ref{eq:sumdiv2}) as
\begin {equation}
  I = \int_{C_1} d(\Delta t) \> \frac{\Omega}{\Delta t}
    + \int_{C_2} d(\Delta t) \> \frac{-\Omega}{\Delta t} \,,
\label {eq:contours}
\end {equation}
where the contours $C_1$ and $C_2$ are shown in fig.\ \ref{fig:contours}.
The integrals indeed cancel {\it except}\/ for the contributions from the
tiny region around $\Delta t=0$, where the contours differ.
The integrals around each quarter-circle give $\pm i \pi/2$ times
the residue at the origin, which brings (\ref{eq:sumdiv2}) to the
same result $I=i \pi \Omega$ as (\ref{eq:Iwarmup}).

A quick mnemonic for all of this is to use the standard relation
\begin {subequations}
\label{eq:poleformula}
\begin {equation}
   \frac{1}{\Delta t \mp i\epsilon} = 
   \operatorname{P.P.}\frac{1}{\Delta t} \pm i \pi \, \delta(\Delta t) ,
\end {equation}
where ``$\operatorname{P.P.}$'' indicates the principal part
prescription, and
with the understanding that
\begin {equation}
   \int_0^R d(\Delta t) \, \delta(\Delta t) = \tfrac12
\end {equation}
\end {subequations}
integrates over only half of the $\delta$ function.
The principal part pieces then cancel between the two terms
in (\ref{eq:sumdiv2}), and the $\delta$ function pieces
account for the
answer $I = i\pi\Omega$.  (In this context, the principal part
prescription does not really do anything since we are not
integrating over negative $\Delta t$.)

% ........................................................................

\subsubsection {Double splitting}
\label{sec:epsilon2}

Let us now apply (\ref{eq:poleformula}) to the $1/\Delta t$ divergences
found previously for double splitting.  For the $xy\bar y\bar x$ result,
this divergence was given in (\ref{eq:div1}), where
$\Delta t \equiv t_\ybx-t_\yx$.  Since times in the conjugate
amplitude should get a $-i\epsilon$, we take
$\Delta t \to \Delta t - i\epsilon$ in (\ref{eq:div1}).  After then
applying (\ref{eq:poleformula}), we will have a $1/\Delta t$ piece
and a $+i\pi\delta(\Delta t)$ piece.  The $1/\Delta t$ pieces will
cancel between diagrams, as already noted in the discussion of
fig.\ \ref{fig:cancel}.  The $+i\pi\delta(\Delta t)$
piece will give an additional contribution to the $xy\bar y\bar x$
interference that we need to retain, given by
\begin {multline}
   \left[ \frac{d\Gamma}{dx\>dy} \right]^{\rm pole}_{x y \bar y \bar x}
   =
     -\frac{\CA^2 \alphas^2}{32\pi} \, 
       \bigl( \Omega_\ix \sgn M_\ix + \Omega_\fx \sgn M_\fx \bigr)
       \hat x_1^2 \hat x_2 \hat x_3^2 \hat x_4
       (\hat x_1+\hat x_4)^2(\hat x_3+\hat x_4)^2
\\ \times
       \left[
         (\alpha + \beta)
         - \frac{(\alpha + \gamma) \hat x_2 \hat x_4}
                {(\hat x_1 {+} \hat x_4)(\hat x_3 {+} \hat x_4)}
       \right]
    .
\label {eq:pole1}
\end {multline}

For $x\bar y y\bar x$, we have a similar situation, but
$\Delta t$ in (\ref{eq:div2b}) is
$\Delta t \equiv t_\yx - t_\ybx \to \Delta t + i\epsilon$, and so the
$i \pi \delta(\Delta t)$ in (\ref{eq:poleformula}) has the opposite
sign.  As a result, the corresponding contribution arising from
(\ref{eq:div2b}) is
\begin {multline}
   \left[ \frac{d\Gamma}{dx\>dy} \right]^{\rm pole}_{x\bar y y\bar x}
   =
   +\frac{\CA^2 \alphas^2}{32\pi} \, 
       \bigl( \Omega_\ix \sgn M_\ix + \Omega_\fx \sgn M_\fx \bigr)
       \hat x_1^2 \hat x_2 \hat x_3^2 \hat x_4
       (\hat x_1+\hat x_4)^2(\hat x_3+\hat x_4)^2
\\ \times
       \left[
         - (\alpha + \beta)
         - \frac{(\beta + \gamma) \hat x_2 \hat x_4}
                {\hat x_1 \hat x_3}
       \right]
    .
\label {eq:pole2}
\end {multline}

\begin {figure}[t]
\begin {center}
  \includegraphics[scale=0.8]{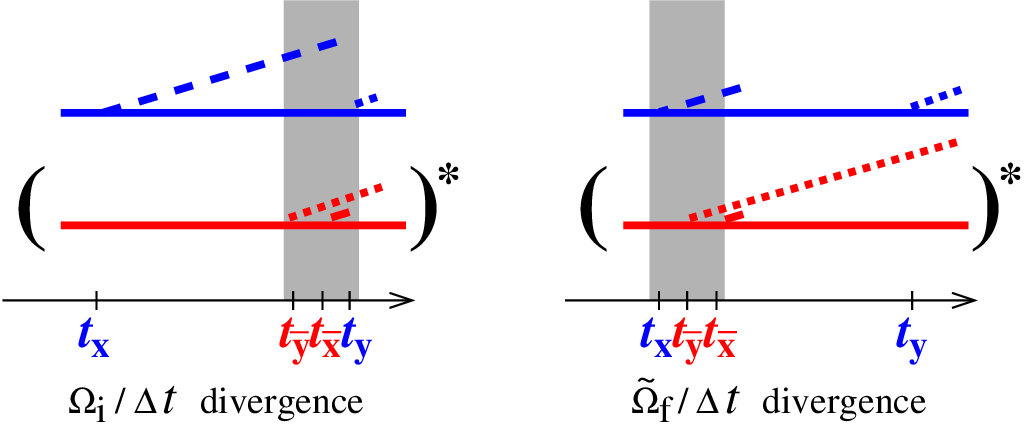}
  \caption{
     \label{fig:xyxysing}
     As fig.\ \ref{fig:xyyxsing} but instead for the $x\bar y\bar x y$
     interference of fig.\ \ref{fig:xyxyrate}.
  }
\end {center}
\end {figure}

The $x\bar y \bar x y$ contribution of fig.\ \ref{fig:xyxyrate} is
a little more subtle because there $\Delta t \equiv t_\xbx-t_\ybx$ does
not pick up a $\pm i\epsilon$ prescription from what we have discussed
so far.  Here, we will indicate how one might guess the answer, and we
leave a more precise argument to appendix \ref{app:epsilon2}.
Analogous to the earlier discussion of fig.\ \ref{fig:xyyxsing}
for $xy\bar y\bar x$,
the $\Omega_\ix/\Delta t$ and $\tilde\Omega_\fx/\Delta t$ divergences
of (\ref{eq:div3b}) arise from the situations depicted in
fig.\ \ref{fig:xyxysing}.  Looking at the figure for
the $\tilde\Omega_\fx/\Delta t$ divergence, remember that, in
the derivations of our expressions (section \ref{sec:firstlast} in particular),
we have already
integrated the first time $t_\xx$ all the way up to the second
time (in this case $t_\ybx$).  And so one might guess that
the $t_\xbx-t_\ybx$ will inherit the $i\epsilon$ prescription of
$t_\xbx-t_\xx$, which is $t_\xbx-t_\xx-i\epsilon$.  Our guess
is that $\Delta t \equiv t_\xbx-t_\ybx \to \Delta t-i\epsilon$
will give the correct result for the $\tilde\Omega_\fx/\Delta t$
divergence.

In contrast, for the $\Omega_\ix/\Delta t$ divergence in
fig.\ \ref{fig:xyxysing}, $t_\xx$ is not close to $t_\ybx$,
but instead $t_\yx$ is close to $t_\xbx$.  One might guess that
$t_\xbx-t_\ybx$ then inherits the $i\epsilon$ prescription of
$t_\yx-t_\ybx$, which is $t_\yx-t_\ybx+i\epsilon$.  So our guess
is that $\Delta t \to \Delta t + i\epsilon$ will give the correct
result for the $\Omega_\ix/\Delta t$ divergence of $x\bar y\bar x y$.
These guesses are correct (see appendix \ref{app:epsilon2})
and put together with
(\ref{eq:div3b}) and (\ref{eq:poleformula}) give
\begin {multline}
   \left[ \frac{d\Gamma}{dx\>dy} \right]^{\rm pole}_{x\bar y\bar x y}
   =
   \frac{\CA^2 \alphas^2}{32\pi} \, 
       \bigl( \Omega_\ix \sgn M_\ix - \tilde \Omega_\fx \sgn \tilde M_\fx \bigr)
       \hat x_1^2 \hat x_2 \hat x_3^2 \hat x_4
       (\hat x_1+\hat x_4)^2(\hat x_3+\hat x_4)^2
\\ \times
       \left[
         \frac{(\alpha + \gamma) \hat x_2 \hat x_4}
                {(\hat x_1 {+} \hat x_4)(\hat x_3 {+} \hat x_4)}
         + \frac{(\beta + \gamma) \hat x_2 \hat x_4}
                {\hat x_1 \hat x_3}
       \right]
    .
\label {eq:pole3}
\end {multline}

The sum of the $\Omega_\ix$ terms of
(\ref{eq:pole1}--\ref{eq:pole3}) is
\begin {equation}
     \frac{\CA^2 \alphas^2}{16\pi} \, 
       \Omega_\ix
       \hat x_1^2 \hat x_2 \hat x_3^2 \hat x_4
       (\hat x_1+\hat x_4)^2(\hat x_3+\hat x_4)^2
       \left[
         - (\alpha + \beta)
         + \frac{(\alpha + \gamma) \hat x_2 \hat x_4}
                {(\hat x_1 {+} \hat x_4)(\hat x_3 {+} \hat x_4)}
       \right]
\end {equation}
(where we've finally used the fact that $\sgn M_\ix = +1$ since we
will no longer be considering variable substitutions that will change
the sign).
If we add (\ref{eq:pole1}--\ref{eq:pole3}) to their conjugates
with $x \leftrightarrow y$, as in the subset of diagrams of
fig.\ \ref{fig:cancel}, then we get
\begin {multline}
   \left[ \frac{d\Gamma}{dx \> dy}
   \right]^{\rm pole}_{\rm fig.\>\ref{fig:cancel}}
   =
     \biggl\{
       \frac{\CA^2 \alphas^2}{16\pi} \, 
       (\Omega_\ix + \tilde\Omega_\fx) \,
       \hat x_1^2 \hat x_2 \hat x_3^2 \hat x_4
       (\hat x_1+\hat x_4)^2(\hat x_3+\hat x_4)^2
\\ \times
       \left[
         - (\alpha + \beta)
         + \frac{(\alpha + \gamma) \hat x_2 \hat x_4}
                {(\hat x_1 {+} \hat x_4)(\hat x_3 {+} \hat x_4)}
       \right]
     \biggr\}
     + \{ x \leftrightarrow y \}^* .
\label {eq:Apole0}
\end {multline}
(where we have now used the fact that $\sgn \tilde M_\fx = -1$).
So, even though the $1/(\Delta t)$ terms naively cancel between
the diagrams of fig.\ \ref{fig:cancel}, the pole contributions
arising from the $\pm i\epsilon$'s do not.

Recalling that the $(\alpha,\beta,\gamma)$ of (\ref{eq:abc}) are
symmetric under $x\leftrightarrow y$, (\ref{eq:Apole0}) can be written
more explicitly as
\begin {subequations}
\label {eq:Apole}
\begin {multline}
   \left[ \frac{d\Gamma}{dx \> dy}
   \right]^{\rm pole}_{\rm fig.\>\ref{fig:cancel}}
   =
       - \frac{\CA^2 \alphas^2}{16\pi} \, 
       [\Omega_{-1,1-x,x} + \Omega_{-(1-y),1-x-y,x}
        + \Omega_{-1,1-y,y}^* + \Omega_{-(1-x),1-x-y,y}^*]
\\ \times
       x y (1{-}x)^2 (1{-}y)^2(1{-}x{-}y)^2
       \left[
         (\alpha + \beta)
         + \frac{(\alpha + \gamma) xy}{(1{-}x)(1{-}y)}
       \right]
     \biggr\}
\end {multline}
where
\begin {equation}
   \Omega_{x_1,x_2,x_3} \equiv
   \sqrt{ 
     -\frac{i \hat q_{\rm A}}{2E}
     \left( \frac{1}{x_1} 
            + \frac{1}{x_2} + \frac{1}{x_3} \right)
   } .
\end {equation}
\end {subequations}

% =========================================================================

\section {Summary of Result}
\label {sec:summary}

We now collect our final result for the sum of all crossed diagrams,
depicted by fig.\ \ref{fig:subset} or \ref{fig:subset2}, for
gluons in the large-$\Nc$ limit in a form appropriate for
(convergent) numerics.  The result is
\begin {equation}
   \left[ \frac{d\Gamma}{dx\>dy} \right]_{\rm crossed}
   =
   A(x,y) + A(1{-}x{-}y,y) + A(x,1{-}x{-}y) ,
\label {eq:summary1}
\end {equation}
where $A(x,y)$ is the result for the six diagrams of
fig.\ \ref{fig:cancel} plus their conjugates, with vacuum
result subtracted.  We will write this as
\begin {align}
   A(x,y) &\equiv
   2 \Re\left\{
      \left[ \frac{d\Gamma}{dx\>dy} \right]_{\rm fig.\>\ref{fig:cancel}}
      - \lim_{\hat q\to 0}
        \left[ \frac{d\Gamma}{dx\>dy} \right]_{\rm fig.\>\ref{fig:cancel}}
   \right\}
\nonumber\\
   &=
   2 \Re\left[ \frac{d\Gamma}{dx\>dy} \right]^{\rm pole}_{\rm fig.\>\ref{fig:cancel}}
   + \int_0^{+\infty} d(\Delta t) \> 
     2 \Re \bigl[ B(x,y,\Delta t) + B(y,x,\Delta t) \bigr] ,
\label {eq:summaryA}
\end {align}
where $B(x,y)$ is the $\Delta t$ integrand for the first three
diagrams of fig.\ \ref{fig:cancel} (with vacuum results subtracted).
This is
\begin {align}
   B(x,y,\Delta t) &=
       C(\{\hat x_i\},\alpha,\beta,\gamma,\Delta t)
       + C(\{x'_i\},\beta,\alpha,\gamma,\Delta t)
       + C(\{\tilde x_i\},\gamma,\alpha,\beta,\Delta t)
\nonumber\\
   &=
       C({-}1,y,1{-}x{-}y,x,\alpha,\beta,\gamma,\Delta t)
       + C\bigl({-}(1{-}y),{-}y,1{-}x,x,\beta,\alpha,\gamma,\Delta t\bigr)
\nonumber\\ &\qquad\qquad
       + C\bigl({-}y,{-}(1{-}y),x,1{-}x,\gamma,\alpha,\beta,\Delta t\bigr) ,
\end {align}
where the three terms correspond to
$xy\bar y\bar x$, $x\bar y y\bar x$, and $x\bar y\bar x y$ respectively,
and $\alpha(x,y)$, $\beta(x,y)$, and $\gamma(x,y)$ are given by
(\ref{eq:abc}).
$C$ is defined to have the vacuum result subtracted, so write
\begin {equation}
   C = D - \lim_{\hat q\to 0} D .
\label {eq:summaryC}
\end {equation}
$D$, defined as the integrand for $xy\bar y\bar x$, is given
by (\ref{eq:IGamma}) as
\begin {align}
   D(x_1,&x_2,x_3,x_4,\alpha,\beta,\gamma,\Delta t) =
\nonumber\\ &
   \frac{\CA^2 \alphas^2 M_\ix M_\fx}{32\pi^4 E^2} \, 
   ({-}x_1 x_2 x_3 x_4)
   \Omega_+\Omega_- \csc(\Omega_+\Delta t) \csc(\Omega_-\Delta t)
\nonumber\\ &\times
   \Bigl\{
     (\beta Y_\bx Y_\Ax + \alpha \Ybar_{\bx\Ax} Y_{\bx\Ax}) I_0
     + (\alpha+\beta+2\gamma) Z_{\bx\Ax} I_1
\nonumber\\ &\quad
     + \bigl[
         (\alpha+\gamma) Y_\bx Y_\Ax
         + (\beta+\gamma) \Ybar_{\bx\Ax} Y_{\bx\Ax}
        \bigr] I_2
     - (\alpha+\beta+\gamma)
       (\Ybar_{\bx\Ax} Y_\Ax I_3 + Y_\bx Y_{\bx\Ax} I_4)
   \Bigl\}
\label {eq:summaryD}
\end {align}
with the $I_n$ defined by (\ref{eq:I});
the $(X,Y,Z)$'s defined by (\ref{eq:XYZdef});
the 4-particle normal modes and frequencies used in
their definition given by
(\ref{eq:Omegapm}--\ref{eq:Npm}), (\ref{eq:af}), (\ref{eq:ai}),
and (\ref{eq:uOmega});
and
the 3-particle $M$'s and $\Omega$'s in these formulas
defined specifically in terms
of the 4-particle $x_i$ (the arguments of $D$ above) as
\begin {equation}
   M_\ix = x_1 x_4 (x_1{+}x_4) E ,
   \qquad
   M_\fx = x_3 x_4 (x_3{+}x_4) E
\end {equation}
and
\begin {equation}
   \Omega_\ix
   = \sqrt{ 
     -\frac{i \hat q_{\rm A}}{2E}
     \left( \frac{1}{x_1} 
            + \frac{1}{x_4} - \frac{1}{x_1{+}x_4} \right)
   } ,
   \qquad
   \Omega_\fx
   = \sqrt{ 
     -\frac{i \hat q_{\rm A}}{2E}
     \left( \frac{1}{x_3} + \frac{1}{x_4}
            - \frac{1}{x_3 + x_4}
     \right)
   } .
\end {equation}
The pole contribution in (\ref{eq:summaryA}) is given by
(\ref{eq:Apole}).

The $\hat q\to 0$ limit for the vacuum piece in (\ref{eq:summaryC})
corresponds to taking all $\Omega$'s to zero.
The vacuum piece in (\ref{eq:summaryC}) could be worked out
algebraically, but it is simpler to just use the same numerical
code as for (\ref{eq:summaryD}) but with the replacements
\begin {equation}
   \Omega_\ix \to 0,
   \qquad
   \Omega_\fx \to 0,
   \qquad
   \uOmega \cot(\uOmega\,\Delta t) \to (\Delta t)^{-1} ,
   \qquad
   \uOmega \csc(\uOmega\,\Delta t) \to (\Delta t)^{-1}
\end {equation}
[and so also $\Omega_\pm \csc(\Omega_\pm \Delta t) \to (\Delta t)^{-1}$
in the prefactor of (\ref{eq:summaryD})].

  It is possible to scale out the factors of $\hat q_{\rm A}$ and
$E$ from the above expressions by replacing
$\Delta t$ by the dimensionless variable
$\Delta\mathsf{t} \equiv (\hat q_{\rm A}/E)^{1/2} \Delta t$.
For numerics, it is convenient
to work in units where $\hat q_{\rm A}{=}1$ and $E{=}1$, which then gives
the result for the rate
$d\Gamma/dx\,dy$ in units of $(\hat q_{\rm A}/E)^{1/2}$.

% =========================================================================

\section {Behavior of Result}
\label {sec:behavior}

Though this paper calculates only a subset of the diagrams contributing
to $d\Gamma/dx\,dy$, it is interesting to examine the behavior of the
results in the soft bremsstrahlung limit.

% -------------------------------------------------------------------------

\subsection {Comparison to earlier results for \boldmath$y \ll x \ll 1$}
\label {sec:logs}

As mentioned before, refs.\ \cite{Blaizot,Iancu,Wu} have
studied the effect of double splitting on energy loss when one
of two soft bremsstrahlung gluons becomes extremely soft compared
to the other ($y \ll x \ll 1$).  They find that the double-splitting
correction to energy loss is enhanced by a double logarithm that,
at leading logarithmic order, can be absorbed into the
single splitting result (\ref{eq:BDMSggg}) for energy loss by a
redefinition of
$\hat q$.  This energy-dependent
correction $\delta\hat q$
to $\hat q$ is seemingly universal and was computed yet
earlier by Liou, Mueller, and Wu \cite{Wu0}, with result
\begin {equation}
   \delta \hat q_{\rm A} \simeq
   \frac{\alphas \CA}{2\pi} \, \hat q_{\rm A}
   \ln^2 \!\left( \frac{L}{l_0} \right) .
\label {eq:dqhat}
\end {equation}
In the current context of a thick medium, the relevant distance
scale $L$
is the formation time associated with emission of the $x$ gluon:
\begin {equation}
   L \sim \sqrt{\frac{xE}{\hat q_{\rm A}}} \,.
\label {eq:Ldef}
\end {equation}
[The fact that this is a parametric statement ($\sim$) rather than an
exact equality does not matter at the level of identifying the leading
logarithmic result.]  The $l_0$ above characterizes the small distance
scale, characteristic of the medium,
at which the multiple scattering (harmonic oscillator) approximation
breaks down.  A discussion of resumming the leading logarithmic terms
at all orders may be found in refs.\ \cite{Wu0,Blaizot,Wu,Iancu2}.
The results are very interesting because they can change the power
law dependence of how energy loss depends on energy.
For instance, the stopping distance of a high-energy parton
due to repeated single-splitting processes scales with energy
as $E^{1/2}$, but Blaizot and Mehtar-Tani \cite{Blaizot} have discussed
how resumming the double logarithm above can modify the scaling
to $E^{\frac12 - \# \sqrt{\alphas}}$.

These earlier results provide a useful check on our own results:
Do they agree if we take the $y \ll x \ll 1$ limit?  We will focus
on a comparison with Blaizot and Mehtar-Tani \cite{Blaizot} and
Wu \cite{Wu}.  Comparison is
slightly complicated by the fact that in this paper we have only
calculated certain contributions to $d\Gamma/dx\,dy$ but have not
calculated virtual corrections like fig.\ \ref{fig:virtual} that
are needed for the energy loss calculation.  On the flip side,
the processes studied by refs.\ \cite{Blaizot,Wu} do not
include all possible time orderings of the splittings
(presumably because only certain orderings are relevant to
the double logarithm).  So, we will need to compare a subset of
our diagrams to a subset of theirs.
In our terminology, this subset is
$x y \bar y \bar x + x \bar y y \bar x + z y \bar y \bar z + z \bar y y \bar z$
(plus their conjugates).
For reasons explained in Appendix \ref{app:xyyxWu}, the
$x y \bar y \bar x + x \bar y y \bar x$ contributions are difficult to
disentangle from the virtual processes also included in the analysis
of refs.\ \cite{Blaizot,Wu}, and so we shall content ourselves with
checking the contribution from
$z y \bar y \bar z + z \bar y y \bar z$, which are
just the $x y \bar y \bar x$ and $x \bar y y \bar x$
diagrams analyzed earlier with the substitution
$x \to z \equiv 1{-}x{-}y$.
An alternative depiction of $z y\bar y \bar z$ was given earlier
in fig.\ \ref{fig:zyyz}.
(In Wu \cite{Wu}, our $z y \bar y \bar z$ and $z \bar y y \bar z$
are respectively the second diagram of his (30) and the
first diagram of his (32).%
\footnote{%
\label{foot:Wuref}%
   Here and throughout, we refer to Wu's equations \cite{Wu}
   by their numbering in
   the arXiv version of the paper (versions 1 or 2) rather than the
   JHEP version of the paper, in part to make it
   a little easier to distinguish references to our equations from
   references to his.
}
In Blaizot and Mehtar-Tani \cite{Blaizot},
they are called $A_3$ and $C_1$.)

The work of Wu \cite{Wu} is closest in technique to our own.
In appendix \ref{app:dbllog}, we briefly sketch the relationship
between Wu's expressions and the
$y \lesssim x \ll 1$ limit of our formal expressions.
Wu then goes on to extract the double log behavior
(getting the same result as others)
to which we now turn.

The contributions from $z y \bar y \bar z + z \bar y y \bar z$ that
give rise to double log corrections turn out to be
\begin {equation}
   \sqrt{\frac{y\vphantom{E}}{x\vphantom{\hat q}}} \> \sqrt{\frac{yE}{\hat q}}
   \ll \Delta t \ll
   \sqrt{\frac{yE}{\hat q}} ,
\label {eq:region}
\end {equation}
which is equivalent to the regions found to give the double
logarithms in refs.\ \cite{Blaizot,Wu} (see appendix \ref{app:dbllog}
for details).
When we expand our own final results (from section \ref{sec:summary})
in this range, also taking
$y \ll x \ll 1$, we find
\begin {equation}
   2\Re \left[\frac{d\Gamma}{dx\,dy}\right]_{
     z y \bar y\bar z + z \bar y y \bar z
   }
  =\left[
     (2+0+0) \,
      \frac{1 + \ln\left( \frac{y}{\sqrt{2x} \, \Delta t} \right)}
           {\pi^2 x y (\Delta t)^2}
     + \frac{(12-3+1)}{8\pi^2 y x^{3/2} \Delta t}
     + \cdots
   \right]
   \CA^2\alphas^2 \sqrt{\frac{\hat q_{\rm A}}{E}}
   .
\label {eq:dG1}
\end {equation}
Here, we have distinguished which contributions
arise from interaction of the $y$ gluon with the medium, using notation
that we will explain in a moment.
Specifically, the evolution of the $y$ gluon for
$z y \bar y \bar z + z \bar y y \bar z$ is only relevant during
the 4-particle part of the evolution, during which non-zero
$\hat q$ only enters the calculation via non-zero values of
$\Omega_\pm$.
In the $y \ll x \ll 1$ limit, the relevant values of $\Omega_\pm$
(\ref{eq:Omegapm}) approach
\begin {equation}
   \Omega_+^2 \to \Omega_y^2 \equiv \frac{-i\hat q_{\rm A}}{2 y E} ,
   \qquad
   \Omega_-^2 \to \Omega_x^2 \equiv \frac{-3i\hat q_{\rm A}}{8 x E}
\end {equation}
for $z y \bar y \bar z$ and
\begin {equation}
   \Omega_+^2 \to \Omega_x^2 ,
   \qquad
   \Omega_-^2 \to - \Omega_y^2
\end {equation}
for $z \bar y y \bar z$.
The ``$(12-3+1)$'' in (\ref{eq:dG1}) means that the contributions that
do not depend at all on $\Omega_\pm$ (and so correspond to no
interactions of the $y$ gauge boson with the medium) give ``$12$'' there;
contributions that are proportional to $|\Omega_x|^2$ give $-3$;
and contributions that are proportional to $|\Omega_y|^2$ give $+1$.
Similarly for the ``(2+0+0).''

The $1/(\Delta t)^2$ and $1/(\Delta t)$ pieces of (\ref{eq:dG1}) have
nothing to do with the small-$\Delta t$ divergences discussed
back in sections \ref{sec:smalldt1} and \ref{sec:smalldt}.  Those
divergences and their cancellations
arise in the limit of arbitrarily small $\Delta t$
rather than
the $\Delta t$ range (\ref{eq:region}) being considered here.
(Specifically, the small-$\Delta t$ cancellations in
$[z y \bar y \bar z + z \bar y y \bar z + z \bar y \bar z y]
  + [z \leftrightarrow y]^*$
require  $\Delta t \ll \sqrt{y/x} \sqrt{yE/\hat q}$.)

The double log effects which are found in the energy loss calculations of
refs.\ \cite{Blaizot,Wu} arise from single interactions of the
soft $y$ gluons with the medium, and so, when comparing
similar diagrams, correspond to just the
pieces
\begin {equation}
    \frac{(0-3+1)}{8\pi^2 y x^{3/2} \Delta t} \,
    \CA^2\alphas^2
    \sqrt{\frac{\hat q_{\rm A}}{E}}
\end {equation}
of (\ref{eq:dG1}).
Now integrate over $y$ and $\Delta t$,
and compare to the
single splitting result (\ref{eq:BDMSggg}) for $d\Gamma/dx$
in the limit $x \ll 1$, which is
\begin {equation}
   \frac{d\Gamma}{dx}
   \simeq
   \frac{\CA \alphas}{\pi x^{3/2}} \,
   \sqrt{\frac{\hat q_{\rm A}}{E}} \,.
\label {eq:BDMSsoft}
\end {equation}
The result is that the double-splitting contribution considered above
can be absorbed (up to higher order corrections)
into the single splitting result (\ref{eq:BDMSsoft})
by $\hat q_{\rm A} \to \hat q_{\rm A} + \delta \hat q_{\rm A}$ with
(see Appendix \ref{app:details})
\begin {equation}
   \delta \hat q_{\rm A} \simeq
   \frac{({-}3{+}1)}{4} \,
   \frac{\alphas \CA}{\pi} \, \hat q_{\rm A}
   \int \frac{dy \> d(\Delta t)}{y \> \Delta t}
   \simeq
   \frac{({-}3{+}1)}{4} \,
   \frac{\alphas \CA}{2\pi} \, \hat q_{\rm A}
   \ln^2 \!\left( \frac{L}{l_0} \right)
\label {eq:dqhat99}
\end {equation}
[where now we show only the contributions arising from $\Omega_\x$ and
$\Omega_y$ in the notation ``$({-}3{+}1)$''].
Even after adding in $xy\bar y\bar x + x\bar y y \bar x$
(and conjugate) contributions,
this will not exactly match (\ref{eq:dqhat}) because in this paper we
have not included the virtual corrections relevant to energy loss.
However, we have checked that
the contribution to $\delta \hat q_{\rm A}$ from just the
$z y \bar y\bar z + z \bar y y\bar z$ (and conjugate) diagrams in
refs.\ \cite{Blaizot,Wu} would have led to the same result for
the double log pieces, i.e.
\begin{equation}
    \left[ \frac{\delta \hat q_A}{\hat q_A} \right]_{
         \substack{z y \bar y \bar z + z \bar y y \bar z \\
                   + \rm conjugates}
    }
    = \frac{({-}3{+}1)}{4} \, \frac{\delta \hat q_A}{\hat q_A}
\end {equation}
with $\delta\hat q_A/\hat q_A$ on the right-hand side
representing
the total
$\delta\hat q_A/\hat q_A \simeq (\alphas/2\pi) \ln^2(L/l_0)$
of (\ref{eq:dqhat}).

% -------------------------------------------------------------------------

\subsection {Small \boldmath$y$ behavior of full result}

\begin {figure}[t]
\begin {center}
  \includegraphics[scale=0.5]{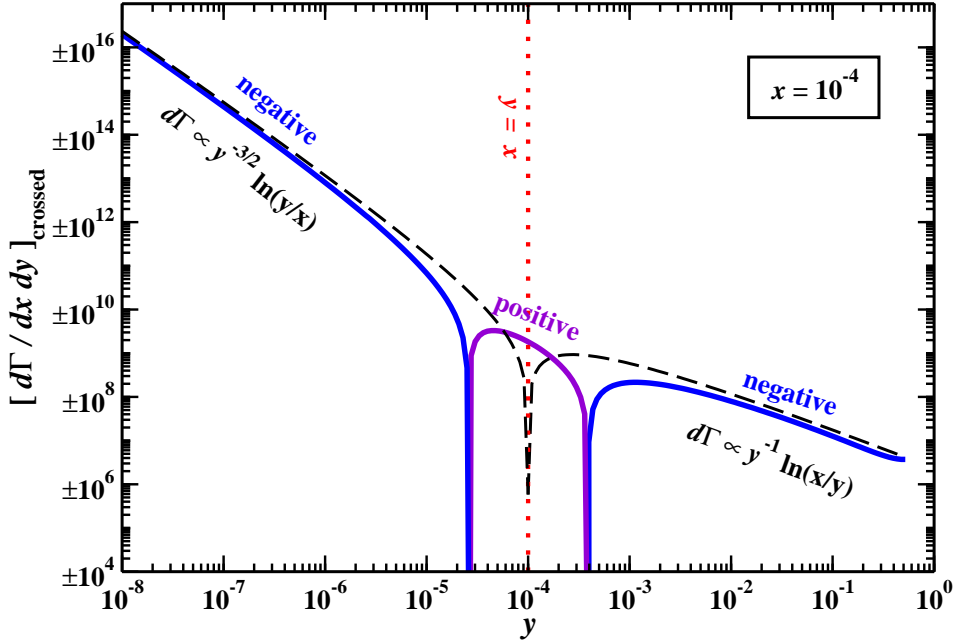}
  \caption{
     \label{fig:xyscale}
     Our numerical results (solid lines) for the
     total crossed diagram contribution
     to $d\Gamma/dx\,dy$ [in units of
     $\CA^2\alphas^2\sqrt{\hat q_{\rm A}/E}\,$] vs.\
     $y$ for fixed $x = 10^{-4}$.
     The solid violet and blue curves depict positive and negative
     numerical values, respectively.
     The $d\Gamma \propto y^{-3/2} \ln(y/x)$
     dashed line shows the $y^{-3/2} \ln(y/x)$ behavior of the
     $y \ll x \ll 1$ power law quoted in
     (\ref{eq:dGsoft}).
     The $d\Gamma \propto y^{-1} \ln(x/y)$ dashed line shows the
     $x^{-1} \ln(x/y)$
     behavior of the same power law if one switches the
     labels $x$ and $y$.
     We have only shown results for $y \le 0.5$; results for
     $y \ge 0.5$ are given by the permutation symmetry
     $y \leftrightarrow z \equiv 1{-}x{-}y$ of the problem.
     [{\it Note added.} This plot does {\it not} reflect the
     correction to the pole terms found in ref.\ \cite{dimreg},
     which change the result to be everywhere negative.]
  }
\end {center}
\end {figure}

Our calculation contains other diagrams and effects
besides the ones considered
in the double log works of \cite{Blaizot,Wu}.  Numerically taking the
$y \ll x \ll 1$ limit of our full result for the crossed diagram
contributions to $d\Gamma/dx\,dy$, we find that the result behaves
as%
\footnote{
  See Appendix \ref{app:details} for a discussion of how the pole terms
  (\ref{eq:Apole}) make a partial but incomplete cancellation of
  the leading $y^{-3/2}$ behavior of (\ref{eq:dGsoft}).
}
\begin {equation}
   \left[\frac{d\Gamma}{dx\,dy}\right]_{\rm crossed}
   \propto \frac{\ln(y/x)}{x y^{3/2}} \,,
\label {eq:dGsoft}
\end {equation}
as shown, for example, by the numerical results of fig.\ \ref{fig:xyscale}.%
\footnote{
  As an additional
  numerical check for anyone who ever desires to implement our formulas of
  section \ref{sec:summary}, our value of
  $[d\Gamma/dx\,dy]_{\rm crossed}$ is
  $11.463126196457 \> \CA^2 \alphas^2 \sqrt{\hat q_{\rm A}/E}$
  for $(x,y)=(0.3,0.6)$.
  [{\it Note added.} The correction to the pole terms found in
  ref.\ \cite{dimreg} changes this to
  ${-}10.892657927744 \> \CA^2 \alphas^2 \sqrt{\hat q_{\rm A}/E}$.]
}
The crossed contribution to $d{\Gamma}/dx\,dy$ is mildly
divergent in the infrared.  If we cut off $x$ and $y$ at some lower
value $x_{\rm min}$, (\ref{eq:dGsoft}) implies that the
crossed contribution to the total
rate $\Gamma$ for double splitting into non-virtual daughters scales
as $x_{\rm min}^{-1/2} \ln^2 x_{\rm min}$.
For comparison, the total rate for single splitting
diverges as a relatively mild $x_{\rm min}^{-1/2}$.

Keep in mind that we have only computed a subset of diagrams
(fig.\ \ref{fig:subset2}).
Our crossed interference diagrams do not sum to be the square of something
and so, unlike total $d\Gamma/dx\,dy$, the result for
$[d\Gamma/dx\,dy]_{\rm crossed}$ need not be positive.
Also, with regard to applications to
energy loss, remember that we have not included virtual
corrections to single bremsstrahlung.

% =========================================================================

\section {Conclusion}
\label {sec:conclusion}

%% [[remove below for journal version]]
This paper is long.  Congratulations if you read (or even skimmed)
this far! \medskip

A complete calculation of $d\Gamma/dx\,dy$ outside of the $x \ll 1$
and/or double log approximation will require a little further work.
But the analysis in this paper of all of the crossed diagrams
provides most of the techniques that will be necessary for the full
analysis, and we look forward to filling in the rest.

The reason that we have saved un-crossed diagrams such as
fig.\ \ref{fig:uncrossed} for later work is that their
$1/\Delta t$ divergences do not cancel as simply as
fig.\ \ref{fig:cancel}.
One must instead take great care in separating
the overlapping formation time corrections for double bremsstrahlung
from the naive calculation of two independent, consecutive
single-bremsstrahlung processes.  But that is a story for another day.

% =========================================================================
% =========================================================================

\begin{acknowledgments}

This work was supported, in part, by the U.S. Department
of Energy under Grant No.~DE-SC0007984.
We thank Han-Chih Chang, Gabriel Wong, and Bin Wu for useful discussions.

\end{acknowledgments}

% =========================================================================
\appendix
% =========================================================================

\section{More details on some formulas}
\label{app:details}

\paragraph*{Eq.\ (\ref{eq:1brem1}) and (\ref{eq:1brem2}):}
We use non-relativistic normalization convention for $|\p\rangle$, so
that $\langle \p' | \p \rangle = (2\pi)^2 \delta^{(2)}(\p-\p')$ and
$\int_\p \equiv \int d^2p/(2\pi)^2$.
The factor of $V_\perp^{-1}$ is just the square of the
normalization factor associated with the initial state $|\p_\ix\rangle$,
since $\langle \p_\ix|\p_\ix\rangle = V_\perp$ in our normalization.
[In these formulas, we have been a little sloppy about factors of
$V_\perp$ associated with the normalization of the
photon state because we wanted to emphasize that $\k_\fx = \k_\xx$.
If concerned about the details of all factors of $V_\perp$, refer
to the more general formula (\ref{eq:1brem2qcd}) instead.  If you want to
see how engineering dimensions match up, then also see
(\ref{eq:dHamp}--\ref{eq:bcalP1})
and the discussion of them in this appendix.]
The factor of $E/2\pi$ in (\ref{eq:1brem1})
comes from rewriting the longitudinal phase space factor
$dk_z/2\pi$ in
\begin {equation}
  d\Gamma = \frac{d^3k}{(2\pi)^3} \, \cdots
\end {equation}
as $dk_z/2\pi = (E/2\pi)\, dx$ and dividing both sides by $dx$.
Evolution of the final state $|\p_\fx\k_\fx\rangle$ after $t_\xbx$
can be ignored by unitarity of time evolution (before taking
the statistical average $\dlangle \cdots \drangle$)
and the completeness relation
$\int_{\p_\fx,\k_\fx}|\p_\fx \k_\fx\rangle \langle\p_\fx \k_\fx| = \openone$
within the two-particle sector.  See section \ref{sec:N5} for
a more detailed argument in a different but closely related context.
(The same hidden assumption discussed in footnote \ref{foot:scales2}
applies to the argument here.)
A similar argument can be made for time evolution of the initial state
before $t_\xx$ by invoking
rotational invariance.  Rotational invariance of the rate
we are computing means that we can
average the initial $\p_\ix$ over small rotations, which in the
context of our large-$p_z$ approximation is discussed in detail
in appendix \ref{app:cancel}, with the result that one may
replace $|\p_\ix\rangle \langle \p_\ix |$ by something proportional
to $\int_{\p_\ix} |\p_\ix\rangle \langle \p_\ix |$.  The same
unitarity and completeness arguments used for the final state
for $t > t_\xbx$
then imply that we also
need not follow the evolution of the initial state for $t < t_\xx$ in
(\ref{eq:1brem1}) and (\ref{eq:1brem2}).

% .......................................................................

\medskip
\paragraph*{Eq.\ (\ref{eq:VqcdHO}):}
Here is a more general argument for (\ref{eq:VqcdHO}).
In general, any translation-invariant harmonic approximation for
3 particles can be put into the form
\begin {equation}
   V(\b_1,\b_2,\b_3) =
   A_{21} (\b_2-\b_1)^2 + A_{32} (\b_3-\b_2)^2 + A_{13} (\b_1-\b_3)^2
\label {eq:VHOgeneral}
\end {equation}
with some coefficients $A_{ij}$.
In the special case of $\b_1=\b_2$, this gives
\begin {equation}
   V(\b_1,\b_1,\b_3) =
   (A_{32} + A_{13}) (\b_3-\b_1)^2.
\label {eq:Vconstraint1}
\end {equation}
However, in this case we have
color charge $T_3$ at $\b_3$ and total color charge
$T_1+T_2=-T_3$ at $\b_1=\b_2$.  This is effectively a two-particle
problem with separation $\b = \b_3-\b_1$, and the small-$\b$ limit
of that problem is precisely
what is used in proposals
to non-perturbatively define (using real-time Wilson
loops \cite{WilsonLoops}) the parameter $\hat q$ for color representation
$T_3$, which means that
\begin {equation}
  V(\b_1,\b_1,\b_3) = - \tfrac{i}4 \hat q_3 (\b_3-\b_1)^2 .
\end {equation}
Combined with (\ref{eq:Vconstraint1}), this gives a constraint on
the values of the coefficients $A_{ij}$.  Permuting the particle
labels on this argument gives three constraints on the three unknown
coefficients $A_{ij}$, which determines (\ref{eq:VHOgeneral}) to be
(\ref{eq:VqcdHO}).

% .......................................................................

\medskip
\paragraph*{Eq.\ (\ref{eq:qhat}):}
In weak coupling, $\hat q$ defined by (\ref{eq:qhat}) is UV logarithmically
divergent, $\propto \alphas^2 n \int d^2q_\perp/q_\perp^2$
where $n$ is the density
of partons in the medium.  In the harmonic approximation, $\hat q$ then
has a mild logarithmic dependence on $\b$, and the harmonic approximation
is effectively a leading-log approximation (to which sub-leading-log
corrections may be systematically computed%
\footnote{
  See, for example, refs.\ \cite{ArnoldDogan,stop}.
}%
).  However, if the
coupling $\alphas$ is replaced by running $\alphas(q_\perp)$,
$\hat q$ defined by (\ref{eq:qhat}) is convergent.  This form of
$\hat q$ is only relevant if the total momentum transfer
$Q_\perp$ during a formation time is large enough
that the integral (\ref{eq:qhat}) is not significantly modified
if restricted to $q_\perp \lesssim Q_\perp$.  This happens when
$\alphas(Q_\perp) \ll \alphas(\mD)$, where $\mD$ is the inverse
screening distance in the medium.  For thick media,
$Q_\perp \sim (\hat q E)^{1/4}$ for hard bremsstrahlung, and note that
our assumption throughout this paper is that $\alphas(Q_\perp)$ can be
treated as small.  Finally, $\hat q L$ describes the average
$Q_\perp^2$ picked up from the medium over a distance $L$ and
characterizes a Gaussian probability distribution for $Q_\perp$
after a large number of collisions.  However,
power-law tails of the probability distribution for $Q_\perp$,
representing rare collisions, can dominate certain types of
calculations.%
\footnote{
  See, for example, the discussion of power law tails in
  sec. 3.1 of BDMPS \cite{BDMPS3}; Zakharov \cite{ZakharovTail};
  Arnold \cite{HO}; appendix A of Peigne and Smilga \cite{PeigneSmilga};
  and D'Eramo et al.\ \cite{KrishnaTail}.
}

% .......................................................................

\medskip
\paragraph*{Eq.\ (\ref{eq:IIxyyx1}):}
In this formula, we make use of $\sum_i x_i = 0$ to occasionally
rewrite, for example, ${-}\hat x_3{-}\hat x_4$ as $\hat x_1 {+} \hat x_2$.
Other than that, the expression follows from the rules of
fig.\ \ref{fig:dH} together with the overall factor of
$(E/2\pi)^2$ of (\ref{eq:xyyx2}), which comes from
rewriting $dk_z/2\pi = (E/2\pi)\,dx$ and $d\kappa_z/2\pi = (E/2\pi)\,dy$,
similar to the single factor of $E/2\pi$ in (\ref{eq:1brem1}).
The one thing to be careful about in applying the rules
is the signs of the ${\cal B}_{ij}$.
For instance, consider the rule associated with the vertex at
$t=t_\xx$ in fig.\ \ref{fig:xyyx}.  Using the top rule
in fig.\ \ref{fig:dH}, the vertex is associated with a factor
of $\grad\delta^{(2)}(\bcalB_{ji})$ according to the labeling of
that rule, but we may just as well write
$\grad\delta^{(2)}(\bcalB_{ik})$ using the identity
$\bcalB_{ji} = \bcalB_{kj} = \bcalB_{ik}$ noted in the caption.
Comparing fig.\ \ref{fig:dH} to fig.\ \ref{fig:xyyx},
this $\grad\delta^{(2)}(\bcalB_{ik})$ translates to
$\grad\delta^{(2)}(\B_{14})$ in the index labeling used in the
latter figure.  In (\ref{eq:IIxyyx1}), however, we have used
$\C_{41}^\yx$ at the start $t = t_\yx$ of the 4-particle evolution.
As a result of this choice, it
was then convenient to define our $\B$ for the initial
3-particle evolution $t_\xx < t < t_\yx$ as $\B \equiv \B_{41}$
instead of $\B \equiv \B_{14}$.  So our factor of
$\grad\delta^{(2)}(\B_{14})$ at $t=t_\xx$ is $-\grad\delta^{(2)}(\B^\xx)$.
In contrast, the $\B$ defined for the final 3-particle evolution
$t_\ybx < t < t_\xbx$ is $\B \equiv \B_{34}$, which is the same
$\B$ that comes from applying the top rule of fig.\ \ref{fig:dH}
to the splitting at $t=t_\xbx$.  So the corresponding factor
for that splitting is $+\grad\delta^{(2)}(\B^\xbx)$.
In a similar vein, the rules of fig.\ \ref{fig:dH} give factors of
$\grad\delta^{(2)}(\C_{21}^\ybx) = - \grad\delta^{(2)}(\C_{12}^\ybx)$ at $t=t_\ybx$
and
$\grad\delta^{(2)}(\C_{23}^\ybx)$ at $t=t_\yx$.
We then integrate all the $\delta$-functions (integrating by parts
as necessary) to get (\ref{eq:IIxyyx1}).

% .......................................................................

\medskip
\paragraph*{Eqs.\ (\ref{eq:dHamp}--\ref{eq:bcalP1}):}
We offer here a quick note on the scaling dimensions of our formulas.
Since we use non-relativistic normalization
$\langle \p | \p' \rangle = (2\pi)^2 \delta^{(2)}(\p-\p')$,
then $|\p_\ix\rangle$ has dimensions {\sf mass}$^{-1}$ and
$\langle\p_j,\p_k|$ has dimension {\sf mass}$^{-2}$, and so one
would expect $\langle\p_j,\p_k|\delta H|\p_i\rangle$
to have dimension {\sf mass}$^{-2}$.  The right-hand side of
(\ref{eq:dHamp}) has an implicit $(2\pi)^2 \delta^{(2)}(\p_i-\p_j-\p_k)$.
Including this factor gives the right-hand side dimensions of
{\sf mass}$^{-5/2}$.  The discrepancy has to do with the normalization of
the states of the longitudinal momentum.  Here are two ways to handle
the normalization.  Method 1: If we normalize states so that
$\langle \p,x|\p',x'\rangle = (2\pi)^2 \delta^{(2)}(\p-\p') \, \delta_{x,x'}$,
with a Kronecker $\delta$ for the $x$'s, then there is really a
factor of $V_z^{-1/2}$ on the right-hand side of (\ref{eq:dHamp}),
where $V_z$ is the (infinite) length of the $z$-direction.
(This $V_z$ should not be confused with the finite
physical width of the medium in cases where the medium is finite.)
This fixes up the dimensions of (\ref{eq:dHamp}).
With this normalization convention for states of longitudinal momentum,
the sum over the longitudinal momentum of the bremsstrahlung particle
in single splitting, for example, becomes $\int V_z \, dp_z/2\pi$
instead of $\int dp_z/2\pi$, and so the single-splitting formula
(\ref{eq:1brem1}) should have included an overall factor of
$V_z$.  This $V_z$ then cancels the two factors of $V_z^{-1/2}$
that we get from the two $\delta H$ matrix elements.
Similar cancellations occur in the double-splitting case.
We have not bothered to include any of these canceling factors of
$V_z$ in our formulas.
Method 2: Alternatively, one could normalize states as
$\langle \p,x|\p',x'\rangle
 = (2\pi)^3 \delta^{(2)}(\p-\p') \, \delta(p_z-p'_z)$
and avoid ever introducing $V_z$.  This normalization would change
the dimensional analysis to make the two sides of (\ref{eq:dHamp})
agree dimensionally, without any additional factor of $V_z^{-1/2}$,
if we take there to be an implicit
$(2\pi)^3\delta^{(2)}(\p_i{-}\p_j{-}\p_k) \, \delta(p_{iz}{-}p_{jz}{-}p_{kz})$
on the right-hand side.

% .......................................................................

\medskip
\paragraph*{Eq.\ (\ref{eq:3ints}):}
If $\Omega$ has a negative imaginary part, we get
$\cot(\Omega\infty) = i$ and (\ref{eq:3int0}).  If $\Omega$ had a
positive imaginary part, we would instead get $\cot(\Omega\infty) = -i$
and so (\ref{eq:3int0}) with $\exp(-\frac12\,M\Omega B^2)$ replaced
by $\exp(+\frac12\,M\Omega B^2)$.  We could therefore write the exponential
in the general case as $\exp\bigl(\frac12\,M\Omega B^2\sgn(\Im\Omega)\bigr)$.
However, there is a correlation between the sign of $M$ and the sign of
$\Im\Omega$ that makes for a more compact expression.
First note that if the 3-particle effective
Hamiltonian ${\cal H}$
of (\ref{eq:calH3}) makes any sense for our problem, then
it should have a negative imaginary part, corresponding to
exponential decay rather than exponential growth.  This means
it should have $M \Im \Omega_0^2 < 0$, and so,
also using (\ref{eq:Omega0B}),
\begin {equation}
   \sgn \Im\Omega = \sgn \Im\Omega^2 = - \sgn M .
\end {equation}
This allows us to
rewrite our general-case exponential above in the form of
(\ref{eq:3intsA}), as promised.

Readers may wonder whether the condition that $M \Im \Omega_0^2 < 0$
could ever be violated.
Using $x_1+x_2+x_3=0$ for 3-particle evolution,
(\ref{eq:M3}),
and (\ref{eq:Omega0B}),
\begin {equation}
  M \Im\Omega^2
    = -\tfrac14 \bigl(
         (\hat q_2{+}\hat q_3{-}\hat q_1)x_1^2
         + (\hat q_3{+}\hat q_1{-}\hat q_2)x_2^2
         + (\hat q_1{+}\hat q_2{-}\hat q_3)x_3^2
     \bigr)
    .
\label {eq:MOm2}
\end{equation}
Requiring the right-hand side to be negative for all allowed values of
$x_i$ satisfying $x_1+x_2+x_3=0$ is equivalent to requiring that the
largest value of $\sqrt{\hat q_i}$ be smaller than the sum of the
two others.%
\footnote{
  One (perhaps inelegant) way
  to derive this condition is to plug $x_3 = -(x_1+x_2)$ into
  (\ref{eq:MOm2}) and then rewrite the condition in terms of
  the ratio $r \equiv x_1/x_2$, which gives
  $\hat q_2 r^2 + (\hat q_1{+}\hat q_2{-}\hat q_3)r+\hat q_1 > 0$,
  which we require for all values of $r$.
  Requiring the minimum with
  respect to $r$ to be positive gives a condition that can
  be written in the form
  $\hat q_1^2 + \hat q_2^2 + \hat q_3^2 <
   2(\hat q_1 \hat q_2 + \hat q_2 \hat q_3 + \hat q_3 \hat q_1)$.
  By investigating what happens if the ``$<$'' is replaced by an
  equal sign and then solving for any one of the $\hat q_i$,
  this condition can then be translated into the
  one quoted in the main text above.
}
So, if $\hat q_3$ is the largest, the condition would be
\begin {equation}
   \sqrt{\hat q_3} < \sqrt{\hat q_1} + \sqrt{\hat q_2} .
\label {eq:qcondition}
\end {equation}
In the weakly-coupled case, this would be
$\sqrt{C_3} < \sqrt{C_1} + \sqrt{C_2}$, which can be rewritten
as
\begin {equation}
   \lVert T_1{+}T_2 \rVert < \lVert T_1 \rVert + \lVert T_2 \rVert ,
\label {eq:Tinequality}
\end {equation}
where $T_i$ are the color generators associated with each
particle, acting on a 3-particle color
singlet.
The inequality (\ref{eq:Tinequality})
follows from the triangle inequality, given
that the action of $T_1$ and $T_2$ on the 3-particle color singlet
are not simply proportional to each other.  [We are not sure how
to argue mathematically that (\ref{eq:qcondition}) must also hold
outside of the weak-coupling limit, but a violation would mean
that there is some instability in the problem.]

% .......................................................................

\medskip
\paragraph*{Eq.\ (\ref{eq:CvsAprop}):}
$\langle\C_{34}^\Ax,\C_{12}^\Ax,\Delta t|\C_{41}^\bx,\C_{23}^\bx,0\rangle$
and
$\langle \A_+^\Ax,\A_-^\Ax,\Delta t | \A_+^\bx,\A_-^\bx,0 \rangle$
are the same except for how the states are implicitly normalized: e.g.\
(\ref{eq:C3412norm}) vs.\
$\langle \A_+,\A_- | \A_+',\A_-' \rangle
 = \delta^{(2)}(\A_+-\A_+') \, \delta^{(2)}(\A_--\A_-')$.
The different $\delta$ functions of $\A_\pm$ and of
$\C_{ij}$ are related by the Jacobian of the corresponding
change of variables (\ref{eq:af}) and (\ref{eq:ai}),
giving the determinant factors in
(\ref{eq:CvsAprop}):
$\langle\C_{34}^\Ax,\C_{12}^\Ax| =
 |\det a_\ybx|^{-1} \langle \A_+^\Ax,\A_-^\Ax|$
and
$|\C_{41}^\bx,\C_{23}^\bx\rangle =
 | \A_+^\bx,\A_-^\bx\rangle |\det a_\yx|^{-1}$.
Note that we get
$|\det a_\ybx|^{-1}$ (and similarly $|\det a_\yx|^{-1}$) on
the right-hand side because
there is a factor of $|\det a_\ybx|^{-1/2}$ associated with each
of the two dimensions of the transverse plane.

% .......................................................................

\medskip
\paragraph*{Eq.\ (\ref{eq:WightmanEps}):}
Consider the short-time ($t \to 0$) divergence of any matrix
element $\langle \omega_1 | A(t) \, B(0) | \omega_2 \rangle$,
where the $|\omega_i\rangle$ are energy eigenstates with
energies $\omega_i$.
Insert a complete set of states to write the matrix element as
$\sum_E \langle \omega_1 | A(t) | E \rangle
  \langle E | B(0) | \omega_2 \rangle$.
The short-time divergence will come from intermediate states with
arbitrarily high $E$, and in particular $E \gg \omega_1, \omega_2$.
Fourier transforming from $t$ to $\omega$ gives $\omega = E - \omega_1$,
and so we see that the short-term divergence is associated with purely
positive frequencies $\omega$.  Now return to the original correlator
$\langle \omega_1 | A(t) \, B(0) | \omega_2 \rangle$ in $t$ space,
and consider the Fourier transform integral
$\int dt \> e^{i\omega t}
 \langle \omega_1 | A(t) \, B(0) | \omega_2 \rangle$
taking it to $\omega$ space.
For $\omega<0$, the contour may be closed in the lower-half
complex $t$ plane.
In order for a short-time divergence of the integrand at $t=0$
such as $t^{-n}$ {\it not} to contribute
when $\omega < 0$, the prescription for that divergence
should then be $(t-i\epsilon)^{-n}$, which
is the prescription of (\ref{eq:WightmanEps}).

% .......................................................................

\medskip
\paragraph*{Eq.\ (\ref{eq:dqhat99}):}
Following ref.\ \cite{Blaizot}, for example, the last equality of
(\ref{eq:dqhat99}) is obtained by doing the $y$ integral first
after recasting the range (\ref{eq:region}) as
\begin {equation}
   (\Delta t)^2 \frac{\hat q}{E} \ll y
   \ll \Delta t \, \sqrt{\frac{x\hat q}{E}} \,.
\end {equation}
The result is
\begin {equation}
   \int \frac{dy \> d(\Delta t)}{y \> \Delta t}
   \simeq
   \int \frac{d(\Delta t)}{\Delta t} \,
      \ln\left( \frac{L}{\Delta t} \right) ,
\end {equation}
with $L$ defined parametrically by (\ref{eq:Ldef}) in our case.
Then take the parametric range of the $\Delta t$ integration to be
$\ell_0$ to $L$.

% .......................................................................

\medskip
\paragraph*{Eq.\ (\ref{eq:dGsoft}):}
In both the cases of $A(x,y)$ and $A(1{-}x{-}y,y)$ in (\ref{eq:summary1}),
there are some interesting cancellations between the pole and
the integral pieces in (\ref{eq:summaryA}).  Individually, both of
these pieces scale with $y$ as
\begin {equation}
   O\bigl( y^{1/2}(\alpha{+}\beta) \bigr) + O( y^{3/2} \gamma )
\label {eq:cancel1}
\end {equation}
for small $y$ (and fixed $x$), up to logarithms.
When the two are added together as
in (\ref{eq:summaryA}), there is a cancellation among the leading
contribution from $\alpha$ and $\beta$ to give
\begin {equation}
   \frac{d\Gamma}{dx\,dy} =
   O(y^{3/2}\alpha) + O(y^{3/2}\beta) + O(y^{3/2} \gamma) .
\label {eq:cancel2}
\end {equation}
There is no similar cancellation for the terms involving $\gamma$.  In the
small $y$ limit, $(\alpha,\beta,\gamma)$ scale with $y$ as
\begin {equation}
   \alpha \sim \beta \sim y^{-2} ,
   \qquad
   \gamma \sim y^{-3} .
\end {equation}
In consequence, both (\ref{eq:cancel1}) and (\ref{eq:cancel2}) scale
as $y^{-3/2}$, and so the cancellation of the leading $\alpha{+}\beta$
term does not completely cancel the leading $y^{-3/2}$ behavior.

% ============================================================================

\section{Details on reduction of states from \boldmath$N$ to
         \boldmath$N{-}2$ particles}
\label{app:project}

% --------------------------------------------------------------------------

\subsection {Normalization of projected states}

We first turn to the normalization factors, such as ${\cal N}$ in
the definition (\ref{eq:project}) of $|\{\B_{ij}\}\rangle$, necessary
to define our projection from $N$ particle states to effective
$N{-}2$ particle states.  One of our constraints is that
$\sum_i\p_i = 0$, and in our discussion we will encounter
corresponding $\p$-space $\delta$-functions evaluated at zero argument:
\begin {equation}
   \delta^{(2)}({\textstyle\sum} \p_i) \Bigr|_{\sum\p_i=0} = \int_{\Delta\b} 1
   = V_\perp ,
\end {equation}
where $V_\perp$ is the 2-volume (area) of the transverse spatial directions.
Final results will not depend on $V_\perp$, which is taken to be
infinite.  Our other constraint is that
$\sum_i x_i \b_i = 0$, and it will be convenient to give a name
$\tildeV_\perp$ to the analogous $\b$-space $\delta$-function evaluated at
zero argument:
\begin {equation}
   \tildeV_\perp \equiv
   \delta^{(2)}({\textstyle\sum} x_i \b_i) \Bigr|_{\sum x_i\b_i = 0} .
\label {eq:Vtilde}
\end {equation}
Final results will not depend on $\tildeV_\perp$, which is also infinite.

The projection ${\cal P}$ of an $N$-particle
state $|\b\rangle$ onto the subspace with $\sum_i\p_i = 0$ and
$\sum_i x_i\b_i = 0$ is
\begin {equation}
   {\cal P} |\b_1,\cdots,\b_N\rangle =
     \delta_{\sum x_i\b_i,0} \,
     V_\perp^{-1}
     \int_{\Delta\b}
     |\b_1 {+} \Delta\b, \cdots,\b_N {+} \Delta\b \rangle .
\label {eq:Pb}
\end {equation}
(Here, $\delta_{\sum x_i\b_0,0}$ is a Kronecker $\delta$ rather than Dirac
$\delta$-function and gives $1$, rather than $\tildeV_\perp$,
when $\sum x_i\b_i = 0$.)
One may verify that ${\cal P}^2|\vec\b\rangle = {\cal P}|\vec\b\rangle$.
Because ${\cal P}|\vec\b\rangle$ above depends only on the differences
$\b_i-\b_j$ of transverse positions, it is a state that can be
characterized as depending only
on the values of $\B_{ij} \equiv (\b_i-\b_j)/(x_i{+}x_j)$.
Except for normalization convention, it is the state
$|\{\B_{ij}\}\rangle$ that we defined in (\ref{eq:project}).

The normalization of ${\cal P}|\b\rangle$ is
\begin {align}
   \langle \b'_1,\cdots,\b'_N |{\cal P}| \b_1,\cdots,\b_N\rangle
   &= V_\perp^{-1} 
   \int_{\Delta\b'} \langle \b'_1,\cdots,\b'_N
               | \b_1 {+} \Delta\b, \cdots, \b_N {+} \Delta\b \rangle
\nonumber\\
   &=
   V_\perp^{-1} \int_{\Delta\b} \,
   \prod_{i=1}^N \delta^{(2)}(\b'_i-\b_i-\Delta\b)
\nonumber\\
   &=
   V_\perp^{-1}
   \prod_{i=2}^N \delta^{(2)}(\b'_{i1}-\b_{i1}) ,
\label {eq:realize1B}
\end {align}
where $\b_{ij} \equiv \b_i - \b_j$ and we have assumed that
$\b_i$ and $\b'_i$ have been chosen with
$\sum_i x_i \b_i = 0 = \sum_i x_i\b'_i$.
These last constraints, together with $\sum_i x_i=0$, imply that
\begin {equation}
   \sum_{i=2}^N x_i (\b'_{i1} - \b_{i1}) = 0 ,
\label {eq:bbzero}
\end {equation}
and so the $N{-}1$ $\delta$-functions left in the final expression in
(\ref{eq:realize1B}) are
not independent.
(\ref{eq:realize1B}) is therefore equivalent to $N{-}2$ independent
$\delta$-functions times a factor of
$\delta^{(2)}(0)$.
In order to relate this factor to the normalization
factors we have defined, we must take care how we
write things.
Ignore (\ref{eq:bbzero}) for a moment and
rewrite the $i{=}2$ $\delta$-function in (\ref{eq:realize1B}) as
\begin {align}
  \delta^{(2)}(\b'_{21}-\b_{21}) &=
  \delta^{(2)}\biggl( x_2^{-1} \biggl[
     \sum_{i=2}^N x_i(\b'_{i1}-\b_{i1})
     - \sum_{i=3}^N x_i(\b'_{i1}-\b_{i1})
  \biggr] \biggr)
\nonumber\\
  &=
  x_2^2 \,
  \delta^{(2)} \biggl(
     \sum_{i=1}^N x_i(\b'_{i}-\b_{i})
     - \sum_{i=3}^N x_i(\b'_{i1}-\b_{i1})
  \biggr) .
\end {align}
In the presence of the other $\delta$-functions, this can be replaced by
\begin {equation}
  \delta^{(2)}(\b'_{21}-\b_{21}) \to
  x_2^2 \,
  \delta^{(2)} \biggl(
     \sum_{i=1}^N x_i(\b'_{i}-\b_{i})
  \biggr) .
\end {equation}
Given the constraint (\ref{eq:bbzero}),
this is
\begin {equation}
  \delta^{(2)}(\b'_{21}-\b_{21}) \to
  x_2^2 \tildeV_\perp .
\end {equation}
Combining with (\ref{eq:realize1B}) gives
\begin {equation}
   \langle \b'_1,\cdots,\b'_N |{\cal P}| \b_1,\cdots,\b_N\rangle
   =
   \frac{x_2^2 \tildeV_\perp}{V_\perp}
   \prod_{i=3}^N
   \delta^{(2)}(\b_{i1}-\b'_{i1})
   =
   \frac{x_2^2 \tildeV_\perp}{V_\perp}
   \prod_{i=3}^N
   (x_i+x_1)^{-2} \delta^{(2)}(\B_{i1}-\B'_{i1}) .
\label {eq:normb}
\end {equation}

In order to get rid of the nuisance factors of $V_\perp$ and
$\tilde V_\perp$, we find it useful to define
\begin {equation}
   | \{\B_{ij}\} \rangle \equiv
   \left( \frac{V_\perp}{\tildeV_\perp} \right)^{1/2}
   {\cal P} |\b_1,\cdots,\b_N\rangle ,
\label{eq:normB}
\end {equation}
so that
\begin {equation}
   \langle \{\B'_{ij}\} | \{\B_{ij}\} \rangle =
   x_2^2
   \prod_{i=3}^N
   \delta^{(2)}(\b_{i1}-\b'_{i1})
   =
   x_2^2
   \prod_{i=3}^N
   (x_i+x_1)^{-2} \delta^{(2)}(\B_{i1}-\B'_{i1}) .
\label {eq:BsBs}
\end {equation}
Combining (\ref{eq:Pb}) and (\ref{eq:normB}) determines the normalization
factor ${\cal N}$ of (\ref{eq:project}) in the main text to be
\begin {equation}
   {\cal N} = (V_\perp \tildeV_\perp)^{-1/2} .
\label {eq:calN}
\end {equation}

Eq.\ (\ref{eq:BsBs}) with $N{=}2$ and $N{=}3$
directly gives (\ref{eq:N2norm}) and (\ref{eq:Bsnorm})
for the normalizations in those cases.
For $N{=}4$, it gives
\begin {equation}
  \bigl\langle\{\B_{ij}\}\big|\{\B_{ij}'\}\bigr\rangle
  = x_2^2 (x_1+x_3)^{-2} (x_1+x_4)^{-2} \,
    \delta^{(2)}(\B_{31}-\B'_{31}) \,
    \delta^{(2)}(\B_{41}-\B'_{41}) .
\end {equation}
One may then use the transformations (\ref{eq:brelations}) to
compute the Jacobian for trading variables $(\B_{31},\B_{41})$ in
the $\delta$-functions for $(\C_{34},\C_{12})$, remembering that we
typically rename the $\B_{ij}$ as $\C_{ij}$ in the $N{=}4$ case.
The result of this change of variables is (\ref{eq:Csnorm}).

% --------------------------------------------------------------------------

\subsection {Cancellation of normalization factors}
\label {app:cancel}

As discussed surrounding eqs.\ (\ref{eq:end1}--\ref{eq:end3}), the final
state in the $N$-particle language of $\bar\Hilbert \otimes \Hilbert$
is
\begin {equation}
   |{\rm end}\rangle = \int_{\b_\fx} |\b_\fx,\b_\fx\rangle
   = {\cal N}^{-1}|\rangle .
\label {eq:end}
\end {equation}

Now consider the initial state in this language.  We have not yet
picked a precise direction for the $z$-axis in our calculation---much
of the formalism has been designed precisely to avoid that.  For
simplicity of presentation
in the present argument, pick it for now to be exactly
in the direction of the initial parton at the initial time.  Then the
system starts as $|\p_\ix{=}0\rangle\langle\p_\ix{=}0|$, which in
the language of $\bar\Hilbert \otimes \Hilbert$ is
\begin {equation}
   |{\rm start}\rangle = V_\perp^{-1} |\p_1{=}0,\p_2{=}0\rangle
   = V_\perp^{-1} \int_{\b_1,\b_2} |\b_1,\b_2\rangle .
\end {equation}
Here, the factor of $V_\perp^{-1}$ is the initial state normalization
factor, just as described for (\ref{eq:1brem1}) in
appendix \ref{app:details}.

Because the projection onto the subspace of $\sum_i \p_i = 0$ and
$\sum_i x_i\b_i = 0$ is preserved by time evolution and splitting,
and because the final state $|{\rm end}\rangle$ lies in this subspace,
no harm is done if we also project the initial state onto this
subspace.  That is,
\begin {equation}
   \langle{\rm end}| \cdots |{\rm start}\rangle
   = \langle{\rm end}| {\cal P} \cdots |{\rm start}\rangle
   = \langle{\rm end}| \cdots {\cal P} |{\rm start}\rangle ,
\end {equation}
and so we may replace
\begin {equation}
   |{\rm start}\rangle \to {\cal P} |{\rm start}\rangle
   = V_\perp^{-1} {\cal P} \int_{\b_1,\b_2} |\b_1,\b_2\rangle .
\end {equation}
Using (\ref{eq:Pb}) and (\ref{eq:Vtilde}), this is
\begin {multline}
   |{\rm start}\rangle \to
   V_\perp^{-2} \int_{\b_1,\b_2,\Delta\b} \delta_{x_1\b_1 + x_2\b_2,0}
   |\b_1 {+} \Delta\b, \b_2 {+} \Delta\b\rangle
\\
   =
   V_\perp^{-2} \tildeV_\perp^{-1} \int_{\b_1,\b_2,\Delta\b} \delta(x_1\b_1 + x_2\b_2)
   \, |\b_1 {+} \Delta\b, \b_2 {+} \Delta\b\rangle .
\end {multline}
For the initial particle, $(x_1,x_2)=(-1,1)$, and the above
evaluates to
\begin {equation}
   |{\rm start}\rangle \to
   (V_\perp \tildeV_\perp)^{-1} \int_{\b_\ix} |\b_\ix,\b_\ix\rangle
   = (V_\perp \tildeV_\perp)^{-1} |{\rm end}\rangle
   = (V_\perp \tildeV_\perp)^{-1} {\cal N}^{-1} |\rangle
   = {\cal N} |\rangle ,
\label {eq:start}
\end {equation}
where the last equality uses (\ref{eq:calN}).
Combining (\ref{eq:end}) and (\ref{eq:start}),
\begin {equation}
   \langle{\rm end}| \cdots |{\rm start}\rangle
   = \langle | \cdots |\rangle .
\end {equation}
All the nuisance factors of $V_\perp$ and $\tildeV_\perp$ have
canceled, as promised.

% --------------------------------------------------------------------------

\subsection {\boldmath$\langle B|\delta H|\rangle$ matrix elements}
\label {app:dH3}

\begin {figure}[t]
\begin {center}
  \includegraphics[scale=0.5]{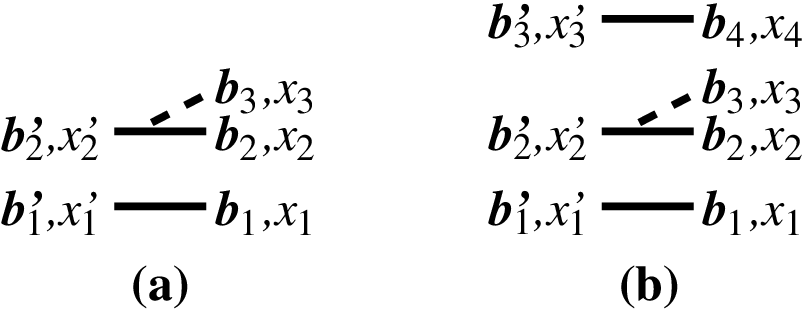}
  \caption{
     \label{fig:dHlabels}
     The notation used in (a) appendix \ref{app:dH3} and
     (b) appendix \ref{app:dH4} to label different particle
     states immediately before and after a splitting.
     (a) is similar to fig.\ \ref{fig:xxrateB}, but it was
     convenient to introduce slightly more general notation here.
     The use of primes ($'$) here is completely unrelated to their
     use in (\ref{eq:xprime}).
  }
\end {center}
\end {figure}

Here we work out the formula for the matrix element
$\langle\B|\delta H|\rangle$.  For simplicity and definiteness,
we will focus on the case where the initial particle in the amplitude
splits into two daughters, using the labeling conventions of
fig.\ \ref{fig:dHlabels}a.  The starting point is the $\delta H$
matrix element in the amplitude, written more conventionally
in terms of the {\it individual}\/ particles in the Hilbert space
$\Hilbert$ (rather than $\bar\Hilbert \otimes \Hilbert$),
given by (\ref{eq:dHamp}).  In $\p$-space, this is
\begin {equation}
   \langle \p_2,\p_3 | \delta H | \p'_2 \rangle
   = g \bcalT_{2' \to 23} \cdot \P_{23}
   = g \bcalT_{2' \to 23} \cdot (x_3\p_2-x_2\p_3) ,
\end {equation}
which in $\b$-space becomes
\begin {equation}
   \langle \b_2,\b_3 | \delta H | \b'_2 \rangle
   = -i g \bcalT_{2' \to 23} \cdot (x_3\grad_{\b_2}-x_2\grad_{\b_3})
   \Bigl[ \delta^{(2)}(\b_2-\b'_2) \, \delta^{(2)}(\b_2-\b_3) \Bigr] .
\label {eq:dHampb}
\end {equation}

Now turn to $\bar\Hilbert \otimes \Hilbert$ notation and our
effective $N-2$ particle description.  Use the definition
(\ref{eq:normB}) and (\ref{eq:Pb}) of the projected states to write
\begin {multline}
   \langle\B | \delta H |\rangle
   =
   \frac{V_\perp}{\tildeV_\perp}
   \langle 1,2,3 | {\cal P} \,\delta H\, {\cal P} | 1',2' \rangle
   =
   \frac{V_\perp}{\tildeV_\perp}
   \langle 1,2,3 | \delta H\, {\cal P} | 1',2' \rangle
\\
   =
   \frac{1}{\tildeV_\perp}
   \int_{\Delta\b}
   \langle
     \b_1, \b_2, \b_3
   | \delta H |
     \b'_1{+}\Delta\b, \b'_2{+}\Delta\b
   \rangle
\end {multline}
(where we have suppressed writing the Kronecker $\delta$-functions---the
initial and final positions are to be understood as both satisfying
the constraint $\sum_i x_i \b_i = 0$).
Using the amplitude matrix element (\ref{eq:dHampb}), this is
\begin {equation}
   \langle\B | \delta H |\rangle
   =
   -\frac{ig}{\tildeV_\perp} \bcalT_{2' \to 23} \cdot (x_3\grad_{\b_2}-x_2\grad_{\b_3})
   \int_{\Delta\b}
   \delta^{(2)}(\b_1{-}\b'_1{-}\Delta\b) \,
   \delta^{(2)}(\b_2{-}\b'_2{-}\Delta\b) \, \delta^{(2)}(\b_2{-}\b_3) .
\label{eq:dHmoo}
\end {equation}
At this point, the simplest way to proceed that will most easily
generalize to other $\delta H$ matrix elements is to note that
(\ref{eq:dHmoo}) can be rewritten, using (\ref{eq:Pb}), in terms of the
$N{=}2$ state normalization as
\begin {equation}
   \langle\B | \delta H |\rangle
   =
   -\frac{ig V_\perp}{\tildeV_\perp}
   \bcalT_{2' \to 23} \cdot (x_3\grad_{\b_2}-x_2\grad_{\b_3})
   \bigl[
     \langle \b_1,\b_2| {\cal P} |
          \b'_1,\b'_2 \rangle_{x'_1,x'_2}
     \, \delta^{(2)}(\b_2{-}\b_3)
   \bigr] ,
\label {eq:trick}
\end {equation}
where the subscripts ($x'_1,x'_2$) on the matrix element show
the two $x_i$ values that are to be used for both the bra and the ket.%
\footnote{
  Readers may reasonably wonder why these $x_i$ values instead of some other.
  The reason has to do with maintaining the consistency of our normalization
  convention (\ref{eq:Vtilde}) for $\delta(\sum x_i\b_i)$ for both
  3-particle and 2-particle states.  In the above expression,
  $\b_2=\b_3$ and (from longitudinal momentum conservation)
  $x'_1 = x_1$ and $x'_2=x_2+x_3$, which means that the
  3-particle expression $x_1 \b_1 + x_2 \b_2 + x_3 \b_3$ for the
  final state is identical, in this context, to the 2-particle
  version $x'_1 \b_1 + x'_2 \b_2$ used for the final state
  in (\ref{eq:trick}).
}
Now we can use the $N{=}2$ version of the
normalization formula (\ref{eq:normb}) to get
\begin {equation}
   \langle\B | \delta H |\rangle
   =
   -ig (x'_2)^2
   \bcalT_{2' \to 23} \cdot (x_3\grad_{\b_2}-x_2\grad_{\b_3}) \,
   \delta^{(2)}(\b_2{-}\b_3) .
\end {equation}
The action of the combination
$x_3 \grad_{\b_2} - x_2 \grad_{\b_3}$ on a function of
$\b_2{-}\b_3$ is the same as the action of $\nabla_\B$,
where $\B \equiv \B_{23} \equiv (\b_2{-}\b_3)/(x_2{+}x_3)$ is the same
3-particle $\B$ we defined
in (\ref{eq:B123}).
Also, we can absorb the overall factor of $(x'_2)^2{=}(x_2{+}x_3)^2$ into
$\delta^{(2)}(\b_2{-}\b_3)$ to get $\delta^{(2)}(\B)$, with result
\begin {equation}
   \langle\B | \delta H |\rangle
   =
   -ig {\bcalT}_{2'\to23} \cdot 
   \grad_{\!\B} \, \delta^{(2)}(\B) .
\label {eq:appdH32}
\end {equation}
This is eq.\ (\ref{eq:dH32}) of the main text.

% --------------------------------------------------------------------------

\subsection {\boldmath$\langle C_{41},C_{23}|\delta H|B\rangle$ matrix elements}
\label {app:dH4}

The analysis proceeds similarly for $\delta H$ matrix
elements between $N{=}3$ and $N{=}4$ particle states, where
we label particles as in fig.\ \ref{fig:dHlabels}b.
The analog of (\ref{eq:trick}) is
\begin {equation}
   \langle\{\C_{ij}\} | \delta H |\B'\rangle
   =
   -\frac{ig V_\perp}{\tildeV_\perp}
   \bcalT_{2' \to 23} \cdot (x_3\grad_{\b_2}-x_2\grad_{\b_3})
   \bigl[
     \langle \b_1,\b_2,\b_4| {\cal P} |
          \b'_1,\b'_2,\b'_3 \rangle_{x'_1,x'_2,x'_3}
     \, \delta^{(2)}(\b_2{-}\b_3)
   \bigr] .
\label {eq:trick4}
\end {equation}
Noting that $\b_4$ is the third $\b_i$ in the bra in
(\ref{eq:trick4}),
the $N{=}3$ version of the
normalization formula (\ref{eq:normb}) then gives
\begin {equation}
   \langle\{\C_{ij}\} | \delta H |\B'\rangle
   =
   -ig (x'_2)^2 (x'_3+x'_1)^{-2} \, \delta^{(2)}(\C_{41}{-}\B') \,
   \bcalT_{2' \to 23} \cdot (x_3\grad_{\b_2}-x_2\grad_{\b_3}) \,
   \delta^{(2)}(\b_2{-}\b_3) .
\end {equation}
The rest follows as before.  Noting that $x'_3+x'_1 = x_4+x_1$ in
the notation of fig.\ \ref{fig:dHlabels}b, we can write the result as
\begin {equation}
   \langle\{\C_{ij}\} | \delta H |\B'\rangle
   =
   -ig (x_4+x_1)^{-2} \, \delta^{(2)}(\C_{41}{-}\B') \, {\bcalT}_{2'\to23} \cdot 
   \grad \delta^{(2)}(\C_{23}) .
\end {equation}
Using an appropriate permutation of the definition (\ref{eq:C3412def}),
\begin {equation}
  |\C_{41},\C_{23}\rangle
  \equiv |x_4+x_1| \, \bigl| \{\C_{ij}\} \bigr\rangle ,
\end {equation}
we have
\begin {equation}
   \langle\C_{41},\C_{23} | \delta H |\B'\rangle
   =
   -ig |x_4+x_1|^{-1} \, \delta^{(2)}(\C_{41}{-}\B') \, {\bcalT}_{2'\to23} \cdot 
   \grad \delta^{(2)}(\C_{23}) ,
\end {equation}
which is equivalent to (\ref{eq:dH43}).
The difference between this and the
previous result (\ref{eq:appdH32}) for $\langle\B | \delta H | \rangle$
is a factor of
\begin {equation}
   |x_4+x_1|^{-1} \delta^{(2)}(\C_{41}{-}\B')
\end {equation}
associated with the spectators.  This is equivalent to the
diagrammatic rule that we presented in the main text at the
bottom of fig.\ \ref{fig:dH}.

% ============================================================================

\section{Connecting our \boldmath$\delta H$ matrix elements
         to standard formulas}
\label{app:dHconnect}

In this appendix, we show how to connect the formulas given in
section \ref{sec:dH} to the usual textbook result
\begin {equation}
   \Bigl| \langle -\p_\perp, (1{-}z) E; \p_\perp, z E
     | \delta H | 
     {\bm 0}, E \rangle_{\rm rel} \Bigr|^2
   = \frac{2 g^2 p_\perp^2}{z(1-z)} \, P(z) ,
\label {eq:Pstandard}
\end {equation}
where the subscript ``rel'' indicates that the states are normalized
using relativistic normalization.
(This use of the notation ``$z$'' is unrelated to our
$z \equiv 1{-}x{-}y$ elsewhere.)

We first note that the $x_i{=}1$ case of
our splitting functions (\ref{eq:Pexplicit})
correspond to the usual DGLAP splitting functions,
with the even more usual helicity-averaged/summed result being
given by
\begin {equation}
   P_{g \to gg}
   = \sum_{h_j,h_k} P_{+\to h_j,h_k}
   = \sum_{h_j,h_k} P_{-\to h_j,h_k} .
\end {equation}
For $x_i$ different than 1, we can factor out the $x_i$ dependence
from (\ref{eq:Pexplicit}) and rewrite these expressions in terms of
the branching fractions
\begin {equation}
   z \equiv z_k \equiv \frac{x_k}{|x_i|}
   \qquad \mbox{and} \qquad
   1-z = z_j \equiv \frac{x_j}{|x_i|}
\label {eq:zdefs}
\end {equation}
of the particular splitting.  The relationship is
\begin {equation}
  P(x_i,x_j,x_k) = |x_i| \, P(-1,z_j,z_k) = |x_i| \, P(z)
\label {eq:Pconvert}
\end {equation}
for each helicity case.

Now consider the square of our matrix element (\ref{eq:dHamp}),
\begin {equation}
   \Bigl| \langle \p_j,\p_k | \delta H | \p_i \rangle \Bigr|^2
   = g^2 |\bcalT_{i \to jk} \cdot \P_{jk}|^2 ,
\end {equation}
implicitly summed/averaged over color here.
Using our definitions (\ref{eq:bcalT}) and (\ref{eq:bcalP1}) of
$\bcalT$ and $\bcalP$, this is
\begin {equation}
   \Bigl| \langle \p_j,\p_k | \delta H | \p_i \rangle \Bigr|^2
   = \frac{g^2 |{\bm e}_{(\pm)} \cdot \P_{jk}|^2}{4 x_i^2 x_j^2 x_k^2 E^3} \,
     P(x_i,x_j,x_k)
\label {eq:dHsquare1}
\end {equation}
(where for brevity we again suppress helicity indices on $P$).
For the choice of axis made in (\ref{eq:Pstandard}),
where $(\p_i,\p_j,\p_k)=({\bm 0},{-}\p_\perp,\p_\perp)$, we have
$\P_{jk} \equiv x_k \p_j - x_j \p_k = - (x_k+x_j) \p_\perp = x_i \p_\perp$.
Because $\P_{jk}$ is real, we have
$|{\bm e}_{(\pm)}\cdot\P_{jk}|^2 = P_{jk}^2$.
Then, converting (\ref{eq:dHsquare1}) to relativistic normalization
by multiplying by $(2 E_i) (2 E_j) (2 E_k) = 8 |x_i x_j x_k| E^3$,
\begin {equation}
   \Bigl| \langle \p_j,\p_k | \delta H | \p_i \rangle_{\rm rel} \Bigr|^2
   = \frac{2 g^2 p_\perp^2 |x_i|}{|x_j x_k|} \,
     P(x_i,x_j,x_k) .
\end {equation}
Using (\ref{eq:zdefs}) and (\ref{eq:Pconvert}) then reproduces
the standard result (\ref{eq:Pstandard}).

% ============================================================================

\section{More details on \boldmath$1/\Delta t$ divergences}
\label{app:smalldt}

% ----------------------------------------------------------------------------

\subsection {\boldmath$\Delta t{\to}0$ limit of \boldmath$xy\bar y\bar x$}
\label{app:dtxyyx}

Here, we will give some more details about how to take the small $\Delta t$
limit of $xy\bar y\bar x$ that gives rise to (\ref{eq:div1}).
For small $\Delta t$,
$\uOmega \cot(\uOmega \Delta t)$ and
$\uOmega \csc(\uOmega \Delta t)$ both become $(\Delta t)^{-1} + O(\Delta t)$.
Then (\ref{eq:XYZdef}) becomes
\begin {subequations}
\begin {align}
   \begin{pmatrix} X_\bx & Y_\bx \\ Y_\bx & Z_\bx \end{pmatrix}
   &=
     -\frac{i}{\Delta t}\, a_\bx^{-1\top} a_\bx^{-1}
     +\begin{pmatrix} |M_\ix|\Omega_\ix & 0 \\ 0 & 0 \end{pmatrix}
     +O(\Delta t) ,
\\
   \begin{pmatrix} X_\Ax & Y_\Ax \\ Y_\Ax & Z_\Ax \end{pmatrix}
   &=
     -\frac{i}{\Delta t}\, a_\Ax^{-1\top} a_\Ax^{-1}
     +\begin{pmatrix} |M_\fx|\Omega_\fx & 0 \\ 0 & 0 \end{pmatrix}
     +O(\Delta t) ,
\\
   \begin{pmatrix} X_{\bx\Ax} & Y_{\bx\Ax} \\ \Ybar_{\bx\Ax} & Z_{\bx\Ax} \end{pmatrix}
   &=
     -\frac{i}{\Delta t}\, a_\bx^{-1\top} a_\Ax^{-1}
     +O(\Delta t) .
\end {align}
\end {subequations}
Except for the $M\Omega$ terms (coming from the evolution before and after
the $\Delta t$ interval), this is the result one would get in vacuum
($\Omega_\pm = 0$).
Now use the simple formulas (\ref{eq:frakM}), (\ref{eq:ai}),
(\ref{eq:aaybar}), which
do not depend on details of the normal mode solutions, to get
\begin {subequations}
\label {eq:XYZsim}
\begin {align}
   \begin{pmatrix} X_\bx & Y_\bx \\ Y_\bx & Z_\bx \end{pmatrix}
   &=
     -\frac{i E (x_1{+}x_4)}{\Delta t}
       \begin{pmatrix} x_1 x_4 & 0 \\ 0 & -x_2 x_3 \end{pmatrix}
     +\begin{pmatrix} |M_\ix|\Omega_\ix & 0 \\ 0 & 0 \end{pmatrix}
     + O(\Delta t) ,
\\
   \begin{pmatrix} X_\Ax & Y_\Ax \\ Y_\Ax & Z_\Ax \end{pmatrix}
   &=
     -\frac{i E (x_3{+}x_4)}{\Delta t}
       \begin{pmatrix} x_3 x_4 & 0 \\ 0 & -x_1 x_2 \end{pmatrix}
     +\begin{pmatrix} |M_\fx|\Omega_\fx & 0 \\ 0 & 0 \end{pmatrix}
     + O(\Delta t) ,
\\
   \begin{pmatrix} X_{\bx\Ax} & Y_{\bx\Ax} \\ \Ybar_{\bx\Ax} & Z_{\bx\Ax} \end{pmatrix}
   &=
     \frac{i E}{\Delta t}
     \begin{pmatrix} x_1 x_3 x_4 &  x_1 x_2 x_4 \\
                     x_2 x_3 x_4 &  x_1 x_2 x_3 \end{pmatrix}
     + O(\Delta t) .
\end {align}
\end {subequations}
Using the definitions (\ref{eq:Mi}) and (\ref{eq:Mf})
of $M_\ix$ and $M_\fx$,
the combination $X_\bx X_\Ax - X_{\bx\Ax}^2$ that appears in the integrals
(\ref{eq:I}) can be written in the form
\begin {equation}
   X_\bx X_\Ax - X_{\bx\Ax}^2
   = \frac{x_1 x_2 x_3 x_4^3 E^2}{(\Delta t)^2}
     - \frac{i M_\ix M_\fx}{\Delta t} \,
            (\Omega_\ix\sgn M_\ix + \Omega_\fx\sgn M_\fx)
     + O\bigl((\Delta t)^0\bigr) .
\label{eq:detXsim}
\end {equation}
The fact that $Y_\bx$ and $Y_\Ax$ vanish at the order of interest
means that (\ref{eq:IGamma}) simplifies to
\begin {align}
   \left[\frac{d\Gamma}{dx\,dy}\right]_{xy\bar y\bar x} = &
   \frac{\CA^2 \alphas^2 M_\ix M_\fx}{32\pi^4 E^2} \, 
   ({-}\hat x_1 \hat x_2 \hat x_3 \hat x_4)
   \int_0^{\infty} \frac{d(\Delta t)}{(\Delta t)^2} \>
\nonumber\\ & \times
   \Bigl[
     \Ybar_{\bx\Ax} Y_{\bx\Ax} (\alpha I_0 + \beta I_2 + \gamma I_2)
     + Z_{\bx\Ax} (\alpha+\beta+2\gamma) I_1
   \Bigl]
   + O\bigl((\Delta t)^0\bigr) .
\label {eq:IGamma5small}
\end {align}
Using (\ref{eq:XYZsim}) and the integrals (\ref{eq:I}),
expanding in $\Delta t$, and using $\sum x_i=0$ to simplify, 
gives eq.\ (\ref{eq:div1}) of the
main text.

% ----------------------------------------------------------------------------

\subsection {Triple time coincidence associated with \boldmath$1/\Delta t$}
\label {app:tripletime}

Here we justify more explicitly
the claim in section \ref{sec:smalldt1} that the
$\Omega_\fx/\Delta t$ divergent terms in
(\ref{eq:div1}) for $xy\bar y\bar x$ arise from the first
three splitting times
$t_\xx$, $t_\yx$, and $t_\ybx$ simultaneously approaching each other.

In deriving results for $xy\bar y\bar x$, we integrated the first
splitting time $t_\xx$ from $-\infty$ up to $t_\yx$ in
(\ref{eq:3int0}) to get
\begin {equation}
   \int_{-\infty}^{t_\yx} dt_\xx \>
   \grad_{\B_\xx} \langle\B_\yx,t_\yx|\B_\xx,t_\xx\rangle
   \biggr|_{\B_\xx=0}
   =
   - \frac{i M_\ix \B_\yx}{\pi B_\yx^2} \,
   \exp\bigl(
      - \tfrac12 |M_\ix| \Omega_\ix B_\yx^2
   \bigr) .
\label {eq:Bint1}
\end {equation}
In order to determine whether the region where
$t_\xx$ is close to $t_\yx$ is important, let us instead integrate
only from $t_\yx-\Delta\tau$ to $t_\yx$ for some
$\Delta\tau$:
\begin {subequations}
\label {eq:varreplace1}
\begin {equation}
   \int_{t_\yx-\Delta\tau}^{t_\yx} dt_\xx \>
   \grad_{\B_\xx} \langle\B_\yx,t_\yx|\B_\xx,t_\xx\rangle
   \biggr|_{\B_\xx=0}
   =
   - \frac{i M_\ix \B_\yx}{\pi B_\yx^2} \,
   \exp\bigl( \tfrac{i}2 M_\ix \Omega_\ix B_\yx^2
              \cot(\Omega_\ix \Delta\tau) \bigr) .
\label {eq:Bint2}
\end {equation}
Comparing (\ref{eq:Bint1}) with (\ref{eq:Bint2}), we see that the
only effect that the change of integration region has on
our results is to replace
\begin {equation}
   \Omega_\ix \to
   \Omega_\ix^{\rm eff}(\Delta\tau)
   \equiv -i \sgn(M_\ix) \, \Omega_\ix \cot(\Omega_\ix \Delta\tau)
\end {equation}
\end {subequations}
in (\ref{eq:XYZdef}) and therefore in the divergent pieces
of (\ref{eq:div1}).
This change has no effect at all on the $\Omega_\fx/\Delta t$ term
in (\ref{eq:div1}), and so the $\Omega_\fx/\Delta t$ term arises
from $t_\xx$ no more than $\Delta\tau$ away from $t_\yx$.
This conclusion breaks down only if we make $\Delta\tau$ so small
that $\Omega_\ix^{\rm eff}$ becomes so large that the small-$\Delta t$
expansion made in (\ref{eq:XYZsim}) breaks down.  That happens
when $\Delta\tau \lesssim \Delta t$.
So the $\Omega_\fx/\Delta t$ term arises from the case where
$t_\xx$, $t_\yx$, and $t_\ybx$ all lie within $O(\Delta t)$ of
each other.

A similar argument holds for the $\Omega_\ix/\Delta t$ divergence,
with the replacement
\begin {subequations}
\label {eq:varreplace}
\begin {equation}
   \int_{t_\ybx}^{+\infty} dt_\xbx
   \to
   \int_{t_\ybx}^{t_\ybx+\Delta\tau} dt_\xbx
\end {equation}
being equivalent to
\begin {equation}
   \Omega_\fx \to
   \Omega_\fx^{\rm eff}(\Delta\tau)
   \equiv -i \sgn(M_\fx) \, \Omega_\fx \cot(\Omega_\fx \Delta\tau)
\label {eq:Omegaeff} .
\end {equation}
\end {subequations}

% ----------------------------------------------------------------------------

\subsection {\boldmath$i\epsilon$ prescription for double splitting
             (especially \boldmath$x\bar y\bar x y$)}
\label {app:epsilon2}

In this section, we will more precisely justify the application of
the $i\epsilon$ prescription in section \ref{sec:epsilon2}, especially
as regards the $x\bar y\bar x y$ calculation.
As discussed above and in the main text, the $1/\Delta t$ divergences
arise when three times become coincident, e.g.\ $(t_\yx,t_\ybx,t_\xbx)$.
In order to sort things out, it will be helpful to know the
explicit dependence on those three times as they approach each
other.  Unfortunately, we have already integrated over one of
those times in our derivation of results like (\ref{eq:div1}),
(\ref{eq:div2b}), and (\ref{eq:div3b}).  So let us step back
and undo that integration.  We could go back and rederive all our
results (including doing all the $\B$ and $\C_{ij}$ integrals)
without integrating over the last (or first) time, but happily there is a
trick that will allow us to extract the time dependence
we want from the integrated results that we already have.

% ........................................................................

\subsubsection{$xy\bar y\bar x$}

We start by discussing $xy\bar y\bar x$, since this was the exemplar
to which we related all other results.  We will focus on the
case where the last three times approach each other, giving rise
to the $\Omega_\ix/\Delta t$ divergence.
In (\ref{eq:varreplace}), we saw how to integrate the last time
$t_\xbx$ over the interval $(t_\ybx,t_\ybx+\Delta\tau)$ instead
of the interval $(t_\ybx,\infty)$.
If we take that result and then
differentiate with respect to $\Delta\tau$, it is equivalent
to never having integrated over $t_\xbx$ in the first place.
Using (\ref{eq:Omegaeff}), this operation is equivalent to
\begin {equation}
   \cdots
   = \frac{d}{d\Delta\tau} \int_{t_\ybx}^{t_\ybx+\Delta\tau} dt_\xbx \> \cdots
   = \frac{d}{d\Delta\tau}
     \left[ \int_{t_\ybx}^{\infty} dt_\xbx \> \cdots
     \right]_{\Omega_\fx \to \Omega^{\rm eff}_\fx(\Delta\tau)} ,
\end {equation}
with the understanding that $\Delta\tau$ should be re-interpreted as
$t_\xbx - t_\ybx$ after all of these operations have been completed.
We are interested in what happens when the last three times
approach each other, which is the case
where both $\Delta\tau$ and $\Delta t$ are small.
So we may use the small $\Delta\tau$ limit
of (\ref{eq:Omegaeff}), giving
\begin {equation}
   \cdots \simeq
   \frac{d}{d\Delta\tau}
     \left[ \int_{t_\ybx}^{\infty} dt_\xbx \> \cdots
     \right]_{\Omega_\fx \to -i \sgn(M_\fx)/\Delta\tau} .
\end {equation}
Now carry out this procedure on (\ref{eq:IGamma}) to recover the
$t_\xbx$ integrand.  Then expand the integrand for simultaneously
small $\Delta t$ and small $\Delta\tau$,
treating
them as the same order.  We find the result%
\footnote{
  Note that here $[d\Gamma/dx\>dy]^{\rm vac}$, which is the
  $\hat q \to 0$ limit of $d\Gamma/dx\>dy$, does {\it not} correspond
  to all $\Omega$'s $\to 0$ because $\Omega_\fx$ has been replaced
  by $-i \sgn(M_\fx)/\Delta t$, which does not vanish as $\hat q\to 0$.
}
%\begin {align}
%   &
%   \left[\frac{d\Gamma}{dx\,dy}\right]_{xy\bar y\bar x} -
%   \left[\frac{d\Gamma}{dx\,dy}\right]^{\rm vac}_{xy\bar y\bar x}
%\nonumber\\ & \qquad
%   \simeq
%   \frac{\CA^2 \alphas^2}{16\pi^2} \, 
%   i \Omega_\ix \sgn(M_\ix)
%   \hat x_1^2 \hat x_2^2 \hat x_3^2 \hat x_4^2
%  (\hat x_1{+}\hat x_4)^2 (\hat x_3{+}\hat x_4)^2
%   \int_0^\infty d(\Delta t) \int_0^\infty d(\Delta\tau) \>
%\nonumber\\ & \qquad\quad \times
%       \biggl[
%          \frac{4(\Delta t+\Delta\tau)}{D^3} \,
%             \hat x_2 \hat x_4 (\hat x_1{+}\hat x_4)(\hat x_3{+}\hat x_4)
%             (\alpha{+}\beta{+}\gamma)
%          + \frac1{D^2}(\hat x_1 \hat x_3 - 2 \hat x_2 \hat x_4)
%             (\alpha{+}\beta{+}\gamma)
%\nonumber\\ & \qquad\quad\qquad
%          + \frac1{D^2} ((\hat x_1{+}\hat x_4)(\hat x_3{+}\hat x_4) \gamma
%                        - \hat x_2 \hat x_4 \beta)
%       \biggr] ,
%\label {eq:triple1int}
%\end {align}
\begin {subequations}
\label {eq:triple1}
\begin {align}
   &
   \left[\frac{d\Gamma}{dx\,dy}\right]_{xy\bar y\bar x} -
   \left[\frac{d\Gamma}{dx\,dy}\right]^{\rm vac}_{xy\bar y\bar x}
\nonumber\\ & \qquad
   \simeq
   \frac{\CA^2 \alphas^2}{16\pi^2} \, 
   i \Omega_\ix \sgn(M_\ix)
   \hat x_1^2 \hat x_2^2 \hat x_3^2 \hat x_4^2
  (\hat x_1{+}\hat x_4)^2 (\hat x_3{+}\hat x_4)^2
   \int_0^\infty d(\Delta t) \int_0^\infty d(\Delta\tau) \>
\nonumber\\ & \qquad\quad \times
       \biggl[
          \frac{4(\Delta t{+}\Delta\tau)}{D^3} \,
             \hat x_2 \hat x_4 (\hat x_1 \hat x_3 {-} \hat x_2 \hat x_4)
             (\alpha{+}\beta{+}\gamma)
          + \frac{\hat x_1 \hat x_3}{D^2} (\alpha{+}\beta{+}2\gamma)
%\nonumber\\ & \qquad\quad\qquad
          - \frac{\hat x_2 \hat x_4}{D^2} (2\alpha{+}3\beta{+}3\gamma)
       \biggr] ,
\label {eq:triple1int}
\end {align}
where $\Delta t \equiv t_\ybx-t_\yx$,
$\Delta\tau \equiv t_\xbx - t_\ybx$, and
\begin {equation}
   D \equiv -\hat x_2 \hat x_4 \, \Delta\tau
            + (\hat x_1+\hat x_4)(\hat x_3+\hat x_4) \, \Delta t .
\label {eq:Denom1}
\end {equation}
\end {subequations}
One may check that doing the $\Delta\tau$ integral
indeed recovers the $\Omega_\ix/\Delta t$ divergence in (\ref{eq:div1}).
The details of (\ref{eq:triple1int}) will be mostly
irrelevant to the present discussion: All we need to take
away is that the only singularities of the integrand are associated
with inverse powers of the denominator $D$ defined by (\ref{eq:Denom1}).
$D$ is proportional to the leading behavior of
$(X_\yx X_\ybx - X_{\yx\ybx}^2) (\Delta t)^2\Delta\tau$
[with $\Omega_\fx \to -i \sgn(M_\fx)/\Delta\tau$].
And so one could have gotten what we will need for discussing
$i\epsilon$ prescriptions simply by looking
at the structure of $X_\yx X_\ybx - X_{\yx\ybx}^2$ rather than
deriving (\ref{eq:triple1int}) in detail.

% ........................................................................

\subsubsection{Permutations, and the $\Omega_\ix/\Delta t$ divergence
               of $x\bar y \bar x y$}

We may write (\ref{eq:triple1}) more explicitly in terms of the
last three
times by writing (\ref{eq:Denom1}) as
\begin {align}
   D
   &= -(\hat x_1+\hat x_4)(\hat x_3+\hat x_4) t_\yx
      + \hat x_1 \hat x_3 t_\ybx
      - \hat x_2 \hat x_4 t_\xbx
\nonumber\\
   &= \hat x_1 \hat x_3 (t_\ybx-t_\yx) - \hat x_2 \hat x_4 (t_\xbx-t_\yx)
\label {eq:Denom}
\end {align}
and rewriting the integration $\int d(\Delta t) \> d(\Delta\tau)$
as $\int dt_\yx \> dt_\ybx \> dt_\xbx$ divided by a factor of total
time (and restricted to the appropriate ordering of the three times).
In this form, the result turns out to describe the $\Omega_\ix/\Delta t$
divergences of not only $xy\bar y\bar x$ but also of $x\bar y y\bar x$
and $x\bar y\bar x y$ as well.  As a check, one may integrate
the result over the last time in the
$x\bar y y\bar x$ and $x\bar y\bar x y$ cases ($t_\xbx$ and $t_\yx$)
appropriately and obtain (\ref{eq:div2b}) and (\ref{eq:div3b})
with $\Delta t \equiv t_\yx - t_\ybx$ and $\Delta t \equiv t_\ybx - t_\yx$
respectively.%
\footnote{
  On a related note, the formula (\ref{eq:Denom}) for $D$ is
  invariant, up to an overall sign,
  under (i) $\hat x_i \to x'_i$ with
  $(t_\yx,t_\ybx,t_\xbx) \to (t_\ybx,t_\yx, t_\xbx)$
  and under (ii) $\hat x_i \to \tilde x_i$ with
  $(t_\yx,t_\ybx,t_\xbx) \to (t_\ybx,t_\xbx, t_\yx)$.
  The $\int dt_\yx \> dt_\ybx \> dt_\xbx$ integral representation
  of (\ref{eq:triple1})
  is invariant under the same transformations with additionally
  (i) $(\alpha,\beta,\gamma) \to (\beta,\alpha,\gamma)$ and
  (ii) $(\alpha,\beta,\gamma) \to (\gamma,\alpha,\beta)$
  respectively, as in section \ref{sec:other}.
}

Now let's consider $i\epsilon$ prescriptions in the denominator
$D$ of (\ref{eq:Denom}).  The times $t_\ybx$ and $t_\xbx$ in the
conjugate amplitude should have
small negative imaginary parts compared to the time $t_\yx$ in
the amplitude, as explained in section \ref{sec:smalldt}.
For completeness, however, we should also consider the $i\epsilon$
prescriptions of $t_\ybx$ and $t_\xbx$ relative to each other.  Two
operators associated with times $t_1$ and $t_2$
both in the
amplitude should be time-ordered, and so
\begin {equation}
  t_2-t_1 \to t_2-t_1 - i\epsilon \sgn(t_2{-}t_1) .
\label {eq:timeorder}
\end {equation}
Two operators with times $\bar t_1$ and $\bar t_2$
in the conjugate amplitude should
be anti-time-ordered, corresponding to the conjugate
of (\ref{eq:timeorder}),
\begin {equation}
  \bar t_2-\bar t_1 \to \bar t_2-\bar t_1 + i\epsilon \sgn(\bar t_2{-}\bar t_1) .
\label {eq:tbareps}
\end {equation}
A quick way to remember all of the $i\epsilon$ prescriptions
is to think of a Schwinger-Keldysh
contour that runs from $t=-\infty$ to $t=+\infty$ for times
in the amplitude and then turns around and runs back again
from $t=+\infty$ to $t=-\infty$ for the times in the
conjugate amplitude.  The rule is that any time $t_2$
that appears further along this contour than another time
$t_1$ should have a more negative (infinitesimal) imaginary part.

So, for instance, we can implement all of the relative imaginary parts
for the relevant times in $x \bar y \bar x y$
(for which $t_\ybx < t_\xbx < t_y$) by
\begin {equation}
  (t_\ybx, t_\xbx, t_\yx)
  \to \bigl( t_\ybx - i(\epsilon_1+\epsilon_2),
             t_\xbx - i\epsilon_2,
             t_\yx \bigr) .
\label {eq:tttprescription}
\end {equation}
Using the explicit values (\ref{eq:xhat}) of the $\hat x_i$,
(\ref{eq:Denom}) can be put in the form
% \hat x_1 \hat x_3 (t_\ybx-t_\yx) - \hat x_2 \hat x_4 (t_\xbx-t_\yx) .
% (\hat x_1,\hat x_2,\hat x_3,\hat x_4) = (-1,y,1{-}x{-}y,x) .
% D = -(1-x-y) (t_\ybx-t_\yx) - xy (t_\xbx-t_\yx)
\begin {equation}
  D = (1-x-y) (t_\xbx-t_\ybx) + (1-x)(1-y) (t_\yx - t_\xbx) ,
\end {equation}
for which the prescription (\ref{eq:tttprescription}) gives
\begin {equation}
  D \to (1-x-y) (t_\xbx-t_\ybx+i\epsilon_1) + (1-x)(1-y) (t_\yx-t_\xbx+i\epsilon_2)
  = D + i\epsilon .
\end {equation}
The important point here is that we could get the same prescription
by instead writing
\begin {equation}
  D \to (1-x-y) (t_\xbx-t_\ybx+i\epsilon) + (1-x)(1-y) (t_\yx-t_\xbx) ,
\end {equation}
which is $\Delta t \to \Delta t + i\epsilon$ for the
$\Delta t \equiv t_\xbx-t_\ybx$ relevant to $x\bar y\bar x y$.
This agrees with the prescription guessed in
section \ref{sec:epsilon2} of the main text for
the $\Omega_\ix/\Delta t$ divergences of $x\bar y\bar x y$.
[In this case, it happens to also be the same
as the prescription one would get from simply
applying (\ref{eq:tbareps}) in isolation
to $\Delta t \equiv t_\xbx - t_\ybx$.  However, that is an
accident, as we will see in a moment.]

One may similarly check the $i\epsilon$ prescriptions used
for all of the other $\Omega_\ix/\Delta t$ terms in section
\ref{sec:epsilon2}.

% ........................................................................

\subsubsection{$\tilde\Omega_\fx/\Delta t$ divergence of $x\bar y\bar x y$}

Now consider the $\tilde\Omega_\fx/\Delta t$ divergence of
$x\bar y\bar x y$, which arises from the first
three times $(t_\xx,t_\ybx,t_\xbx)$
approaching each other.  The corresponding value
of $D \propto (X_\yx X_\ybx - X_{\yx\ybx}^2) (\Delta t)^2\Delta\tau$
[now with $\Omega_\ix \to -i \sgn(M_\ix)/\Delta\tau$] is
\begin {equation}
   \tilde D \equiv -\tilde x_2 \tilde x_4 \, \Delta\tau
            + (\tilde x_1+\tilde x_4)(\tilde x_3+\tilde x_4) \, \Delta t ,
\end {equation}
where $\Delta t \equiv t_\xbx - t_\ybx$ and
$\Delta\tau \equiv t_\ybx-t_\xx$.
This can be written explicitly as
\begin {equation}
   \tilde D = (1-x)(1-y)(t_\ybx-t_\xx) + (1-x-y)(t_\xbx-t_\ybx) .
\label {eq:Dtilde}
\end {equation}
The relative $i\epsilon$ prescriptions for
$t_\xx < t_\ybx < t_\xbx$ in
$x\bar y\bar x y$ can be organized as
\begin {equation}
  (t_\xx, t_\ybx, t_\xbx)
  \to \bigl( t_\xx,
             t_\ybx - i(\epsilon_1+\epsilon_2),
             t_\xbx - i\epsilon_2 \bigr) .
\label {eq:tttilde}
\end {equation}
It is convenient to reorganize (\ref{eq:Dtilde}) as
\begin {equation}
   \tilde D = (1-x)(1-y)(t_\xbx-t_\xx) + xy(t_\ybx-t_\xbx) ,
\end {equation}
from which is it particularly easy to read off that
(\ref{eq:tttilde}) gives
\begin {equation}
  \tilde D \to \tilde D - i\epsilon .
\end {equation}
This is the same prescription as just taking
$\Delta t \to \Delta t - i\epsilon$ in (\ref{eq:Dtilde}),
which matches the guess that we made back in 
section \ref{sec:epsilon2} of the main text.
[Note that this is {\it not}\/ the prescription we
would have gotten by
applying (\ref{eq:tbareps}) in isolation
to $\Delta t \equiv t_\xbx - t_\ybx$.]

One may similarly check all of the other $1/\Delta t$ divergences of
section \ref{sec:epsilon2}.

% ============================================================================

\section{Relating different crossed interference contributions}
\label{app:relate}

In this appendix, we show more explicitly how to relate the
$x \bar y y\bar x$ and $x\bar y \bar x y$ contributions to the
$x y \bar x\bar y$ contribution.

% ----------------------------------------------------------------------------

\subsection {\boldmath$x \bar y y\bar x$}

\subsubsection {The basic discussion}

For $x\bar y y\bar x$, the analog of the $xy\bar y\bar x$ formula
(\ref{eq:IIxyyx1}) is
\begin {align}
   \left[\frac{dI}{dx\,dy}\right]_{x\bar y y\bar x}
   &=
   d_R^{-1} \tr(T_R^a T_R^b T_R^a T_R^b)
   \frac{\alphas^2 }{4 E^4}
   |x'_1+x'_4|^{-1} |x'_3+x'_4|^{-1}
   \int_{t_\xx < t_\ybx < t_\yx < t_\xbx}
   \sum_{h_\xx,h_\yx,h_\zx,h,\bar h}
   \int_{\B^\Ax,\B^\bx}
\nonumber\\ &\times
   \bcalP^*_{\bar h \to h_\zx,h_\xx}\bigl(1{-}y \to 1{-}x{-}y,x\bigr)
       \cdot \grad_{\B^\Bx}
   \langle\B^\Bx,t_\Bx|\B^\bx,t_\bx\rangle
   \Bigr|_{\B^\Bx=0}
\nonumber\\ &\times
   \bigl( \bcalP^*_{h_\ix \to \bar h, h_\yx}\bigl(1 \to 1{-}y,y\bigr)
          \cdot \grad_{\C_{12}^\Ax} \bigr)
   \bigl( \bcalP_{h \to h_\zx,h_\yx}\bigl(1{-}x \to 1{-}x{-}y,y\bigr)
          \cdot \grad_{\C_{23}^\bx} \bigr)
\nonumber\\ &\qquad\qquad
  \langle\C_{41}^\bx,\C_{23}^\bx,t_\bx|\C_{34}^\Ax,\C_{12}^\Ax,t_\Ax\rangle
   \Bigr|_{\C_{12}^\Ax=0=\C_{23}^\bx; ~ \C_{34}^\Ax=\B^\Ax; ~ \C_{41}^\bx=\B^\bx}
\nonumber\\ &\times
   \bcalP_{h_\ix \to h,h_\xx}\bigl(1 \to 1{-}x,x\bigr)
   \cdot \grad_{\B^\xx}
   \langle\B^\Ax,t_\bx|\B^\ax,t_\ax\rangle
   \Bigr|_{\B^\ax=0}
\end {align}
[where we have used the notation of (\ref{eq:calPsum2})
rather than (\ref{eq:calPsum})
for the splitting functions $\bcalP$].
Substituting (\ref{eq:abcdef}) for (\ref{eq:calPsum2}) and then
using (\ref{eq:3ints}), the analog of (\ref{eq:IIxyyx2}) is
\begin {align}
   \left[\frac{d\Gamma}{dx\,dy}\right]_{x\bar y y\bar x}
   = &
   - \frac{\CA^2 \alphas^2 M_\ix M_\fx}{8 \pi^2 E^4}
   \frac{
     ( \alpha \delta^{\bar n n} \delta^{\bar m m}
     {+} \beta \delta^{\bar n \bar m} \delta^{nm}
     {+} \gamma \delta^{\bar n m} \delta^{n \bar m} )
   }{
     |x'_1 + x'_4| |x'_3 + x'_4|
   }
\nonumber\\ & \times
   \int_0^{\infty} d(\Delta t)
   \int_{\B^\Ax,\B^\bx}
   \frac{B^{\yx}_{\bar n}}{(B^\yx)^2} \,
   \frac{B^{\ybx}_{m}}{(B^\ybx)^2} \,
   \exp\bigl(
      - \tfrac12 |M_\fx| \Omega_\fx (B^\yx)^2
      - \tfrac12 |M_\ix| \Omega_\ix (B^\ybx)^2
   \bigr)
\nonumber\\ &\qquad\times
   \nabla_{\C_{12}^\Ax}^{\bar m}
   \nabla_{\C_{23}^\bx}^n
   \langle\C_{41}^\bx,\C_{23}^\bx,\Delta t|\C_{34}^\Ax,\C_{12}^\Ax,0\rangle
   \Bigr|_{\C_{12}^\Ax=0=\C_{23}^\bx; ~ \C_{34}^\Ax=\B^\Ax; ~ \C_{41}^\bx=\B^\bx}
\label {eq:xyyxswitch}
\end {align}
with $\Delta t \equiv t_\bx - t_\Ax$.
The $\C_{ij}$ and 4-particle propagator
are now implicitly defined in terms of
$x_i = x_i'$ instead of $x_i = \hat x_i$.
The derivation of (\ref{eq:Cprop}) is symmetric in such a way that
\begin {equation}
   \langle\C_{41}^\bx,\C_{23}^\bx,\Delta t|\C_{34}^\Ax,\C_{12}^\Ax,0\rangle
   =
   \langle\C_{34}^\Ax,\C_{12}^\Ax,\Delta t|\C_{41}^\bx,\C_{23}^\bx,0\rangle
   .
\end {equation}
Using this relation, the $x\bar y y\bar x$ result
(\ref{eq:xyyxswitch}) is superficially the same as making
the substitutions summarized in section \ref{sec:xyyx2} on
the $xy\bar y\bar x$ expression (\ref{eq:IIxyyx2}).
We say superficially only because there is a sign
issue that must be addressed when replacing $\hat x_i \to x'_i$
in the formula for
$\langle\C_{34}^\Ax,\C_{12}^\Ax,\Delta t|\C_{41}^\bx,\C_{23}^\bx,0\rangle$,
which we discuss next.

% ...........................................................................

\subsubsection {A sign issue}
\label {app:sign}

When the 4-particle $x_i$ are the $\hat x_i$ (\ref{eq:xhat}) appropriate
for $x y\bar y\bar x$, then there are a number of important and related
properties of the 4-particle normal mode formulas in
section  \ref{sec:normal}:
\begin {itemize}
\item
  $\lambda_\pm \equiv
   \frac1{x_1} + \frac1{x_2} + \frac1{x_3} + \frac1{x_4} \pm \sqrt\Delta$
  is always positive, so that the normal mode frequencies
  $\Omega_+$ and $\Omega_-$ given by (\ref{eq:Omegapm})
  are both proportional to $\sqrt{-i}$.
\item
  $x_1 x_3/2N_+ E$ and $x_1 x_3/2N_- E$ are both positive,
  so that the normal modes
  $(C^+_{34},C^+_{12})$ and $(C^-_{34},C^-_{12})$ of
  (\ref{eq:Cmodes}) are normalized to be real.
\item
  As a result, the normal mode degrees of freedom $\A_\pm$ defined
  by (\ref{eq:Adef}), which were implicitly assumed to be real,
  are indeed real.
\end {itemize}
In contrast,
when the 4-particle $x_i$ are the $x'_i$ (\ref{eq:xprime}) appropriate
for $x \bar y y\bar x$, the situation changes:
\begin {itemize}
\item
  $\lambda_+$ defined above is still positive but $\lambda_-$ is
  negative, so that $\Omega_+ \propto \sqrt{-i}$ but
  $\Omega_- \propto \sqrt{+i}$.
\item
  $x_1 x_3/2N_+E$ is positive but $x_1 x_3/2N_-E$ is negative,
  so that the normal mode
  $(C^+_{34},C^+_{12})$ has been normalized to be real by
  (\ref{eq:Cmodes}), but the normal mode $(C^-_{34},C^-_{12})$
  has been normalized to be imaginary.
\item
  As a result, the definition (\ref{eq:Adef}) makes
  $\A_-$ imaginary instead of real.
\end {itemize}
Imaginary values for $(C^-_{34},C^-_{12})$ confuse the analysis that
follows in section \ref{sec:GeneralForm}, especially the normalization
of the relation (\ref{eq:CvsAprop}).
The easiest way to clear this up is to keep to real normalizations for
the normal modes:
\begin {equation}
   \begin {pmatrix} C^-_{34} \\ C^-_{12} \end{pmatrix}^{\!\rm new}
   \equiv i \begin{pmatrix} C^-_{34} \\ C^-_{12} \end{pmatrix}
\label {eq:Cnew}
\end {equation}
and
\begin {equation}
   (\A_+,\A_-)^{\rm new} \equiv (\A_+,-i \A_-) .
\label {eq:Anew}
\end {equation}
This means that the Lagrangian (\ref{eq:AL}) for $\A_\pm$ is then
normalized as
\begin {equation}
   L =
   \sum_\pm \pm \left[
      \tfrac12 \, (\dot\A_\pm^{\rm new})^2
      - \tfrac12 \, \Omega_\pm^2 (\A_\pm^{\rm new})^2
   \right] .
\end {equation}
Now the $\A_\pm^{\rm new}$ are both real, but the Lagrangian
for $\A_-^{\rm new}$ has an unusual overall sign.  To understand the
effect of the sign, it is useful to look at the corresponding
Hamiltonian
\begin {equation}
   {\cal H} =
   \sum_\pm \pm \left[
      \tfrac12 \, \P_{A_\pm}^2
      + \tfrac12 \, \Omega_+^2 \A_\pm^2
   \right]^{\rm new} .
\end {equation}
In time evolution $\exp(-i H t)$, negating a Hamiltonian has the
same effect as negating time $t$.
So the $\A_-$ part of the time evolution (\ref{eq:Aprop}) changes from
\begin {equation}
   \langle \A_-,t | \A_-',0 \rangle
   =
   \frac{\Omega_-\csc(\Omega_- t)}{2\pi i} \,
      \exp\Bigl(
        i \bigl[ \tfrac12 (\A_-^2+\A_-'^2) \Omega_-\cot(\Omega_- t)
        - \A_-\cdot\A_-' \Omega_- \csc(\Omega_- t) \bigr]
      \Bigr)
\end {equation}
to its conjugate
\begin {multline}
   \langle \A_-,t | \A_-',0 \rangle^{\rm new}
   =
\\
   - \frac{\Omega_-\csc(\Omega_- t)}{2\pi i} \,
      \exp\Bigl(
        - i \bigl[ \tfrac12 (\A_-^2+\A_-'^2)^{\rm new} \Omega_-\cot(\Omega_- t)
        - (\A_-\cdot\A_-')^{\rm new} \Omega_- \csc(\Omega_- t) \bigr]
      \Bigr) .
\label {eq:AAnew}
\end {multline}
The one change here that matters is the appearance of an overall minus
in the prefactor,
which we will address in a moment.  The other change is to the
sign of the argument of the exponential,
which we now show is inconsequential.  Combined with (\ref{eq:Cnew}),
its effect is to change
expressions of the form
\begin {equation}
  a^{-1\top} \uOmega \cot(\uOmega\,\Delta t) \, a^{-1}
  \qquad \mbox{and} \quad
  a^{-1\top} \uOmega \csc(\uOmega\,\Delta t) \, a^{-1}
\label {eq:aaold}
\end {equation}
in the propagator (\ref{eq:Cprop}) to
\begin {equation}
  a_{\rm new}^{-1\top}
  \left( \begin {smallmatrix} 1 & \\[4pt] & -1 \end{smallmatrix} \right)
  \uOmega \cot(\uOmega\,\Delta t) \, a_{\rm new}^{-1}
  \qquad \mbox{and} \quad
  a_{\rm new}^{-1\top}
  \left( \begin {smallmatrix} 1 & \\[4pt] & -1 \end{smallmatrix} \right)
  \uOmega \csc(\uOmega\,\Delta t) \, a_{\rm new}^{-1} ,
\label {eq:aanew}
\end {equation}
where, based on (\ref{eq:af}), (\ref{eq:ai}) and (\ref{eq:Cnew}),
\begin {equation}
  a_{\rm new} = a
  \left( \begin {smallmatrix} 1 & \\[4pt] & i \end{smallmatrix} \right)
  .
\end {equation}
Using the fact that $\uOmega$ is a diagonal matrix (\ref{eq:uOmega}),
the new expressions (\ref{eq:aanew}) are the same as the old (\ref{eq:aaold}).

The one important change is the overall minus sign in the prefactor of
(\ref{eq:AAnew}), but we have already accounted for it in
the main text when we decided in the formula (\ref{eq:Cprop})
for the $N{=}4$ propagator to rewrite the
overall factor $|x_1 x_2 x_3 x_4|$ as
$-x_1 x_2 x_3 x_4$.  For $xy\bar y\bar x$, for which
the $x_i$ are $\hat x_i$ (\ref{eq:xhat}), this had no effect because
\begin {equation}
   -\hat x_1 \hat x_2 \hat x_3 \hat x_4 =
   |\hat x_1 \hat x_2 \hat x_3 \hat x_4| .
\end {equation}
For $x\bar y y \bar x$, however, the $x_i$ are $x'_i$ (\ref{eq:xprime}),
and
\begin {equation}
   -x'_1 x'_2 x'_3 x'_4 =
   - |x'_1 x'_2 x'_3 x'_4| ;
\end {equation}
so using $-x_1 x_2 x_3 x_4$ instead of $|x_1 x_2 x_3 x_4|$ introduces
precisely the additional overall sign that we need to account for
in the prefactor of (\ref{eq:AAnew}).

% ----------------------------------------------------------------------------

\subsection {\boldmath$x \bar y \bar x y$}

A similar analysis applies to $x\bar y\bar x y$.
The analog of the $xy\bar y\bar x$ formula
(\ref{eq:IIxyyx1}) is
\begin {align}
   \left[\frac{dI}{dx\,dy}\right]_{x\bar y \bar x y}
   &=
   d_R^{-1} \tr(T_R^a T_R^b T_R^a T_R^b)
   \frac{\alphas^2 }{4 E^4}
   |\tilde x_1+\tilde x_4|^{-1} |\tilde x_3+\tilde x_4|^{-1}
   \int_{t_\xx < t_\ybx < t_\xbx < t_\yx}
   \sum_{h_\xx,h_\yx,h_\zx,h,\bar h}
   \int_{\B^\Ax,\B^\Bx}
\nonumber\\ &\times
   \bcalP_{h \to h_\zx,h_\yx}\bigl(1{-}x \to 1{-}x{-}y,y\bigr)
       \cdot \grad_{\B^\bx}
   \langle\B^\bx,t_\bx|\B^\Bx,t_\Bx\rangle
   \Bigr|_{\B^\bx=0}
\nonumber\\ &\times
   \bigl( \bcalP_{h_\ix \to \bar h, h_\yx}\bigl(1 \to 1{-}y,y\bigr)
          \cdot \grad_{\C_{12}^\Ax} \bigr)
   \bigl( \bcalP^*_{\bar h \to h_\zx,h_\xx}\bigl(1{-}y \to 1{-}x{-}y,x\bigr)
          \cdot \grad_{\C_{23}^\Bx} \bigr)
\nonumber\\ &\qquad\qquad
  \langle\C_{41}^\Bx,\C_{23}^\Bx,t_\Bx|\C_{34}^\Ax,\C_{12}^\Ax,t_\Ax\rangle
   \Bigr|_{\C_{12}^\Ax=0=\C_{23}^\xbx; ~ \C_{34}^\Ax=\B^\Ax; ~ \C_{41}^\xbx=\B^\xbx}
\nonumber\\ &\times
   \bcalP_{h_\ix \to h,h_\xx}\bigl(1 \to 1{-}x,x\bigr)
        \cdot \grad_{\B^\ax}
   \langle\B^\Ax,t_\bx|\B^\ax,t_\ax\rangle
   \Bigr|_{\B^\ax=0} ,
\end {align}
and the analog of (\ref{eq:IIxyyx2}) is
\begin {align}
   \left[\frac{d\Gamma}{dx\,dy}\right]_{x\bar y \bar x y}
   = &
   - \frac{\CA^2 \alphas^2 M_\ix \tilde M_\fx}{8 \pi^2 E^4}
   \frac{
     ( \alpha \delta^{\bar n n} \delta^{\bar m m}
     {+} \beta \delta^{\bar n \bar m} \delta^{nm}
     {+} \gamma \delta^{\bar n m} \delta^{n \bar m} )
   }{
     |x'_1 + x'_4| |x'_3 + x'_4|
   }
\nonumber\\ & \times
   \int_0^{\infty} d(\Delta t)
   \int_{\B^\Ax,\B^\Bx}
   \frac{B^{\xbx}_{n}}{(B^\xbx)^2} \,
   \frac{B^{\ybx}_{m}}{(B^\ybx)^2} \,
   \exp\bigl(
      - \tfrac12 |\tilde M_\fx| \tilde \Omega_\fx (B^\xbx)^2
      - \tfrac12 |M_\ix| \Omega_\ix (B^\ybx)^2
   \bigr)
\nonumber\\ &\qquad\times
   \nabla_{\C_{12}^\Ax}^{\bar m}
   \nabla_{\C_{23}^\Bx}^{\bar n}
   \langle\C_{41}^\Bx,\C_{23}^\Bx,\Delta t|\C_{34}^\Ax,\C_{12}^\Ax,0\rangle
   \Bigr|_{\C_{12}^\Ax=0=\C_{23}^\Bx; ~ \C_{34}^\Ax=\B^\Ax; ~ \C_{41}^\Bx=\B^\Bx}
\end {align}
with $\Delta t \equiv t_\Bx - t_\Ax$.  This result is related to
the other diagrams by the transformations given in
section \ref{sec:xyxy}.

There is also exactly the same sign issue as the one discussed above
for $x \bar y y\bar x$, which is accounted for in the same way
by our choice of using $- x_1 x_2 x_3 x_4$ instead of
$|x_1 x_2 x_3 x_4|$ in (\ref{eq:Cprop}).

% ============================================================================

\section{More details on comparison with Refs.\ \cite{Blaizot,Wu}}
\label{app:dbllog}

Here we give a few details on making contact with
Blaizot and Mehtar-Tani \cite{Blaizot} and Wu \cite{Wu}, which
explored energy loss in the limiting case
$y \lesssim x \ll 1$.  We will focus mostly on the comparison with
Wu, which uses language closest to our own.

% ---------------------------------------------------------------------------

\subsection {The double log region (\ref{eq:region})}

In Blaizot and Mehtar-Tani \cite{Blaizot} (50),%
\footnote{
   Similar to our convention for Wu
   (our footnote \ref{foot:Wuref}), we number equations from
   Blaizot and Mehtar-Tani \cite{Blaizot} according to the
   arXiv version of their paper (version 2).
}
the double log arises,
in their notation, from
$\hat q\tau \ll q^2 \ll p^2$.
Using their definition%
\footnote{
  They write $\tau \equiv \omega/q^2$ just before their (50), but the
  relevant frequency in their later discussion of radiation in section 4
  is what they call $\omega'$ there.
}
$\tau \equiv \omega'/q^2$, these conditions may be
rewritten as
\begin {equation}
   \frac{\omega'}{p^2} \ll \tau \ll \sqrt{\frac{\omega'}{\hat q}} \,.
\label {eq:regionBM}
\end {equation}
The relevant $p^2$ is identified as $\simeq \sqrt{\omega \hat q}$
after their eq.\ (77).
Their $\omega$, $\omega'$, and $\tau$ correspond to
our $xE$, $y E$, and $\Delta t$.  With this change of notation,
(\ref{eq:regionBM}) above becomes our (\ref{eq:region}).

In Wu \cite{Wu}, the second inequality of our (\ref{eq:region}) is
given explicitly in his (38).  The first inequality is
used implicitly in the expansion of his (43) into (44), which
assumes [in addition to the explicit conditions of Wu's (38)]
that $\omega_2 x_2^2/t_2 \ll 1$ (in Wu's notation).  Given that
$x_2^2 \sim 1/\hat q L$, the two conditions can be written as
\begin {equation}
   \frac{\omega_2}{\hat q L} \ll t_2 \ll \sqrt{\frac{\omega_2}{\hat q}} .
\label{eq:regionWu}
\end {equation}
In our notation, $\omega_2$ and $t_2$ are $y E$ and $\Delta t$.
In our thick-media approximation, the relevant $L$ is the formation
length $\sim \sqrt{xE/\hat q}$ of the $x$ gluon, which then makes
(\ref{eq:regionWu}) above equivalent to our (\ref{eq:region}).

In our own work, the second inequality of (\ref{eq:region})
corresponds to $\max(\Omega_+\,\Delta t, \Omega_-\,\Delta t) \ll 1$
and so corresponds to the small-$\Delta t$ expansions of
(\ref{eq:XYZsim}) [but one must include more terms of the
expansion than shown there to obtain (\ref{eq:dG1})].
In contrast, the first inequality of (\ref{eq:region})
guarantees that $X_\yx X_\ybx - X_{\yx\ybx}^2$ is {\it not}\/
given by its small-$\Delta t$ expansion (\ref{eq:detXsim})
in the cases of
\begin {equation}
   (x_1,x_2,x_3,x_4) = (-1,y,1{-}z{-}y,z) = (-1,y,x,1{-}x{-}y)
\label{eq:x1234zyyz}
\end {equation}
for $zy\bar y\bar z$
[given by $x \to z \equiv 1{-}x{-}y$ in (\ref{eq:xhat})]
and
\begin {equation}
   (x_1,x_2,x_3,x_4) = \bigl(-(1{-}y),-y,1{-}z,z\bigr)\
   = \bigl(-(1{-}y),-y,x{+}y,1{-}x{-}y\bigr)
\end {equation}
for $z\bar y y\bar z$
[similarly from (\ref{eq:xprime})].
The different behavior of $X_\yx X_\ybx - X_{\yx\ybx}^2$ is due
to an approximate cancellation of the leading terms when
$y \ll x \ll 1$.

% ---------------------------------------------------------------------------

\subsection {\boldmath$y \lesssim x \ll 1$ formulas of Ref.\ \cite{Wu}}
\label {app:Wu}

Let's focus on our $zy\bar y \bar z$, which
corresponds to the second diagram in Wu's eq.\ (30) for his $e_1$
contribution to energy loss.  Wu's formula for our $d\Gamma/dx\,dy$,
extracted from his (33),
would be
\begin {align}
   \left[ \frac{dI}{dx\,dy} \right]_{zy\bar y\bar z}
   =
   \frac{\alphas^2 \Nc \CA E^2}{2\omega_1^3 \omega_2^3}
   &
   \int_{z_1<z_2<z_3<z_4} \int d^2\x_2\,d^2\x_3
\nonumber\\ &\quad
   \grad_{\x_1}\cdot\grad_{\x_4}
   \Bigl[
     G^{(3)}(\x_4,z_4;\x_3,z_3;\omega_1) \,
     G^{(3)}(\x_2,z_2;\x_1,z_1;\omega_1)
   \Bigr]^{\x_1=0=\x_4}
\nonumber\\ &\quad \times
   \grad_{\B_2}\cdot\grad_{\B_3}
   G^{(4)}(\x_3,\B_3,z_3;\x_2,\B_2,z_2;\omega_1,\omega_2)
   \Bigr|^{\B_2=\x_2,\B_3=0} ,
\label {eq:Wu}
\end {align}
where the notation on the left-hand side is ours and the notation on the
right-hand side is his, except that the first integral sign is just our
way of writing the $\int dz_4\,dz_3\,dz_2\,dz_1$ integration.
$G^{(3)}$ and $G^{(4)}$ are given by Wu (14--16) and (20).
A translation
table between Wu's notation and the $y \lesssim x \ll 1$ limit of
our notation is given in table \ref{tab:Wu}.  When comparing to
our formulas in the main text, keep in mind that the 4-particle $x_i$ for
$zy\bar y\bar z$ are given by (\ref{eq:x1234zyyz}) and not (\ref{eq:xhat}).
Using the translations of table \ref{tab:Wu}, one may verify that
Wu's formula (21) for relating $(\tilde\x,\tilde\B)$ to $(\x,\B)$ is
equivalent to the $y \lesssim x \ll 1$ limit of
our formula (\ref{eq:changef})
[with (\ref{eq:Cmodes}) and (\ref{eq:af})]
for relating $\A_\pm$ to $(\C_{34},\C_{12})$.

\begin {table}
\tabcolsep 10pt
\begin {tabular}{|c|l|}
\hline
  Wu & $y \lesssim x \ll 1$ limit of our $zy\bar y\bar z$ \\
\hline
  $z$ & $t$ \\
  $\omega_1$ & $xE$ \\
  $\omega_2$ & $yE$ \\
  $(K_1,K_2)$ & $(\Omega_-,\Omega_+)$ \\
  $(\Omega_1,\Omega_2)$ &
      $(\frac{2}{\sqrt3}\,\Omega_-,\Omega_+)$ when $y \ll x \ll 1$ \\
  $(\x,\B)$ &
      $(\b_{34},\b_{24}) =
       \bigl((x_3{+}x_4)\C_{34},-x_1\C_{12}{+}x_3\C_{34}\bigr)
       \simeq (\C_{34},\C_{12})$  \\
  $m_1^{1/2} \tilde\x$ & $\A_-$ \\
  $m_2^{1/2} \tilde\B$ & $\A_+$ \\
  $G^{(3)}$ & $\langle \B,t | \B',0 \rangle$ \\
  $(m_1 m_2)^{-1} G^{(4)}$ &
      $\langle \A_+,\A_-,t | \A'_+,\A'_-,0 \rangle$ \\
\hline
\end {tabular}
\caption
    {
    \label {tab:Wu}
    Some translations between Wu \cite{Wu} and the
    $y \lesssim x \ll 1$ limit of the $zy\bar y z$ case
    of this paper.
    }
\end {table}

With these translations, we find that the $y \ll x \ll 1$ limit
for the $zy\bar y\bar z$ version of (\ref{eq:IIxyyx}),
combined with (\ref{eq:Cprop1}),
reproduces (\ref{eq:Wu}).
However, for $y \sim x \ll 1$, we find some differences.
One difference is that Wu (20) is missing a factor of the Jacobean
of the transformation between the variables $(\x,\B)$ and
$(\tilde\x,\tilde\B)$ [as can be seen by taking the $z_3 \to z_2$ limit,
in which all the Green functions become $\delta$-functions].
The corrected relation can be written in the form
\begin {multline}
   G^{(4)}(\x_3,\B_3,z_3;\x_2,\B_2,z_2;\omega_1,\omega_2)
   \simeq
\\
   \frac{m_1 m_2}{\omega_1 \omega_2} \,
   G(\tilde\x_3,\tilde\x_2,z_3{-}z_2,m_1,K_1) \,
   G(\tilde\B_3,\tilde\B_2,z_3{-}z_2,m_2,K_2) .
\end {multline}
Another difference appears to be in the treatment of the triple
gluon vertex in Wu (30).  We find multiple terms, associated with the
different helicity amplitudes in our section \ref{sec:dH}
and their combination into (\ref{eq:abcdef}).
The derivative associated with that vertex (our $\grad_{\C_{23}^\yx}$)
is also more complicated than Wu's $\grad_{\B_2}$.
Finally, the two 3-particle evolution frequencies in
(\ref{eq:Wu}) should not both be $\omega_1$ when $x \sim y$.
Altogether, our result
for $y \lesssim x \ll 1$ replaces
(\ref{eq:Wu}) by
\begin {align}
   \left[ \frac{dI}{dx\,dy} \right]_{zy\bar y\bar z}
   \simeq
   &
   \frac{\alphas^2 \Nc \CA E^2}{2\omega_1^2 \omega_2^2 (\omega_1{+}\omega_2)^2}
   \left(
     \frac{\delta^{\bar n n} \delta^{\bar m m}}{\omega_1}
     - \frac{\delta^{\bar n \bar m} \delta^{nm}}{\omega_1{+}\omega_2}
     + \frac{\delta^{\bar n m} \delta^{n \bar m}}{\omega_2}
   \right)
   \int_{z_1<z_2<z_3<z_4} \int d^2\x_2\,d^2\x_3
\nonumber\\ &
   \nabla_{\x_1}^m \nabla_{\x_4}^{\bar n}
   \Bigl[
     G^{(3)}(\x_4,z_4;\x_3,z_3;\omega_1) \,
     G^{(3)}(\x_2,z_2;\x_1,z_1;\omega_1{+}\omega_2)
   \Bigr]^{\x_1=0=\x_4}
\nonumber\\ &\times
   (\omega_1 \nabla_{\B_2}^n - \omega_2 \nabla_{\x_2}^n) \nabla_{\B_3}^{\bar m}
   G^{(4)}(\x_3,\B_3,z_3;\x_2,\B_2,z_2;\omega_1,\omega_2)
   \Bigr|^{\B_2=\x_2,\B_3=0} .
\end {align}
Note that, in momentum space, the combination
$\omega_1 \grad_{\B_2} - \omega_2 \grad_{\x_2}$ above is
(proportional to) simply one of our rotationally-invariant
combinations $\P_{ij} \equiv x_j\p_i-x_i\p_j$ of the transverse
momenta associated
with that vertex.

%[[Give any discussion how to extract the corresponding piece of
%$\delta\hat q$ from Wu??]]

% ---------------------------------------------------------------------------

\subsection {Problem with comparing the double log contribution
             of \boldmath$xy\bar y\bar x+x\bar y y\bar x$}
\label {app:xyyxWu}

\begin {figure}[t]
\begin {center}
  \includegraphics[scale=0.3]{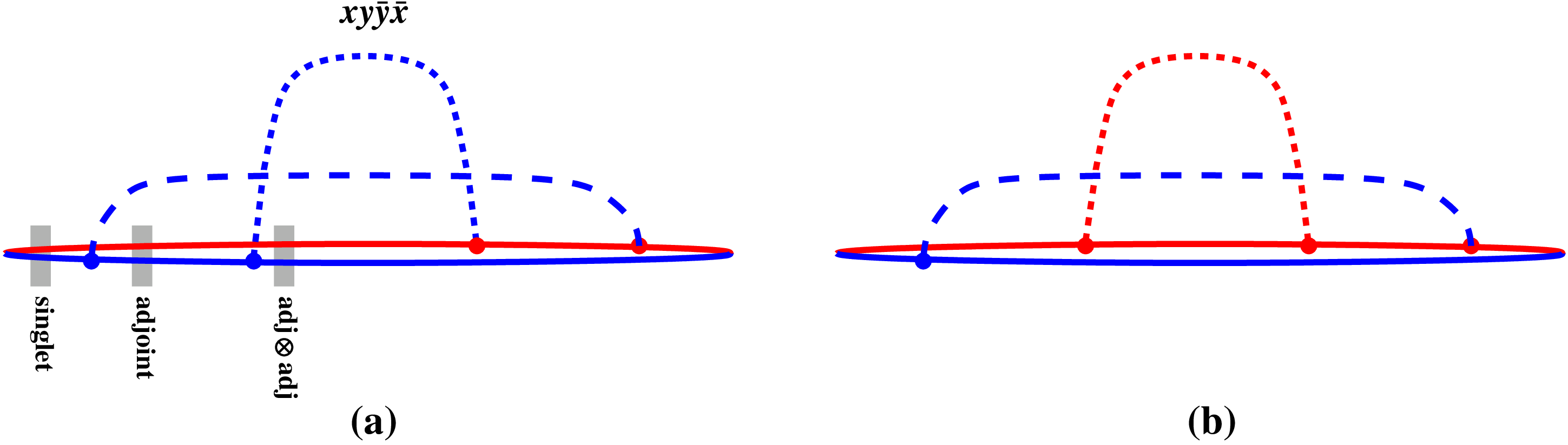}
  \caption{
     \label{fig:color}
     (a) A depiction of possible total dipole color at different times for
     $xy\bar y\bar x$ for the case $y \lesssim x \ll 1$.
     (b) A virtual correction to single bremsstrahlung
     that conspires to suppress
     the range of dipole color possibilities in energy loss calculations.
  }
\end {center}
\end {figure}

There is a subtle problem comparing (i) our results for
the $xy\bar y\bar x+x\bar y y\bar x$ contribution to the double
log to (ii) corresponding pieces of refs.\ \cite{Blaizot,Wu}.
(These contributions correspond respectively
to the second diagram in Wu (31) \cite{Wu}
and the third diagram in Wu (32), and they correspond to
$B_3$ and $C_2$ of Blaizot and Mehtar-Tani \cite{Blaizot}.)
A simple indication of the problem can be found by comparing
(i) our potential (\ref{eq:Vlarge4}) applied to the $x,y \ll 1$ limit
of $xy\bar y\bar x$ or $x\bar y y\bar x$ to (ii) the potential of
Wu (17).  Consider $xy\bar y\bar x$, where our notational convention
(\ref{eq:xhat}) was that $\b_2$ is the position of the $y$ gluon and
$\b_4$ is the position of the $x$ gluon.
In the $x,y \ll 1$ limit, the two harder particles stay extremely close
together (relative to other transverse distance scales), corresponding
to $\b_1\to\b_3$.  Adopting Wu's convention of using $\x$ for the
separation of the $x$ parton from the hard partons
(our $\b_{41} \simeq \b_{43}$ in this case), and
$\B$ for the $y$ parton from the hard partons
(our $\b_{21} \simeq \b_{23}$ here),
the corresponding limit of (\ref{eq:Vlarge4}) is
\begin {equation}
   V \simeq
  - \frac{i \hat q_{\rm A}}{8}
  ( 2\x^2 + 2\B^2 ) .
\end {equation}
This does not agree with Wu's
\begin {equation}
   V \simeq
  - \frac{i \hat q_{\rm A}}{8}
  \bigl[ \x^2 + \B^2 + (\x-\B)^2 \bigr] .
\label {eq:VWu}
\end {equation}
The reason for the discrepancy is that Wu assumes that the two hardest
particles form a color dipole with total color in
the {\it adjoint}\/ representation A.
When those particles are gluons, however, they could in principle
form a dipole in any color state contained in
A$\otimes$A, which contains many more possibilities than
A itself (especially in the large $\Nc$ limit).
This situation is depicted in fig.\ \ref{fig:color}a (which is
drawn by distorting fig.\ \ref{fig:xyyxrate} into a style
similar to Wu's diagrams to depict relative transverse separations).
In this picture, the initial singlet dipole first emits a gluon and must
become an adjoint dipole to balance color.
The emission of the second gluon
in principle allows the dipole to become anything in A$\otimes$A.  However,
all of these possibilities other than A itself
will be suppressed by the small dipole
size once we also add in virtual processes where the $y$ boson
connects to the other particle in the dipole, as in fig.\ \ref{fig:color}b.
Since in this paper we are just computing $d\Gamma/dx\>dy$ (for fixed
$x$ and $y$) and not
energy loss, we have not included virtual processes,
and so the other color possibilities are not suppressed.
The color dynamics of our calculation is therefore
different than Wu's, which
makes it difficult to directly compare partial results.
(The same holds for comparison to Blaizot and Mehtar-Tani.)

\begin {figure}
\begin {center}
  \includegraphics[scale=0.3]{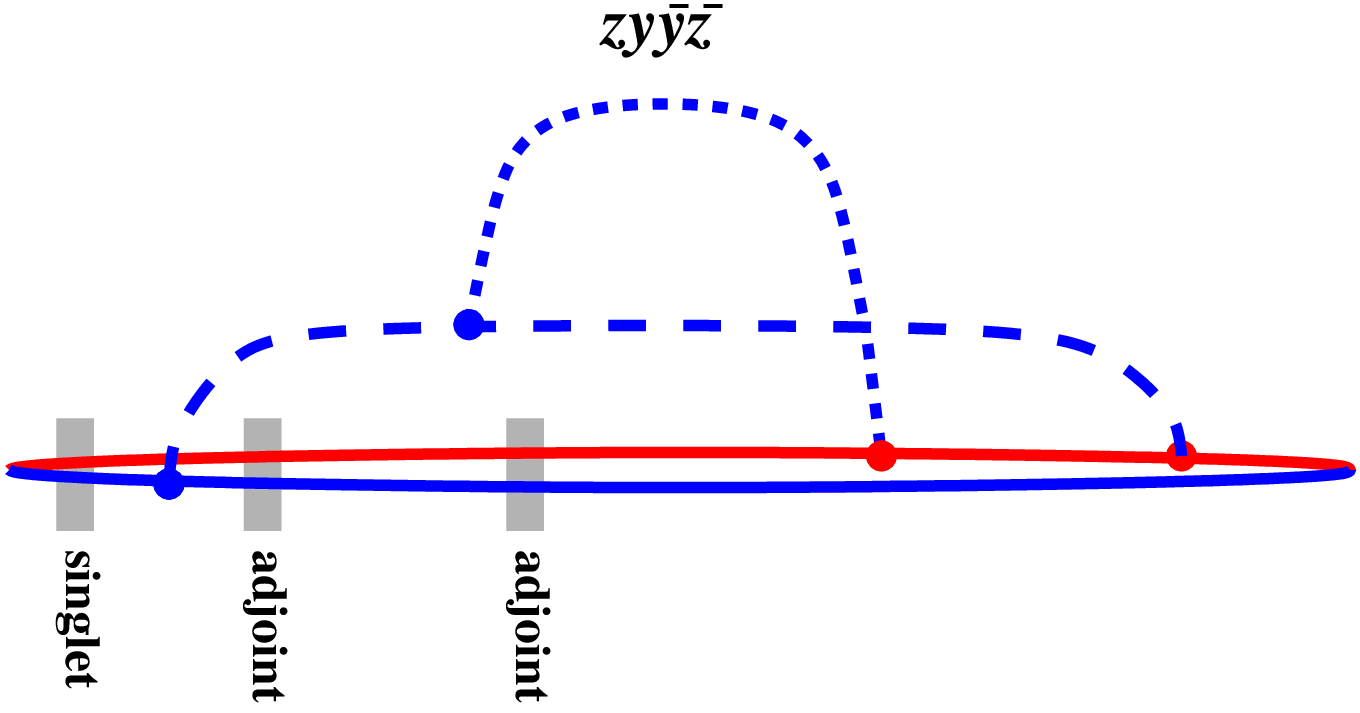}
  \caption{
     \label{fig:colorz}
     Similar to fig.\ \ref{fig:color}a but for $zy\bar y\bar z$
     (fig.\ \ref{fig:zyyz}).
  }
\end {center}
\end {figure}

For $z y\bar y\bar z$, however, all is well, as depicted in
fig.\ \ref{fig:colorz}: In the $x,y \ll 1$ limit, the
calculation of $[d\Gamma/dx\,dy]_{z y\bar y\bar z}$ is also restricted to an
adjoint color dipole at intermediate times.
On a related note, following
the numbering convention of (\ref{eq:x1234zyyz}),
$\b_4 \to \b_1$ for $x,y \ll 1$, and our $x$ and $y$ gluons
correspond to $\b_3$ and $\b_2$ respectively.
Wu's $\x$ is then our $\b_{34} \simeq \b_{31}$, and his $\B$
is our $\b_{21} \simeq \b_{24}$.
In this limit, our potential
(\ref{eq:Vlarge4}) then indeed agrees with Wu's potential (\ref{eq:VWu}).

%%%%%%%%%%%%%%%%%%%%%%%%%%%%%%%%%%%%%%%%%%%%%%%%%%%%%%%%%%%%%%%%%%%%%%%%%%%%%%

\end {document}